\documentclass[12pt]{article}
\usepackage[letterpaper,margin=1in]{geometry}
\usepackage{newtxtext,newtxmath}
\usepackage{graphicx}
\usepackage{scicite}
\usepackage{siunitx}
\usepackage{xcolor}

\renewenvironment{abstract}{\begin{quote}\bfseries}{\end{quote}}

\newif\ifsubmission
\submissiontrue

\newcommand{\papertitle}{Splenic retention of echinocytes in stored blood and disease}
\title{\papertitle}
\author{\normalsize Zhaojie Chai$^{1\dagger}$, Jianlu Zheng$^{2\dagger}$, He Li$^{3}$, Ming Dao$^{2}$,
George Em Karniadakis$^{1}$\\[4pt]
\normalsize{$^{1}$Division of Applied Mathematics, Brown University, Providence, RI 02912, USA.}\\
\normalsize{$^{2}$Department of Materials Science and Engineering, Massachusetts Institute of Technology,}\\
\normalsize{Cambridge, MA 02139, USA.}\\
\normalsize{$^{3}$School of Chemical, Materials and Biomedical Engineering, University of Georgia,}\\
\normalsize{Athens, GA 30602, USA.}\\
\normalsize{$^\dagger$These authors contributed equally to this work.}}
\date{}

\begin{document}

\maketitle

\begin{abstract}
Stored red blood cells become echinocytes, growing spicules before losing membrane; the spleen clears the resulting
small, spherical cells. Whether spicules alone hinder splenic passage is unknown. Shape and area change together. In a
two-component model whose shrinking cytoskeleton raises spicules, we drove a discocyte and stage I--III echinocytes
of equal area and volume through a model splenic slit. The pressure drop needed to pass rose with stage; stage III
echinocytes were held at 37 to 50 times that of the discocyte. Bilayer--cytoskeleton links stretched near full
extension held these cells; lengthening them about twofold released a held cell. Breakable links failed in held stage
III echinocytes at pressure drops rising with strength; discocytes lost none. Capillary-sized channels separated
the stages far less. At high pressure drops the model held only storage-induced microerythrocytes, which perfused
spleens clear most. Spicules, besides membrane loss, may decide which stored cells the spleen retains.
\end{abstract}

\noindent\textbf{Teaser:} Stage III echinocytes stay held in a model splenic slit by stretched bilayer--cytoskeleton
links; stages I and II pass.

\section*{Introduction}
Red blood cells stored for transfusion turn into echinocytes~\cite{Roussel2017}. Echinocytes progress from discs
with an irregular contour (stage I) through crenated discs (stage II) to compact cells covered with spicules (stage III)
and spheroechinocytes~\cite{Reinhart1986,Wen2025}. The spicules come before the loss of membrane:
lysophosphatidylcholine, which first accumulates in the outer leaflet of the bilayer, turns red cells into echinocytes
and then, as they release vesicles, into spherocytes that have lost membrane area at nearly constant
volume~\cite{Safeukui2012}. Vesicles of stored blood bud from the tips of the spicules and carry proteins of the
bilayer but not spectrin, the main protein of the cytoskeleton~\cite{Greenwalt2006}, so a patch of bilayer must come
free of the cytoskeleton before it can leave the cell. In stored blood, stage III echinocytes, spheroechinocytes and spherocytes form a
subpopulation of small cells, the storage-induced microerythrocytes (SMEs), whose projected area is 15 to 20\% below
that of the other cells~\cite{Roussel2017}. Counted by imaging flow cytometry, their share rose from about 5\% of the
cells after 3 days of storage to about a quarter after 42 days~\cite{Roussel2017,Roussel2021}, and microparticles
carrying glycophorin A, a red-cell membrane protein, accumulated in the stored blood with the same time
course~\cite{Roussel2017}. Perfused through isolated human spleens, 61\% of these cells were cleared within 70
minutes, against 15\% of all the stored cells~\cite{Roussel2021}. In 31 healthy volunteers, units with a larger share
of these cells tended to leave fewer of the transfused cells in circulation 24 hours later (Spearman $r =
-0.42$)~\cite{Roussel2021}. Metabolic rejuvenation of stored cells lowered their share from 28\% to 9.5\%, mostly by
restoring the shape of the cells~\cite{Marin2021}. Spicules and membrane loss thus arrive together in the cells that
the spleen clears. Echinocytes also circulate in disease, as in pyruvate kinase deficiency after
splenectomy~\cite{Leblond1978a} and in sickle cell anemia~\cite{Warth1984} (Fig.~\ref{fig:exp}A).

The spleen removes red cells that can no longer squeeze through the interendothelial slits of its venous sinuses. On
electron micrographs these slits are about \SI{1.9}{\micro\meter} long, with apertures of 0.25 to
\SI{1.2}{\micro\meter}~\cite{Deplaine2011}. Cells held at the slits can then be taken up by
macrophages~\cite{Li2018,chai2026multiscale}.
In isolated perfused human spleens, the retention of red cells rose with the area they had lost, and more than 90\% of
the cells that had lost more than 18\% of their area were retained~\cite{Safeukui2012}. Filters of packed microbeads
that mimic the slits retain heated, parasitized and spherocytic cells, and retained parasitized cells at the rate of
perfused spleens~\cite{Deplaine2011}. In microfluidic slits, red cells passed apertures as narrow as
\SI{0.28}{\micro\meter} at \SI{37}{\degreeCelsius}, provided that their area was large enough, for their volume, to
form two equal spheres joined by a tether~\cite{Moreau2023}. Simulations of single cells in a model slit tie passage
to the area and volume of
the cell. A cell that straddles the slit must hold its volume in two lobes bounded by its membrane, which gives a
smallest area for a given volume~\cite{Pivkin2016,Li2018}. In the same slit, Du et al.~\cite{Du2026} gave model cells
the sphericity index and the shear modulus measured on red cells fatigued by repeated extrusion. These cells passed or
were held according to their sphericity index, which is set by area and volume, and a fourfold rise in the modulus
hardly mattered.

These limits treat the cell as a smooth envelope of given area and volume. In an echinocyte, however, part of the
bilayer is folded into spicules over a cytoskeleton that has shrunk below the area of the bilayer~\cite{Wen2025}. In
filters with \SI{2.6}{\micro\meter} pores, echinocytes of stages II and III did not differ from control cells, and
Reinhart and Chien concluded that sphericity, more than shape, appears to determine passage through narrow
channels~\cite{Reinhart1986}. In a network of microchannels 5 to \SI{70}{\micro\meter} wide, red cells turned into
stage III echinocytes by sodium salicylate flowed a quarter more slowly than discocytes, which Piety et al.\
attributed to the flow in the wider channels and not to the passage of single cells through the narrowest
ones~\cite{Piety2016}. The openings of both tests were at least twice as wide as the apertures of the slits.
Pivkin et al.\ noted that the coupling between the bilayer and the cytoskeleton calls for a two-component model of the
membrane~\cite{Pivkin2016}; with such a model, Li et al.\ found that cells with weakened links between the two layers
shed membrane as they cross the slit, and cells with a weakened cytoskeleton break up~\cite{Li2018}. In a
coarse-grained model, the periodic spectrin--actin skeleton of axons~\cite{chai2022periodic} is likewise severed under
extension once its actin--spectrin links dissociate~\cite{chai2023dynamics}. In experiments, shape and membrane area
change together~\cite{Reinhart1986,Safeukui2012,Du2026}. Whether the spicules alone hold a cell in the slits is
therefore not known.

A second question concerns the links between the two layers, which have a finite strength. A tether pulled from a red
cell carries bilayer but no cytoskeleton~\cite{WaughBauserman1995}, and separating the two layers in this way takes an
energy of 0.05 to \SI{0.078}{\milli\joule\per\meter\squared}~\cite{Hwang1997,Butler2008}. The bond between spectrin and
ankyrin, which joins the cytoskeleton to band 3 in the bilayer, broke at a mean force of 21 to \SI{48}{\pico\newton} as
the loading rate rose a hundredfold~\cite{Saito2015}. Zhu et al.\ let the links fail above a critical stress, but in
healthy cells the stress that pulled the two layers apart in a slit stayed far below it~\cite{Zhu2017}, and Amir Aslan
Pour and Mao let links fail in healthy cells in an extensional flow~\cite{AmirAslanPour2026}. To our knowledge, no study
has let the links of an echinocyte fail as the cell meets a slit. In an echinocyte the links are
sparse~\cite{Wen2025}, so they might fail at pressure drops that leave a discocyte intact.

We have combined experiments with particle models to quantify stomatocytes and reticulocytes in health and disease,
the red cells of Gaucher disease and the vacuoles of pocked
cells~\cite{chai2026stomatocyte,chai2026quantifying,chai2025silico,chai2026gaucher,chai2026adhesion,chai2026pocked}. Here we ask
whether, at fixed area and volume, the echinocyte stage changes the passage of red cells through the splenic slits,
and how the loss of membrane adds to it (Fig.~\ref{fig:exp}). We build the discocyte (D) and the echinocyte stages I
to III (E-I to E-III) with the two-component model of Wen et al.~\cite{Wen2025} at the area and volume of the
discocyte, and remove bilayer area from smooth and spiculated cells. The model cells are first driven through
micropores like the filter pores of Reinhart and Chien and then through a model interendothelial slit, with the
pressure drop raised in steps. In the micropores the stages need different pressure drops, but all pass below those
measured across such filters. At fixed area and volume, the pressure drop that a cell needs to pass the slit rises
along the sequence, and stage III echinocytes stay held at pressure drops far above those that pass the discocyte.
Spiculated cells are held well below the area-to-volume limit that holds smooth cells, by links between the bilayer
and the cytoskeleton that are already close to full extension; when these links are lengthened in a held stage III
echinocyte, the cell passes. In a model of the constriction channels of our device, which are of capillary size, the
stages keep their order but differ far less. Weighted by the shape classes of stored blood, the
slit holds at higher pressure drops only the storage-induced microerythrocytes, and reversing the spicules eases
passage only for cells that have not lost enough membrane to be held as smooth cells. Finally, we let the links fail
once they are stretched past a set length, at three strengths around the one we estimate from the separation energy.
The links of held stage III echinocytes then fail at pressure drops that rise with their strength, the weakest even at
rest, while discocytes lose none.

\section*{Results}

\subsection*{Discocytes and echinocytes of every stage reach the constriction channels together}
\begin{figure}[htbp]
\centering
\includegraphics[width=\textwidth,height=0.5\textheight,keepaspectratio]{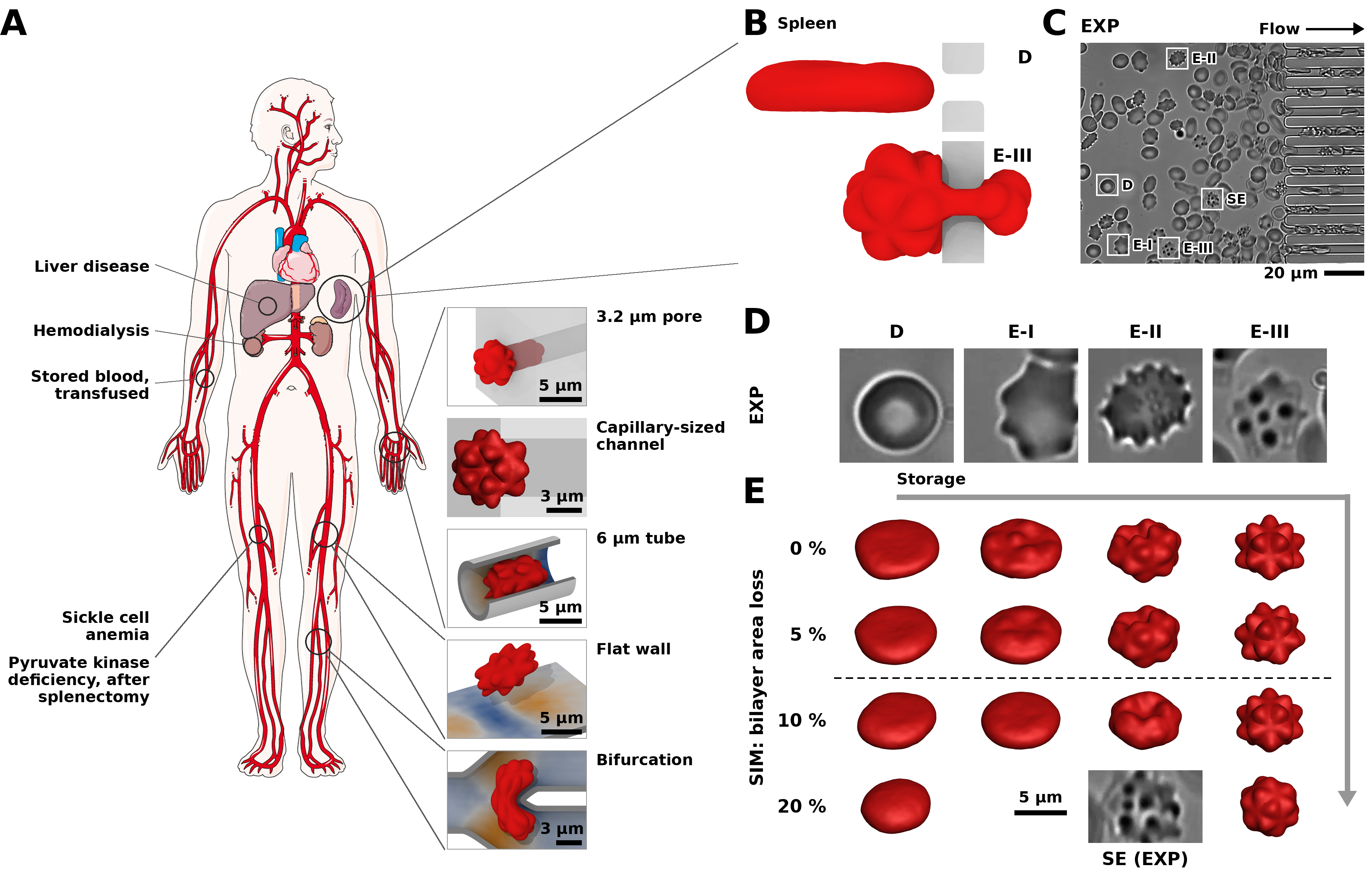}
\caption{\textbf{Stored red cells grow spicules and then lose membrane.}
(A) Echinocytes in stored blood and in disease, and the model set-ups (B, Figs.~\ref{fig:micropore}
and~\ref{fig:chip}, Supplementary Text), each with E-III (at the bifurcation, with the modulus of D); colored walls,
the time-averaged wall shear stress (orange, above the value without a cell; blue, below). Body and organ drawings
adapted from Servier Medical Art (smart.servier.com), CC BY 4.0. (B) The model slit, \SI{4.0}{\micro\meter} wide and
\SI{1.2}{\micro\meter} high,
with rounded edges: D at the start of a run, and E-III held at \SI{75}{\pascal} (side views, wall cut at
the mid-plane of the slit; Fig.~\ref{fig:slit}). (C) Bright-field frame of the device (EXP): chamber and channel
entrances (5 $\times$ \SI{5}{\micro\meter}); boxes, the cells of D and SE. Scale bar,
\SI{20}{\micro\meter}. (D) Cells enlarged from C: a discocyte (D) and echinocytes of stages I to III (E-I to E-III),
classified by eye. (E) The model cells (SIM; oblique view), stage in columns and bilayer area removed in rows: at the
membrane area and volume of D (0\%) and with 5, 10 and 20\% less bilayer area. Gray arrow, the order of the two
changes: spicules first, then membrane loss as vesicles are released~\cite{Safeukui2012,Roussel2017}. Dashed line,
$V/V_2 = 1$; cells above it have $V/V_2 < 1$. SE, a spheroechinocyte from the same frame, beside the model
E-III with 20\% less bilayer area. D, E and SE share one scale; scale bar, \SI{5}{\micro\meter}.}
\label{fig:exp}
\end{figure}

Blood from patients with sickle cell anemia (Materials and Methods), driven through a microfluidic device, carried discocytes and crenated cells side by side upstream of its constriction channels, and
crenated cells were seen inside the channels (Fig.~\ref{fig:exp}C). The crenated cells ranged from discs with an
irregular contour (E-I) and crenated discs (E-II) to compact cells covered with spicules (E-III) and smaller,
spherical ones, the spheroechinocytes (SE) (Fig.~\ref{fig:exp}, C to E). The model cells of the next section follow the same sequence at the same scale (Fig.~\ref{fig:exp}E, top
row). At the area and volume of D, the longest axis of the face-on outline shortens from \SI{8.5}{\micro\meter} in D
to \SI{6.9}{\micro\meter} in E-III, as the spicules lengthen and the cell thickens. E-III with 20\% of its bilayer
area removed is \SI{6.1}{\micro\meter} across.

In storage, a cell moves along the top row of Fig.~\ref{fig:exp}E as its spicules grow, and down its column as it
loses membrane~\cite{Safeukui2012,Roussel2017}. Losing membrane at constant volume raises $V/V_2$, where $V_2 = 2
\cdot \tfrac{4\pi}{3} (A/8\pi)^{3/2}$ is the largest volume that two spheres of total area $A$ can hold. A cell that
straddles a thin wall with half of its volume on each side needs $V \le V_2$ unless its membrane stretches. Every
stage crosses $V/V_2 = 1$ between 5 and 10\% loss (dashed line in Fig.~\ref{fig:exp}E). We asked which of the two
changes holds a cell in a model splenic slit (Fig.~\ref{fig:exp}B).

\subsection*{Shrinking the cytoskeleton raises spicules and stretches the bonds beneath them}
In the two-component model of Wen et al.~\cite{Wen2025}, the bilayer and the cytoskeleton are two networks joined by
connecting bonds, or links (Fig.~\ref{fig:model}A; Materials and Methods). With every vertex pair connected, the two networks
relax together into a discocyte \SI{8.4}{\micro\meter} wide in its mid-plane section (Fig.~\ref{fig:model}E), which
is \SI{2.2}{\micro\meter} thick at the rim and \SI{1.7}{\micro\meter} at the center. The echinocytes keep only the
bonds on the edges of an icosahedron, which leaves 20 patches of bilayer free, and their cytoskeleton shrinks. With
the shrinkage and the shear moduli of stages E-I to E-III, each free patch lifts off the cytoskeleton into a spicule
(Fig.~\ref{fig:model}B). The spicules lengthen with the stage. Their tips stand on average 0.47, 0.88 and
\SI{1.28}{\micro\meter} above the cytoskeleton, along its normal (Fig.~\ref{fig:model}C). The bilayer area meanwhile
stays within 0.3\% of $A_0$. E0, which has the connection pattern and the settings of the echinocytes but neither their
cytoskeleton shrinkage nor their shear modulus, stays discoid. No bilayer vertex lies more than
\SI{0.34}{\micro\meter} from its partner on the cytoskeleton.

The shrinkage also stretches the connecting bonds before the cell meets any constriction. A connecting bond cannot be
longer than its contour length, $l_c = 2.2 \times \SI{0.03}{\micro\meter}$ (Materials and Methods). At rest, the
median bond was $0.53\,l_c$ long in D and $0.51\,l_c$ in E0. It was 0.60, 0.70 and $0.81\,l_c$ in E-I, E-II and
E-III, both with their own shear moduli and with that of D (Fig.~\ref{fig:model}D).

The cytoskeleton shear modulus rises from \SI{2.5e-8}{\newton\per\meter} in D to \SI{9.1e-6}{\newton\per\meter} in
E-III (table~\ref{tab:model}). We therefore also ran the echinocyte stages with the shear modulus of D, which separates
the effect of the shape from that of the modulus. Membrane loss is modeled by removing bilayer area at the same volume
(Fig.~\ref{fig:exp}E and Materials and Methods). The smooth family (Da025 to Da20) has 2.5 to 20\% less bilayer area than D, and the
spiculated family (E3ka05 to E3ka20) 5 to 20\% less than E-III.

\subsection*{A micropore ranks the stages, but all pass below the pressure drops of filters}
Reinhart and Chien filtered red cells along the stomatocyte--echinocyte sequence through filters
\SI{13}{\micro\meter} thick, with pores 2.6, 4.5 and \SI{6.9}{\micro\meter} in diameter~\cite{Reinhart1986}. In the
narrowest pores, the filtration resistance of echinocytes made with 30~mM salicylate, mostly stages II and III, did
not differ significantly from that of control cells. The authors concluded that sphericity, not shape, appears to be
the more important determinant of passage through narrow channels. We asked whether single model cells at fixed area
and volume pass such a pore regardless of their spicules. At fixed area and volume, D and the echinocyte stages cannot
form a capsule narrower than 2.89--\SI{3.00}{\micro\meter}. They therefore cannot pass a \SI{2.6}{\micro\meter} pore
that is \SI{13}{\micro\meter} long. We used a \SI{3.2}{\micro\meter} pore instead. Reinhart and Chien suggested that
cells pass mainly through the wider pores of their filters: 12\% of the nominal \SI{2.6}{\micro\meter} pores are
wider than \SI{2.8}{\micro\meter}.

Stages II and III needed higher pressure drops to pass than D (Fig.~\ref{fig:micropore}A; every run in
fig.~\ref{fig:poreruns}). D and E0 were held at
\SI{1}{\pascal} and passed at every pressure drop tried from 2 to \SI{150}{\pascal}, and E-I was held at 1 and
\SI{2}{\pascal} and passed from 5 to \SI{150}{\pascal} (two runs each at 1 to \SI{20}{\pascal}). At
\SI{200}{\pascal} a bond of the membrane of D
exceeded its maximum length before the cell was held or passed, and the run failed. E-II was held at 20 and \SI{35}{\pascal}, and in three
of four runs at \SI{42}{\pascal}. It passed at \SI{50}{\pascal}. E-III was held in all four runs up to
\SI{125}{\pascal} and in two of four at \SI{150}{\pascal}. It passed at 175 and \SI{200}{\pascal}. Stages II and III
thus needed 42--\SI{50}{\pascal} and about \SI{150}{\pascal} to pass, against 1--\SI{2}{\pascal} for D.

The shape and the shear modulus both contribute to the hold, in different proportions in the two stages. With the
shear modulus of D, E-II was held up to \SI{5}{\pascal} and passed at \SI{10}{\pascal} (two runs each). Its shape
alone thus raises the pressure drop it needs from 1--2 to 5--\SI{10}{\pascal}, and the larger shear modulus of its
cytoskeleton raises it further, to 42--\SI{50}{\pascal}. E-III with the shear modulus of D was still held
at \SI{100}{\pascal} and passed at \SI{125}{\pascal}, so most of its hold comes from its shape. In the pore, the
bilayer of E-III runs ahead of its shrunken cytoskeleton (Fig.~\ref{fig:micropore}, B and D). At \SI{50}{\pascal} the
radial gap between the two layers grows to \SI{0.59}{\micro\meter} toward the front, and the front of the cytoskeleton
lies \SI{0.84}{\micro\meter} behind that of the bilayer, whereas the two layers of D stay together. E-II, which passes
at \SI{50}{\pascal}, opens a gap of at most \SI{0.21}{\micro\meter} (Fig.~\ref{fig:micropore}, B and C). Rounder pore
edges lowered the pressure drops needed by E-II and E-III but kept the order of the cells (fig.~\ref{fig:edge}); with
an edge radius of \SI{1.0}{\micro\meter}, E-II needed 10--\SI{20}{\pascal} and E-III 75--\SI{100}{\pascal}.

Membrane loss holds a cell in another way. A discocyte with 5\% less bilayer area at the same volume, whose
narrowest capsule (\SI{3.23}{\micro\meter}) is about the pore diameter, was held at \SI{10}{\pascal} and passed at
\SI{20}{\pascal}. With 10\% less, the narrowest capsule is \SI{3.48}{\micro\meter}, wider than the pore. This cell
was held at \SI{50}{\pascal} with every edge radius, and at \SI{100}{\pascal} with the edge radius of
\SI{0.3}{\micro\meter} (fig.~\ref{fig:edge}).

These pressure drops lie below those of the filtration experiments. Across a \SI{2.6}{\micro\meter} filter, the
pressure drop of Reinhart and Chien reached about \SI{300}{\pascal} within a few seconds with salicylate-treated cells
and about \SI{500}{\pascal} with control cells, and 0.4--\SI{1.2}{\kilo\pascal} at \SI{30}{\second} (their Fig.~3B).
This does not contradict their finding that stage II and III echinocytes filter like control cells, because the pressure drops
needed to pass are a property of the cells at low drive and do not test the filtration resistances they measured. Nor
do the transit times, which follow the mean drive force on the cell (fig.~\ref{fig:force}). In
\SI{4.5}{\micro\meter} pores, D, E0 and every echinocyte stage passed at \SI{50}{\pascal}, about the initial pressure
drop across \SI{4.5}{\micro\meter} filters with control cells and with cells treated with 120~mM salicylate (their
Fig.~3A).

\subsection*{At fixed area and volume, the echinocyte stage alone raises the pressure drop needed to pass the slit}
The interendothelial slits of the splenic sinus walls are the narrowest passages that red cells meet in the spleen,
and they hold back aged and diseased cells~\cite{Pivkin2016,Li2018}. We drove each model cell through a slit
\SI{4.0}{\micro\meter} wide and \SI{1.2}{\micro\meter} high, the dimensions used in earlier
models~\cite{Pivkin2016,Li2018}, and raised the pressure drop in steps until the cell passed or the run failed
(Fig.~\ref{fig:slit}A, movies S1 to S3 and Materials and Methods). For each run we recorded the highest pressure drop
at which the cell was held and the lowest at which it passed (Fig.~\ref{fig:slit}, B and C; every run in
table~\ref{tab:brackets}).

At the area and volume of D, the pressure drop needed to pass rose along the sequence (Fig.~\ref{fig:slit}B). D
passed at 1.5 or \SI{2}{\pascal}, and E-I at \SI{4}{\pascal}. E-II needed 12.5 or \SI{15}{\pascal}. Each passed
after a hold at the next lower pressure drop tested (three runs each). E-III was held at every pressure drop up to
\SI{75}{\pascal} in all three runs, and each run failed at \SI{100}{\pascal} without the cell passing, so that
\SI{75}{\pascal} is a lower bound for its passage. The four cells have nearly the same $V/V_2$, 0.88 to 0.91, so their
order is set by the stage and not by area and volume. With the shear modulus of D, E-I passed at \SI{3}{\pascal} (four
runs) and E-II at 7 or \SI{10}{\pascal}, while E-III was held up to \SI{50}{\pascal} and failed at
\SI{75}{\pascal} (three runs). The softer cytoskeleton lowered the pressure drop of each stage and kept their order,
and, as in the pore, E-III is held mostly by its shape.

\subsection*{Spiculated cells are held below the area-to-volume limit of smooth cells}
Membrane loss holds a smooth cell once it raises $V/V_2$ past about 1 (Fig.~\ref{fig:slit}C); in perfused human
spleens, too, retention rises with the area a cell has lost~\cite{Safeukui2012}. With 2.5\% less bilayer area
($V/V_2 = 0.95$), the smooth cell passed like D, and with 5\% less (0.99) at 10 or \SI{12.5}{\pascal}. From 7.5\%
loss ($V/V_2 = 1.03$) on, every run was held at every pressure drop until it failed, at 50 or \SI{75}{\pascal} with
7.5\% loss and at up to \SI{300}{\pascal} with 20\%. The switch from passage to retention thus lies between
$V/V_2 = 0.99$ and 1.03.

No spiculated cell passed. With 5 to 20\% less bilayer area, every run was held at every pressure drop until it
failed, and the highest pressure drop at which the cell was held rose with the loss, to \SI{200}{\pascal} with 20\%
(table~\ref{tab:brackets}). With 5\% loss, the spiculated cell has the $V/V_2$ of the smooth cell with 2.5\% loss,
0.95. Yet it was held up to 75 or \SI{100}{\pascal}, whereas the smooth cell passed at 1.5 or \SI{2}{\pascal}. A
spiculated cell is held in the slit well below the area-to-volume limit. With rounder slit edges, neither the order of
the stages nor the switch between $V/V_2 = 0.99$ and 1.03 changed (fig.~\ref{fig:slitedge}).

\subsection*{Bonds stretched near full extension hold the spiculated cells}
At the end of the highest pressure drop at which they were held, the spiculated cells had most of their connecting
bonds near full extension (Fig.~\ref{fig:mechanism}, A to C). In E-III and the spiculated cells with membrane loss,
53 to 77\% of the bonds were longer than $0.9\,l_c$ (16 runs). In E-III with the shear modulus of D, 60 to 67\% were
(three runs). In the 25 runs of Fig.~\ref{fig:mechanism}C that passed at the next pressure drop, at most 2.4\% were.
The smooth cells held beyond the area-to-volume limit, with 7.5 to 20\% loss, kept their bonds slack. At most 0.3\% of
their bonds were that long (12 runs). The stretched bonds were not confined to the slit (Fig.~\ref{fig:mechanism}A;
mean curvature in fig.~\ref{fig:curvature}). In the held spiculated cells, 57 to 84\% of the connecting bonds upstream
of the slit were longer than $0.9\,l_c$. Downstream 75 to 100\% were, and inside the slit 31 to 49\%.

Two controls separate the shrinkage from the connection pattern and from the modulus. E0, with the connection pattern
of the echinocytes and the shear modulus of D but no shrinkage, kept its bonds slack and passed like D, at
\SI{2}{\pascal} after a hold at \SI{1.5}{\pascal} (three runs). E-I with the shear modulus of E-III
(\SI{9.1e-6}{\newton\per\meter}, 31 times its own) passed at \SI{15}{\pascal} after a hold at \SI{10}{\pascal} (two
runs), with none of its bonds longer than $0.9\,l_c$, whereas E-III with the shear modulus of D was held up to
\SI{50}{\pascal}.

A further test asks whether the stretched bonds hold the cell or only accompany the hold. We continued one run of
E-III from its hold at \SI{50}{\pascal}, at the same pressure drop, and raised the
contour length of its connecting bonds in small steps, changing nothing else (fig.~\ref{fig:release}; Supplementary
Text). Half of the runs used a stronger self-repulsion of the bilayer (Materials and Methods). With the contour
length unchanged, the cell stayed
in place for 200 time units (four runs). At $1.5\,l_c$ it advanced by 0.2 to \SI{0.6}{\micro\meter} and did not pass
(four runs). At $2\,l_c$ it passed in six of eight runs, 19 to 143 time units after the lengthening had ended; one run
failed while the cell was passing, and in one the cell advanced by \SI{1.1}{\micro\meter} and had not passed after
200 time units. When the lengthening went on toward $3\,l_c$, the cell passed in all four runs, at 2.3 to
$2.4\,l_c$. While the cell was held, the median bond length stayed at 0.89 to 0.93 of the contour length then in force.

A cell is thus held in the slit in two ways. The smooth cells are held by their area and volume once
$V/V_2$ exceeds about 1, with slack connecting bonds. The spiculated cells, at $V/V_2$ from 0.88 to 1.24, are held by
the bonds that tie their bilayer to the cytoskeleton between the spicules, which the shrinkage of the cytoskeleton has
already brought close to their contour length.

\subsection*{Echinocytes and discocytes in the constriction channels}
The constriction channels of the device are far wider than a slit, and echinocytes entered them. In an earlier frame
of the movie of Fig.~\ref{fig:exp}C, a spiculated cell lies inside a channel between folded discocytes
(Fig.~\ref{fig:chip}A). To see which cells such a channel stops, we drove the model cells into a model of one channel,
5 $\times$ \SI{5}{\micro\meter} in cross-section, with one pressure drop per run (Materials and Methods).

The pressure drop needed to enter the channel rose by about a factor of two from one stage to the next
(Fig.~\ref{fig:chip}, B and C). D entered at \SI{1}{\pascal} in all three runs. At \SI{0.5}{\pascal}, the
lowest pressure drop tested, it did not enter within 1000 time units. One run met the criterion for a hold after 60
time units, and in the other two the cell was still moving slowly into the channel when the run ended. E-I entered at
\SI{2}{\pascal} and E-II at \SI{4}{\pascal}, and each was held at half that pressure drop (three runs each). E-III
entered at \SI{8}{\pascal} in one run and at \SI{16}{\pascal} in the other two. With the shear modulus of D, E-III
entered at \SI{4}{\pascal} (three runs). The stages thus keep the order they have in the slit, but the channel
separates them far less. E-III entered it at 8 to 16 times the pressure drop at which D entered, whereas in the
slit E-III was still held at 37 to 50 times the pressure drop that passed D.

Above these thresholds, the stage did not delay a cell (Fig.~\ref{fig:chip}D). The time from the start of the
pressure drop until the whole cell was inside the channel fell as the pressure drop rose. At the lowest pressure drop
at which they entered, E-I and E-III were slower than D. E-I needed 384 to 506 time units at \SI{2}{\pascal},
against 274 to 286 for D, and the one run of E-III that entered at \SI{8}{\pascal} needed 142, against 80 to 84.
E-III with the shear modulus of D needed 202 to 236 at \SI{4}{\pascal}, against 146 to 154. At higher pressure
drops, and for E-II wherever it entered, the echinocytes took 0.7 to 1.0 times as long as D. The
model has no suspending fluid, so these times compare the cells with one another and are not transit times in the
device. A channel of this size thus stops an echinocyte only at pressure drops within about one order of magnitude
of those that stop a discocyte, which is consistent with the echinocyte seen among the discocytes in the channels of
the device.

\subsection*{At higher pressure drops the slit holds only the storage-induced microerythrocytes}
In red cells stored for transfusion and counted by differential interference contrast in units from six donors,
discocytes fell from 60.1\% of the cells after 3 days of storage to 33.7\% after 42 days~\cite{Roussel2017}
(Fig.~\ref{fig:population}A). The echinocytes of the three stages rose from about a fifth to about half of the cells,
and spheroechinocytes (SE) and spherocytes (S) from almost none to about 8\%. We let a model cell stand for each
class: D and E-I to E-III with their own moduli (Fig.~\ref{fig:slit}B), the spiculated family for spheroechinocytes,
and the smooth family from 7.5\% loss on ($V/V_2 \ge 1.03$) for spherocytes. Stomatocytes and the other cells, 20\% of
the cells on day 3 and 10\% on day 42, were not modeled.

Every run of the model cells of the SME classes (E-III, SE and S) was held at every pressure drop up to
\SI{30}{\pascal}. Every run of D, by contrast, had passed by \SI{2}{\pascal}, of E-I by \SI{4}{\pascal} and of E-II by
\SI{15}{\pascal}. Weighted by the shares of the classes, the slit held a smaller part of the modeled cells at each
higher pressure drop, and from \SI{15}{\pascal} on only the SMEs (Fig.~\ref{fig:population}B). These were 0.25\% of
the modeled cells of day 3 and 18.6\% of those of day 42. Between 4 and \SI{10}{\pascal}, where E-II was still held,
the slit held 1.4\% of the modeled cells of day 3 and 29\% of those of day 42.

Microsphere filters retained none of the red cells stored for 3 weeks or less and 30.6\% of those stored for 3 to 6
weeks~\cite{Deplaine2011}. Perfused human spleens cleared 61\% of the SMEs of long-stored blood within 70 minutes,
four times the 15\% of all its cells~\cite{Roussel2021} (Fig.~\ref{fig:population}C). The model ranks the classes in
the same order. The cells it holds at the higher pressure drops are the SMEs. We compare this order and not the
fractions, because our drive is a force on the part of the cell inside the wall rather than a flow, and its pressure
drops agree with those of other models in order of magnitude only (Discussion).

\subsection*{Reversing the spicules lowers the pressure drop to pass only below the area limit}
We also removed 5 and 10\% of the bilayer area of E-I and E-II (Fig.~\ref{fig:exp}E and Materials and Methods).
Reversing the spicules of a cell at the same bilayer area moves it along its row in Fig.~\ref{fig:exp}E, to the column
of D. Without membrane loss, the reversed cell is D, which passed at 1.5 or \SI{2}{\pascal}. E-I needed
\SI{4}{\pascal} and E-II 12.5 or \SI{15}{\pascal}, and E-III was held until its runs failed.

With 5\% loss, the reversed cell ($V/V_2 = 0.99$) passed at 10 or \SI{12.5}{\pascal}. E-II with the same loss (0.97)
passed at \SI{30}{\pascal}, and the spiculated cell (0.95) was held up to 75 or \SI{100}{\pascal}
(Fig.~\ref{fig:rejuvenation}A). E-I with 5\% loss (0.98) is the exception. It passed at \SI{7}{\pascal}, a pressure
drop at which all three runs of the reversed cell were held, so reversal from E-I did not lower the pressure drop. The
reversed cell lies closer to the limit (0.99 against 0.98), where the pressure drop needed by the smooth cells rises
steeply. With 2.5\% loss ($V/V_2 = 0.95$), the smooth cell passed at 1.5 or \SI{2}{\pascal}.

With 10\% loss, the reversed cell ($V/V_2 = 1.07$) was held up to \SI{75}{\pascal}. So were E-I and E-II with this
loss (both 1.06), up to 50 and \SI{75}{\pascal}, and the spiculated cell (1.03), up to \SI{100}{\pascal}. The three
runs of each of the four cells of E-I and E-II with loss gave the same pressure drops. For all four stages, $V/V_2$
crosses 1 between 5\% loss (0.95 to 0.99) and 10\% loss (1.03 to 1.07) (Fig.~\ref{fig:rejuvenation}B). Below the
limit, the reversed cell passed at a lower pressure drop than E-II and E-III; above it, the reversed cell is held as
well.

\subsection*{Stage III echinocytes lose their links at pressure drops that rise with link strength}
In every run so far, the connecting bonds (links) could not break, and in a held E-III most
of them were stretched past $0.9\,l_c$ (Fig.~\ref{fig:mechanism} and fig.~\ref{fig:heldlinks}A). We therefore
repeated the ladders of Fig.~\ref{fig:slit}B, in which the pressure drop is raised step by step, with links that fail
once they are stretched past a set length (Materials and Methods). We measure the stretch of a link by $x = r/l_c$, its
length $r$ over its contour length. At the failure stretches $x_b = 0.950$, 0.965 and 0.975, a link fails at 30, 62
and \SI{121}{\pico\newton} (fig.~\ref{fig:heldlinks}B). The first lies below, and the others at or above, the strength we estimate from the measured
energy that separates the bilayer from the cytoskeleton, 39 to \SI{61}{\pico\newton} (Materials and Methods).

How close do the links come to these stretches before the slit acts? We kept each cell in front of the slit without a
pressure drop and recorded its links for 20 time units (10 for D). The most stretched link reached $x = 0.917$ in D,
0.914 in E0 and 0.928 in E-I (fig.~\ref{fig:rest}). In E-II it reached 0.933, and 2 of its 462 links passed 0.93. In
E-III, 162 of the 462 links passed 0.93 at least once, and one passed 0.95. No link of any cell reached 0.955. At rest,
the links of E-III are thus stretched to about the weakest failure stretch. E0, with the sparse links of an echinocyte
but no shrinkage, was no more stretched than D, so the preload comes with the shrunken cytoskeleton.

In the slit the stretch grew with the pressure drop (Fig.~\ref{fig:onset}A; unbreakable links; configurations saved
every 2 time units, which miss brief peaks; medians over three runs). The links of D, E0 and E-I stayed at or below 0.922 until these cells
passed. Those of E-II stayed between 0.911 and 0.917 up to \SI{10}{\pascal} and reached 0.934 at \SI{15}{\pascal},
where the cell passed. Those of E-III started at 0.935 at \SI{2}{\pascal}, passed 0.950 between 30 and
\SI{50}{\pascal} and reached 0.965 at \SI{75}{\pascal}, the last step at which the cell was held.

We take as the onset of failure the pressure drop at which 5\% of the links of a cell have failed in all: 24 of the
462 links of an echinocyte, 129 of the 2562 of the discocyte. In every cell these links tie about
\SI{7}{\micro\meter\squared} of bilayer, about the area of one of the 20 free patches of an echinocyte. D, E0 and E-I
never reached the onset (table~\ref{tab:onset}). Their ladders ended step for step as with unbreakable links, and only
E-I, at the weakest strength, lost links at all, two in each of two runs. E-II reached the onset only at the weakest
strength, at \SI{15}{\pascal} in all three runs, the step at which its ladders passed with unbreakable links. E-III
reached it at every strength, and the pressure drop rose with the strength: 5 to \SI{15}{\pascal} at
\SI{30}{\pico\newton}, 50 to \SI{75}{\pascal} at \SI{62}{\pico\newton} and 75 to \SI{100}{\pascal} at
\SI{121}{\pico\newton} (Fig.~\ref{fig:onset}B). With the weakest links, its first link failed in the first step of every
ladder, at \SI{2}{\pascal}. With the strongest, the onset came no later than the \SI{100}{\pascal} step, in which its
runs with unbreakable links had failed.

The weakest links did not need the slit. Held without a pressure drop after it had been pushed into the slit, E-III
with links of \SI{30}{\pico\newton} lost its 24th link 61 to 77 time units after the links could first fail, against 64
to 136 time units in the ladders (fig.~\ref{fig:restcontrols}A). Cells never pushed in, kept \SI{4}{\micro\meter}
upstream of the slit, lost their 24th link after 82 and 69 time units, and at half the time step two held runs lost it
after 66 and 78 time units, while the third had lost 19 links by the end of its 120 time units. At this strength the
onset marks how long the cell had waited, not a load added by the slit. The effect of the slit is read from the
stronger links. Held without a pressure drop for 420 time units, longer than its ladders took to lose 24 links, E-III with links
of \SI{62}{\pico\newton} lost none in three runs (fig.~\ref{fig:restcontrols}B). E-II with links of
\SI{30}{\pico\newton} lost one link in three such runs, whereas its ladders had lost 24 by 340 to 414 time units
(fig.~\ref{fig:restcontrols}C).

With the shear modulus of D, E-III at the weakest strength again lost its first link at \SI{2}{\pascal} and reached the
onset at 7, 10 and \SI{7}{\pascal}, against 7, 5 and \SI{15}{\pascal} with its own modulus. The load on its links
comes with the shrunken cytoskeleton, not with its stiffness. E-II with the modulus of D reached the onset at 5, 7 and
\SI{5}{\pascal}, at or below the pressure drops at which it passed with unbreakable links (table~\ref{tab:brackets}):
a softer cell is pushed through sooner, and its links fail as it goes. E-I with the modulus of D lost no link and passed
at \SI{3}{\pascal}.

\subsection*{Weak links fail in front of the slit, stronger links in and behind it}
Where E-III loses its links also depends on their strength (Fig.~\ref{fig:position}, A to C). With the weakest links,
the first 24 failures of each of the three runs lay upstream of the wall, 23 or 24 of them more than
\SI{0.5}{\micro\meter} from any wall surface. These links fail while the cell waits in front of the slit, as they do
without a pressure drop. With the two stronger strengths, 0 to 7 of the first 24 failures lay upstream in each of the
six runs and 5 to 19 in the slit, the rest downstream. The links of E-II failed mostly downstream, in the part of the
cell that had gone through (16 to 20 of the first 24).

This split follows where the most stretched link lay when the links could not break (Fig.~\ref{fig:position}D). In
the three runs of E-III of Fig.~\ref{fig:slit}B, it lay upstream of the wall in 268 of the 270 configurations saved at
2 to \SI{15}{\pascal}, where the weakest links failed. At 50 and \SI{75}{\pascal}, where the stronger links failed, it
lay in the slit or downstream in 94 of 130, although 68\% of the links stretched above 0.9 still lay upstream. Weak
links thus fail on the part of the cell that waits upstream, strong links where the few most stretched ones are, in and
behind the slit.

The failures also spread differently. With the weakest links, those of E-III failed one at a time: the 24th came 59 to
135 time units after the first, two to five steps later, while the cell was held, and 57 to 73 time units after it
without a pressure drop. Over these steps the bilayer came away from the cytoskeleton over a cap on the upstream part
of the cell, ringed by the failed links (fig.~\ref{fig:burst}A and movie S4). With the stronger links, every failure
after the first came within one step, and the 24th followed the first within 0.9 to 28 time units
(fig.~\ref{fig:burst}B). The membrane warmed as these links failed. Its kinetic temperature was 1.02 to 1.46 times the
set value over the 2 time units before the first failure and, over the half time unit before the 24th, 1.7 to 5.6
times, against 0.98 to 1.04 with the weakest links. Three of these six runs broke down within about one time unit of the
24th failure, stopping with an error, so for the stronger links we report only the onset (Supplementary Text).

In the model only the links tie the cytoskeleton to the bilayer, and they do not keep it inside: even with every link
intact, part of the cytoskeleton lay just outside the bilayer (Supplementary Text). We therefore repeated all 45 runs
with a constraint that keeps every cytoskeleton vertex inside. In 11 of the 12 runs of E-II and E-III that reached the
onset, it came at the same pressure drop as without the constraint, and in the twelfth one step earlier
(table~\ref{tab:onsetcon}). With the weakest links the first 24 failures of E-III again all lay upstream.

\section*{Discussion}
We asked whether spicules alone, at the membrane area and volume of a discocyte, change the passage of a red cell
through a splenic slit. In our model they do. With the shear modulus of D, E-I needed 1.5 to 2 times the pressure drop
at which D passed and E-II 3.5 to 7 times, while E-III was still held at 25 to 33 times that pressure drop, the
highest before its runs failed (Fig.~\ref{fig:slit}B). With their own, stiffer cytoskeletons the factors were 2 to 3
for E-I, 6 to 10 for E-II and 37 to 50 for E-III.

\textit{Area and volume do not decide alone.} A
smooth cell is held once its volume exceeds what two lobes joined through the slit can enclose with its
area~\cite{Pivkin2016,Pivkin2017corr,Li2018}. Red cells driven through microfluidic slits pass only if their area suffices, for their volume, to form two equal spheres joined by a tether, and by the authors'
calculations the cells closest to this limit must also unfold part of their spectrin inside the
slit~\cite{Moreau2023}. In their model of this slit, Li et al.\ found that a cell of
\SI{90}{\micro\meter\cubed} passed with \SI{120}{\micro\meter\squared} of membrane at a pressure gradient of
\SI{5}{\pascal\per\micro\meter}. With \SI{110}{\micro\meter\squared} it was retained, and it stayed retained up to
\SI{20}{\pascal\per\micro\meter}~\cite{Li2018}. Their switch thus lies between $V/V_2 = 1.03$ and 1.17. At this
volume, the analytical model of Pivkin et al.~\cite{Pivkin2016,Pivkin2017corr} gives a critical area of
\SI{113.1}{\micro\meter\squared} ($V/V_2 = 1.13$)~\cite{Li2018}. Our smooth
cells switched at the lower end of this range, between $V/V_2 = 0.99$ and 1.03. At 1.03 the cell of Li et al.\ passed, whereas ours
was held. In the same slit geometry, the model cells of Du et al.\ went from passed to held between sphericity indices
of 0.822 and 0.853~\cite{Du2026}; our switch lies at sphericity indices of 0.79 to 0.81. Their membrane is one network,
given the area, volume and shear modulus measured on each class of fatigued cells, so their echinocytes differ from
their discocytes in these values only. E-III, at $V/V_2 = 0.88$ (sphericity index 0.73), was held well below both
limits, and so was the spiculated cell with 5\% loss, at 0.95 (0.77).

\textit{What holds a spiculated cell.} The shrinkage of the cytoskeleton that raises the spicules brings the bonds
between the bilayer and the cytoskeleton close to their contour length before the cell reaches the slit. In the held
cells, 53 to 77\% of these bonds were near full extension, on both sides of the wall as well as inside the slit
(Fig.~\ref{fig:mechanism}). Neither the connection pattern alone (E0, which passed like D) nor a stiffer cytoskeleton
(E-I with the shear modulus of E-III, which passed at \SI{15}{\pascal} with slack bonds) held a cell in this way. When
the connecting bonds of a held E-III were lengthened at the same pressure drop, the cell passed in most runs once
their contour length was doubled, and in all runs at 2.3 to $2.4\,l_c$ (fig.~\ref{fig:release}). These bonds carry
the hold of a spiculated cell. Of the 37 runs in Figs.~\ref{fig:slit} and~\ref{fig:rejuvenation} that
were held until they failed, 36 stopped after a bond had exceeded its maximum length. This bond joined the bilayer to
the cytoskeleton in 30 runs: 23 of the 25 runs of spiculated cells that failed and 7 of the 12 runs of smooth cells.
In a red cell such loads could
detach the bilayer from the cytoskeleton. With weakened links between the two layers, the model cells of Li et al.\
shed membrane as they crossed the slit~\cite{Li2018,Li2021}. At the exit of microfluidic slits \SI{0.83}{\micro\meter} wide,
the cytoskeleton of red cells from a donor with hereditary spherocytosis seemed detached from the bilayer in the rear
half of the cell~\cite{Gambhire2017}. The links of these runs could not break, so the highest pressure
drops at which they were held are lower bounds for passage without loss of membrane.

\textit{When the links can fail.} Links that can fail give way in held stage III echinocytes, while D, E0 and E-I pass
losing at most two links, and E-II loses links only when they are weak, as it passes. The difference lies in the
preload. The shrunken cytoskeleton of E-III stretches its links at rest to about the weakest failure stretch, so links
that weak fail even without the slit, and stronger links fail once the slit adds the rest. The strength we estimate
from the separation energy, 39 to \SI{61}{\pico\newton}~\cite{Hwang1997,Butler2008}, lies between our two weaker
strengths, and single spectrin--ankyrin bonds broke at mean forces of 21 to \SI{48}{\pico\newton}~\cite{Saito2015},
around the weakest. Links of \SI{62}{\pico\newton} held at rest and failed at 50 to \SI{75}{\pascal}, pressure drops
at which the slit holds E-III (Fig.~\ref{fig:slit}B). Earlier models found that a slit cannot pull the bilayer of a
healthy cell off its cytoskeleton~\cite{Zhu2017,LuPeng2019}, and our D and E0 lost no link either. Here the cell
itself, by shrinking its cytoskeleton, loads its links enough for the slit to break them. Since vesicles of stored
blood carry no spectrin~\cite{Greenwalt2006}, freeing the bilayer comes before the loss of membrane. Red cells passed
repeatedly through narrow constrictions released vesicles~\cite{GarciaHerreros2022}, and in people without a spleen,
old red cells keep more of their hemoglobin, which suggests that the spleen helps the vesicles
leave~\cite{Willekens2003}. If the slit frees the bilayer of stage III echinocytes, it could help the spleen do so for
the SMEs to which these cells belong~\cite{Roussel2017}. In the model a link fails at a fixed stretch, whereas a
molecular bond under force breaks after a time that shortens exponentially with the force~\cite{Bell1978}, at a force
that rises with the loading rate~\cite{Saito2015}, and failed links do not form again. With the stronger links the
membrane warmed as they failed and three runs broke down soon after the onset, so only the onset is quantitative.
Where the cytoskeleton could also tear, springs that break at about a quarter of the lowest force that unfolds spectrin
repeats, scaled to our network, broke before any link had failed in three of nine runs, all in the slit, and in one of them brought the onset
of link failure one pressure step earlier (Supplementary Text). The model does not bud vesicles, so it tells where the
bilayer comes free, not what leaves the cell.

\textit{Pores, channels and slits.} Filtration through micropores did not separate the stages. Echinocytes of stages II and III
filtered through \SI{2.6}{\micro\meter} pores like control cells~\cite{Reinhart1986}. In our \SI{3.2}{\micro\meter}
pore the stages needed different pressure drops, but all passed below the pressure drops across such filters
(Fig.~\ref{fig:micropore}), so at those drives the filters would not have told them apart. In the slit, E-III was held
up to the pressure drop at which its runs failed. The pore keeps the cell in one channel, along which the bilayer of
E-III ran ahead of its cytoskeleton. The slit is \SI{1.2}{\micro\meter} high but only \SI{1.89}{\micro\meter} long,
so the cell straddles the wall and must reshape on both sides of it, and the stretched bonds of the held cells
lay on both sides. The constriction channels of our device lie between the two. In the model channel, 5 $\times$
\SI{5}{\micro\meter} in cross-section, the stages kept their order, but E-III entered at 8 or \SI{16}{\pascal},
about a fifth or less of the pressure drop up to which the slit held it, and in the device an echinocyte lay among
discocytes inside a channel (Fig.~\ref{fig:chip}). Whether stage III echinocytes are held as the model predicts
therefore has to be tested in slits of submicrometer aperture, such as those of Moreau et
al.~\cite{Moreau2023}, and not in channels of capillary size. In a \SI{6}{\micro\meter} model tube, wall shear
stress peaked beside D (\SI{2.73}{\pascal}), E-II (\SI{3.14}{\pascal}) and E-III (\SI{3.41}{\pascal};
fig.~\ref{fig:tubewss}). At a model bifurcation, cells held on the dividing wall raised, then lowered
its wall shear stress (fig.~\ref{fig:bifwss}). An E-III cell sheared at \SI{1.5}{\pascal} near a flat wall
gave local wall shear stresses of 1.14 to \SI{1.85}{\pascal} (fig.~\ref{fig:s6}).

\textit{Stored blood.} In isolated perfused human spleens, retention rose already at small area losses. Red cells treated
with lysophosphatidylcholine that had lost 3\% of their area were retained at 27\%, against 9\% for untreated
cells~\cite{Safeukui2012}. Lysophosphatidylcholine makes red cells echinocytic before it turns them into
spherocytes~\cite{Safeukui2012}, and in our model a spiculated cell was held at the $V/V_2$ at which a smooth cell
passed (0.95). Spiculation may therefore add to the retention of cells that have lost little area.
Microsphere filters
retained 30.6\% of red cells stored for 3 to 6 weeks and none of those stored for 3 weeks or less, close to the 26.5\%
of red cells stored for 25 to 35 days that are cleared within 24 hours of transfusion~\cite{Deplaine2011}. Perfused
human spleens cleared 61\% of the SMEs of long-stored blood and 15\% of all its cells~\cite{Roussel2021}. These are the
classes that our model holds at every pressure drop at which the earlier stages pass, up to \SI{30}{\pascal}
(Fig.~\ref{fig:population}), for two different reasons. The models of spherocytes are held because their loss of area
has brought $V/V_2$ above 1, and E-III because of its spicules, at the area and volume of a discocyte. The spiculated
cells with membrane loss are held for both reasons or, with 5\% loss ($V/V_2 = 0.95$), for the spicules alone.

\textit{Rejuvenation.} Metabolic rejuvenation of stored red cells lowered the share of SMEs from 28\% to 9.5\%, mostly
by restoring their shape, and improved the passage of the cells through a spleen-mimetic filter of microspheres. Marin
et al.\ attributed this to the restored shapes together with a smaller cell volume~\cite{Marin2021}. Our model
separates the two. Reversing the spicules at the same bilayer area let cells of stages II and III through the slit at
lower pressure drops as long as $V/V_2$ was below 1 (spiculated E-III cells were held until their runs failed). It
brought no gain to E-I with 5\% loss and left every cell with $V/V_2$ above 1 held (Fig.~\ref{fig:rejuvenation}). A
cell whose loss of bilayer area has brought $V/V_2$ above 1, between 5 and 10\% loss here, therefore needs a smaller
volume as well to pass; rejuvenation also lowered the volume of the cells~\cite{Marin2021}. Washing in a 1\%
albumin solution likewise returned red cells stored for 6 to 7 weeks from echinocytes toward discocytes and lowered
their volume; it improved their flow through a network of microchannels but not their deformability in
ektacytometry~\cite{Reinhart2015}.

\textit{Echinocytes in disease.} Echinocytes also circulate in disease, and in one disorder their numbers point to
the spleen. In four splenectomized patients with pyruvate kinase deficiency, 30 to 50\% of the red cells were
crenated cells that met the criteria for echinocytes~\cite{Leblond1978a}. No blood smears from before splenectomy
were available, but from the medical charts and the recollection of one of the authors, these shapes had not been
present to any significant degree while the spleen was in place. The authors proposed that such cells survive only
briefly in the presence of the spleen, possibly because they cannot deform enough to travel through its
microcirculation~\cite{Leblond1978a}. A recent case report reads the echinocytes of such a patient in the same
way~\cite{Bassiony2024}. In micropipettes \SI{2.9}{\micro\meter} wide, the crenated cells of each patient
needed 1.3 to 1.8 times the suction that the same patient's discocytes needed to enter, and 10 to 30\% of them
returned abnormally slowly to their shape afterwards~\cite{Leblond1978b}. In sickle cell anemia, the disease of the
blood in our device, echinocytes are found among the dense cells: in nine adults, echinocytes of stages II and III
and spheroechinocytes made up 22\% of these cells during a pain crisis and 12\% once the pain had gone, a change
expected to hinder them in the microcirculation~\cite{Warth1984}. The cells of both diseases differ from discocytes in
more than their spicules, so the observations are consistent with our model but do not isolate the spicules. In
hemodialysis and liver disease, by contrast, the spicules come and go with the plasma. The share of echinocytes on
blood smears rose from 9 to 17--21\% within 45 minutes of dialysis and fell to 5\% by its end~\cite{Agroyannis1997};
normal red cells became echinocytes in plasma drawn 30 minutes into dialysis, and the patients' echinocytes turned
back into discocytes in buffer~\cite{Hasler1998}. In liver disease, abnormal high-density lipoprotein turns
red cells into echinocytes within seconds, and normal high-density lipoprotein reverses the change~\cite{Owen1985}.
Such cells could be driven through slits with and without their spicules, which would separate the effect of the
spicules from the other changes of a diseased cell.

\textit{Limitations.} The model leaves out the fluid, applies the pressure drop as a force on the part of the cell
inside the wall and keeps the membrane area and volume fixed during passage, and it uses one slit, with the largest
aperture height measured on electron micrographs~\cite{Deplaine2011}. Its pressure drops rank the cells under one
drive and are not calibrated against the pressure across splenic slits. D passed at 1.5 or \SI{2}{\pascal}, which over
the wall is a gradient of 0.8 to \SI{1.1}{\pascal\per\micro\meter}. Pivkin et al.\ took about
\SI{1}{\pascal\per\micro\meter}, estimated from microsphere filters and perfused spleens, as sufficient for red cells
of all sizes found in blood to pass the slits~\cite{Pivkin2016}. In the model of Li et al., a cell of
\SI{140}{\micro\meter\squared} and \SI{90}{\micro\meter\cubed} passed from \SI{5}{\pascal\per\micro\meter} on, which
they attributed mainly to their implicit solvent and the hard-core repulsion of their slit walls~\cite{Li2018}. Their
passing gradient rose from 3 to \SI{15}{\pascal\per\micro\meter} as the friction coefficient of the thermostat was
raised from 0.004 to 0.1 in model units~\cite{Li2018}. With 2.5 times the friction of our thermostat, by contrast, no
cell moved by more than one step (Supplementary Text). These values thus agree in order of magnitude only. The
cytoskeleton of D is more than two orders of magnitude softer than micropipette measurements suggest (Materials and
Methods); comparing the stages at the shear modulus of D removes this difference between the cells but not its effect on
the absolute pressure drops.

Further limits concern how the runs were scored and how the cells were built. A step ended as held when the cell
stopped or, at the latest, after 500 time units. Given 540 to 630 time units, E-I and E-I with the shear modulus of D
passed one step lower, at 3 and \SI{2}{\pascal}. The values of the softest cells are therefore upper estimates, but
the order of the stages was the same in every time window tested. With the slit edges rounded with a radius of
\SI{0.5}{\micro\meter} instead of 0.3, every cell that passed did so at the same or a lower pressure drop. The
pressure drop fell most for the smooth cell with 5\% loss, to 1.5 or \SI{2}{\pascal} (fig.~\ref{fig:slitedge}). In 30
of the 31 runs of Fig.~\ref{fig:slit} that were held until they failed, a bond had first exceeded its maximum length,
mostly at the edges or walls of the slit. The highest pressure drops at which these runs were held, 30 to
\SI{200}{\pascal}, are only lower bounds of the pressure drop needed to pass. Finally, the stages were
built, after Wen et al., by shrinking the cytoskeleton under a bilayer of fixed area~\cite{Wen2025}. Spicules also
form when lysophosphatidylcholine accumulates in the outer leaflet of the bilayer~\cite{Safeukui2012}, and such cells
may load the connecting bonds differently. In the model only the connecting bonds tie the cytoskeleton to the
bilayer, and they do not keep it inside. A constraint that does left five of the six ladders of E-II and E-III with
unbreakable links unchanged step for step; in the sixth, E-III failed one step earlier, at \SI{75}{\pascal}, as one of
nine runs did without the constraint (Supplementary Text).

At the area and volume of a discocyte, the echinocyte stage alone raised the pressure drop a model cell needed to
cross the slit. Stage III echinocytes, like the spiculated cells with small area losses, were held below the
area-to-volume limit that holds smooth cells. The hold comes from the links between the bilayer and the cytoskeleton,
which the shrinkage of the cytoskeleton has stretched before the cell reaches the slit. A slit model that describes a
red cell by its area and volume alone would therefore pass echinocytes that ours holds. Links that can fail give way in the held stage III echinocytes, at pressure drops
that rise with their strength, so the slit may free the bilayer of the most shrunken stored cells from their
cytoskeleton while it holds them. Restoring the shape of stored
cells, in turn, can improve their passage only as long as their loss of area has not brought $V/V_2$ above 1. These
predictions can be tested on stored cells sorted by shape. At the same area and volume, stage III echinocytes should
be held at slit-like constrictions that pass discocytes; after rejuvenation, only cells with $V/V_2$ below 1
should regain passage; and held stage III echinocytes stained for spectrin should show patches of bilayer without it in
and behind the slit if their links are strong, which cells kept without a slit should show only if their links are
weak.

\section*{Materials and Methods}

\subsection*{Microfluidic experiments}
Whole blood samples of patients with sickle cell anemia (HbSS) were obtained from Boston Children's Hospital and
Massachusetts General Hospital under excess human material protocols approved by the respective institutional review
boards, the Boston Children's Hospital Institutional Review Board and the Partners HealthCare Institutional Review
Board, with a waiver of consent. The samples were then promptly transported to MIT for experiments under an approved
exempt protocol (Massachusetts Institute of Technology IRB protocol E-1523). As in our earlier work~\cite{chai2026stomatocyte}, the samples were gently washed three times with
phosphate-buffered saline (1$\times$ PBS; CaCl$_2$-free, MgCl$_2$-free; pH 7.4; Gibco) by centrifugation at 1500 rpm
for 3 min at room temperature, and the washed cells were resuspended in PBS containing 1\% (w/v) bovine serum albumin
(BSA; EMD Millipore) at a hematocrit of 2\%. The samples were stored at 4~$^\circ$C and used within 24~h. The
device was made of polydimethylsiloxane (PDMS; base to curing agent 10:1 by weight,
degassed and cast on an SU-8 mold) bonded to a glass slide~\cite{Qiang2019,chai2026stomatocyte}. Its parallel
constriction channels, 14 of which were in the field of view, are \SI{100}{\micro\meter} long with a cross-section of
5 $\times$ \SI{5}{\micro\meter}. The cells were driven by a water column and imaged in bright
field on an inverted microscope (IX71, Olympus) with a 60$\times$ oil-immersion objective (NA 1.25) and a CMOS camera
(The Imaging Source) at 4 frames per second (5472 $\times$ 3648 pixels). The pixel size, \SI{0.040}{\micro\meter},
follows from the channel length.

\subsection*{Simulation method and units}
The membrane model below was developed by Wen et al.~\cite{Wen2025} in dissipative particle dynamics (DPD), one of
the particle methods widely used for blood cells, soft matter and collective
motion~\cite{Pivkin2016,chai2025silico,toscano2026graft,chai2023dynamics,zhang2020deep,zhang2021deep}; we
implemented it in our \textsc{lammps}-based code~\cite{thompson2022lammps}. Except in figs.~\ref{fig:tubewss} to~\ref{fig:s6}, the solvent is implicit. The membrane vertices, each of unit
mass, move under the model forces and a Langevin thermostat at \SI{310}{\kelvin} ($k_BT =
\SI{4.28e-21}{\joule}$, 0.0945 energy units). The friction of the thermostat stands in for the drag of the solvent, so times are given in model
time units without a physical calibration. The length unit is \SI{1}{\micro\meter} and the energy unit
\SI{4.529e-20}{\joule}, so that the units of force, pressure and elastic modulus are \SI{0.04529}{\pico\newton},
\SI{0.04529}{\pascal} and \SI{4.529e-8}{\newton\per\meter}. Wen et al. used an energy unit of \SI{3.960e-21}{\joule}
and thermostat coefficients that give $k_BT = 0.0945$ of that unit (their Tables S1 and S3), a thermal energy about
1/11 of the physical one; we use the physical value.

\subsection*{Two-component RBC model}
Following Wen et al.~\cite{Wen2025}, the lipid bilayer and the cytoskeleton are two triangulated networks with the
topology of a subdivided icosahedron, each with $N_v = 2562$ vertices (table~\ref{tab:model}). The bilayer has a
bending energy $k_b\sum_i[1-\cos(\theta_i-\theta_0)]$ over pairs of adjacent triangles, with $\theta_0 = 0$, and
quadratic penalties on the area of each triangle, the total area and the enclosed volume, with moduli $k_a$, $k_A$
and $k_V$; it has no in-plane springs. The cytoskeleton is a network of worm-like-chain springs with a repulsive power
term (WLC--POW), with a contour length of 2.2 times the stress-free spring length and a persistence length set by the
target shear modulus $\mu$ of the network (Eq.~5 of Wen et al.); in the echinocytes it also has a bending energy and
its own area and volume penalties. Pairs of vertices, one in each network, are joined by connecting WLC bonds (the links between the two layers) with an
equilibrium length of \SI{0.03}{\micro\meter} and an equivalent elastic constant $k_{bs}$. The stress-free radii are
\SI{3.25}{\micro\meter} for the bilayer and \SI{2.8}{\micro\meter} for the cytoskeleton. Each cell starts from
concentric spheres of radii 3.055 and \SI{2.995}{\micro\meter} and relaxes toward the area
$A_0 = \SI{132.6}{\micro\meter\squared}$ and the volume $V_0 = \SI{92}{\micro\meter\cubed}$ for 300\,000 to 600\,000
steps of 0.001 time units, with a thermostat damping time of 0.2 time units.

In the discocyte (D), all 2562 vertex pairs are connected (connecting number $N_c = 16$; table~\ref{tab:model}). In the echinocytes, only the
462 pairs on the edges of an icosahedron are connected ($N_c = 1$), which leaves 20 triangular patches of bilayer
detached from the cytoskeleton, and the cytoskeleton shrinks toward the area $(1-r_{sh})A_0$~\cite{Wen2025} and the
volume $0.94\,V_0(1-r_{sh})^{3/2}$. Stages E-I, E-II and E-III take the target shrinkage ratios $r_{sh} = 0.113$,
0.226 and 0.377 of the echinocyte cases of Wen et al. (their Table~2) and the cytoskeleton shear moduli of their
Table~1. Each echinocyte forms one spicule per detached patch, 20 in all (Fig.~\ref{fig:model}B). The bilayer of E-III
crossed itself during the relaxation, so E-III was relaxed again with the self-repulsion of the bilayer used in the
assays (see Filtration assays).

Three choices are ours. Wen et al. give no volume target for the cytoskeleton; we use 0.94 times the volume at which
the cytoskeleton would have the reduced volume of the bilayer, the largest of the factors tested (1, 0.97 and 0.94) at
which the bilayer no longer crosses the cytoskeleton. Their main text assigns the larger constraint moduli of the
echinocytes to the cytoskeleton and keeps the bilayer at the values of the discocyte, while their supporting material
lists other values; we apply the larger moduli to both networks. We relax the cells with the Langevin thermostat
instead of a DPD solvent, and without the Morse repulsion between the two networks and the improper term on the
connecting bonds, which Wen et al. describe in their supporting material without parameter values. All cells are
within 0.5\% of their target area. Their volume is within 1\% of $V_0$, except in the cells with the shrinkage of
E-III (E-III, E-III with the modulus of D, and the E3ka family below), which are 1.9--2.5\% below $V_0$.

The cytoskeleton shear modulus of the discocyte in this model, \SI{2.5e-8}{\newton\per\meter}, is more than two
orders of magnitude below the 4--\SI{12e-6}{\newton\per\meter} measured by micropipette aspiration~\cite{Wen2025}.
Wen et al. note
this and relate the low value to the shear elasticity obtained from membrane flicker, about 100 times below the
micropipette values, because thermal fluctuations probe small deformations, at which the network behaves as a fluid, while
micropipette aspiration measures it under stress. Above about \SI{4e-8}{\newton\per\meter} their discocyte turns into
a stomatocyte~\cite{Wen2025}. The moduli of E-II and E-III, 2.29 and \SI{9.1e-6}{\newton\per\meter}, lie near or
within the micropipette range. Passage through a pore or a slit is a large deformation under stress, so the cells
with the modulus of D (D, E0 and the Da family) are softer in our assays than the micropipette values suggest. We
therefore also ran the echinocyte stages with the shear modulus of D.

\subsection*{Echinocyte stages and membrane loss}
All cells have the volume target $V_0$. D and the three echinocyte stages have the bilayer area $A_0$ (reduced
volume 0.64; 0.62 for E-III; Fig.~\ref{fig:exp}E). E0 has the settings of the echinocytes ($N_c = 1$, cytoskeleton
bending, constraints on both networks, cytoskeleton volume target $0.94\,V_0$) but no cytoskeleton shrinkage
($r_{sh} = 0$) and the shear modulus of D; it separates the partial detachment of the two layers from the shrinkage.
Membrane loss is modeled by lowering the area targets of the bilayer by a fraction $L$ at the same volume, while the
stress-free cytoskeleton is left unchanged; red cells treated with lysophosphatidylcholine lose area in this way, at
nearly constant volume, as they turn into echinocytes and then spherocytes by shedding vesicles~\cite{Safeukui2012}. The smooth family (Da, settings of D) has $L = 2.5$, 5, 7.5, 10, 15 and
20\% (Da025 to Da20; reduced volume 0.67 to 0.91), and the spiculated family (E3ka, settings of E-III) $L = 5$, 10, 15
and 20\% (E3ka05 to E3ka20; reduced volume 0.67 to 0.88; Fig.~\ref{fig:exp}E). In the spiculated family the cytoskeleton keeps the area
target of E-III, as if the shed membrane carried no cytoskeleton, so its effective shrinkage ratio falls to
$1-(1-r_{sh})/(1-L)$, and its volume target follows from the rule above with this ratio. E-I and E-II with 5 and
10\% loss (E1ka05, E1ka10, E2ka05 and E2ka10; reduced volume 0.69 and 0.75) are built in the same way from their own
settings (Fig.~\ref{fig:exp}E).
In stored blood, the mean cell volume stayed near \SI{90}{\micro\meter\cubed} for 42 days while small cells and
red-cell microparticles accumulated, and the small cells were told apart by a projected area 15 to 20\% below that
of the other cells~\cite{Roussel2017}. In our cells, the area of the silhouette seen along the shortest principal axis
falls by 15 and 23\% from D to Da10 and Da15, but also by 39\% from D to E-III, which has lost no membrane, so for
spiculated cells a projected area does not measure the loss.

\subsection*{Filtration assays}
\textit{Common settings.} The relaxed cell is turned so that its longest principal axis lies along the direction of
passage, placed in front of the plate with its center on the axis of the pore or slit, and equilibrated for 20 time units
while the self-repulsion of the bilayer is switched on. During the assay the global area and volume penalties of the
bilayer are stiffened ($k_A = \SI{4.5e-2}{\newton\per\meter}$, $k_V = \SI{45}{\kilo\pascal}$, $k_a =
\SI{2.27e-4}{\newton\per\meter}$), with the area and volume of the relaxed cell as targets. Bilayer vertices more
than three mesh edges apart repel each other with the soft potential $\epsilon[1+\cos(\pi r/r_c)]$, with $\epsilon = \SI{4.5e-18}{\joule}$
(about 1060~$k_BT$) and $r_c = \SI{0.4}{\micro\meter}$. Because the bilayer has no in-plane springs, its fixed triangulation can
degenerate into needle-shaped triangles in a narrow channel; each edge of the bilayer mesh therefore carries a guard
spring that exerts no force up to three times its initial length and a WLC-like tension above it that diverges at
five times its initial length. The thermostat damping time is 0.5 time units.

\textit{Micropore sieving.} One cylindrical pore of diameter \SI{3.2}{\micro\meter} crosses a rigid plate
\SI{13}{\micro\meter} thick, the thickness of the polycarbonate filters of Reinhart and Chien~\cite{Reinhart1986}; the
pore edges are rounded with radius \SI{0.3}{\micro\meter}. The surfaces of the plate and the pore reflect the
membrane vertices that cross them. As there is no flow, a pressure drop $\Delta p$ is imposed as a force on the
bilayer along the pore axis, equal to $\Delta p$ times the mean cross-section of the part of the cell inside the
plate, which is $\Delta p\,\pi r_p^2$ for a cell that fills a pore of radius $r_p$. After the equilibration, a uniform
body force pushes the cell until the front of its bilayer is \SI{0.15}{\micro\meter} inside the pore; from then on a
single $\Delta p$ acts. A cell is held when its center moves less than \SI{0.05}{\micro\meter} in 20 time units, and
has passed when the rear of its bilayer comes within \SI{0.1}{\micro\meter} of the pore outlet. A run in which a bond
of the membrane exceeded its maximum length, or that stopped, before the cell was held or had passed is counted as
failed (Supplementary Text). Each condition was
run with one to four thermostat seeds, with a time step of 0.0005 time units, reduced to 0.00025 or 0.000125 at
higher pressure drops and near the thresholds; runs in which the bilayer crossed itself were repeated with a stronger
self-repulsion of the bilayer (Supplementary Text). Because the drive is a force on the cell, transit times follow
the force rather than the cell (fig.~\ref{fig:force}), so we compare cells by the pressure drop they need to pass.

\textit{Interendothelial slit.} The slit has the dimensions used by Pivkin et al.\ and Li et al.~\cite{Pivkin2016,Li2018}:
a rectangular hole \SI{4.0}{\micro\meter} wide and \SI{1.2}{\micro\meter} high through a wall \SI{1.89}{\micro\meter}
thick, here with edges rounded with radius \SI{0.3}{\micro\meter}. The wall thickness is the mean slit length measured
on electron micrographs of human spleens, and the height the upper end of the measured slit apertures,
0.25--\SI{1.2}{\micro\meter}~\cite{Deplaine2011}. The cell is also turned so that its shortest principal axis lies
along the height of the slit. Walls, drive and entry are as in the pore, with $\Delta p$ acting on the part of the
cell inside the wall. Instead of one pressure drop per run, each run then raises the pressure drop in steps: 2, 3, 5,
7, 10, 15, 20, 30, 50, 75, 100, 150, 200, 300 and \SI{500}{\pascal}, with 0.5, 1 and \SI{1.5}{\pascal} added below
for D, Da025, E0 and one run of E-I with the shear modulus of D. A step ends when the cell has passed or is held
(criteria as in the pore) or after 500 time units, which counts as held; a cell with less than
\SI{2}{\micro\meter\squared} of bilayer left inside the slit is pushed in again before the next step. Single-step
runs of 200 time units from the state at the end of a held step tested intermediate pressure drops
(\SI{4}{\pascal} for E-I, \SI{12.5}{\pascal} for E-II and Da05) or repeated held steps (the repeats of the two
E-III holds started within the step; Supplementary Text). For each run we report the
highest pressure drop at which the cell was held and the lowest at which it passed (table~\ref{tab:brackets}); a run
that failed gives only a lower bound. Runs are scored by the rules of the pore assay (Supplementary Text); under
these rules, one run each of E3ka05, E-II and E-II with the shear modulus of D was replaced, in whole or in part, by
its repeat with the stronger self-repulsion. The walls do not act on a vertex that is already behind them, and
single vertices slipped behind the rounded inlet edge in some runs; repeated with such vertices moved back to the
surface, six runs changed their outcome by at most one step, and the two runs of E-I with the shear modulus of
E-III, whose cytoskeleton had entered the wall, are reported from these repeats (Supplementary Text). Each
condition was run three times with different thermostat seeds (E3ka05 and E-I with the shear modulus of D four
times, E-I with the shear modulus of E-III twice) and a time step of 0.0005 time units. In a further set of
single-step runs, from the hold of one run of E-III at \SI{50}{\pascal} and at that pressure drop, the
contour length of the connecting bonds was left unchanged or raised to 1.5, 2 or $3\,l_c$, in steps of
$0.011\,l_c$ every 0.5 time units, and the run continued for 200 time units after the last step. Each length was
run with two or four thermostat seeds, with the standard and with the stronger self-repulsion (Supplementary Text).

\textit{Constriction channels.} The model of the device has two parallel channels, each \SI{100}{\micro\meter} long
with a cross-section of 5 $\times$ \SI{5}{\micro\meter} and \SI{10}{\micro\meter} apart from center to center. They
open on a chamber \SI{10}{\micro\meter} deep that shares their floor; the box
is periodic along and across the channels, and the edges of the channels are sharp. The membrane model, the walls,
the drive and the entry are as in the slit, with $\Delta p$ acting on the part of the cell inside the channel, and
vertices found behind a wall are moved back to its surface. The cell starts in front of one channel, on its axis
(E-III \SI{0.3}{\micro\meter} higher, to clear the floor), with its longest principal axis along the channel and its
shortest along the height. Each run applies one pressure drop: 0.5, 1, 2, 4, 8 or \SI{16}{\pascal}. A cell has
entered when every vertex of its bilayer is at least \SI{0.1}{\micro\meter} inside the channel. It is held when its
center moves less than \SI{0.05}{\micro\meter} in 20 time units, and a run that has reached neither state after 1000
time units ends there (two runs of D at \SI{0.5}{\pascal}). A cell that has entered is followed until its rear is \SI{10}{\micro\meter} past the entrance,
for at most 400 time units. D, the three echinocyte stages and E-III with the shear modulus of D were each run with
three thermostat seeds and a time step of 0.0005 time units.

\subsection*{Links that can fail}
In the runs with links that can fail, each link is a WLC spring with a power-law repulsion~\cite{Wen2025}, with an
equilibrium length of \SI{0.03}{\micro\meter} and a contour length $l_c = \SI{0.066}{\micro\meter}$. At a stretch
$x = r/l_c$ its tension is
\begin{equation}
F(x) = \frac{k_BT}{p}\left[\frac{1}{4(1-x)^2} - \frac{1}{4} + x\right] - \frac{k_p}{r^2},
\end{equation}
with $k_BT/p = \SI{0.301}{\pico\newton}$ ($p$, persistence length); $k_p$ makes the tension vanish at the equilibrium length. A link fails, and
is removed for the rest of the run, when $x$ reaches the failure stretch $x_b$. The test is made at every time step,
before the forces are computed. At $x_b = 0.950$, 0.965 and 0.975 a link fails at a tension $f_b$ of 30, 62 and
\SI{121}{\pico\newton} (fig.~\ref{fig:heldlinks}B). Links cannot fail while the cell is equilibrated and pushed into
the slit; they can from the first pressure step on. A run stops once 5\% of the links have failed within one pressure
step (24 of the 462 links of an echinocyte, 129 of the 2562 of the discocyte), at or after the onset, and otherwise ends as with unbreakable
links. The count is checked every 2 time units, so a run can go past it before it stops.

We estimate the strength of a link from the energy needed to separate the bilayer from the cytoskeleton, 0.05 to
\SI{0.078}{\milli\joule\per\meter\squared}, measured by pulling tethers from red cells~\cite{Hwang1997,Butler2008}.
Spread over the bilayer area of one mesh vertex of the discocyte (\SI{0.0518}{\micro\meter\squared}) and done over the
contour length $l_c$, this energy gives a force of 39 to \SI{61}{\pico\newton}, that is $x_b = 0.956$ to 0.965; the
three strengths bracket this range. The estimate sets the force at which a link fails, not the energy it stores:
stretched from its equilibrium length to $x_b = 0.950$, 0.965 and 0.975, a link stores 21, 31 and $44\,k_BT$, whereas
the measured energy over the area of one vertex is 600 to $940\,k_BT$. The echinocytes keep one link per 5.5 vertices,
and we give their links the same strength. The weakest also lies within the mean forces at which single
spectrin--ankyrin bonds broke, 21 to \SI{48}{\pico\newton}~\cite{Saito2015}.

D, E0 and E-I to E-III with their own shear moduli ran at the three strengths with the three thermostat seeds of their
runs in Fig.~\ref{fig:slit}B, 45 runs, so that until its first link failed each run repeated a ladder with unbreakable
links step for step. The same 45 runs were repeated with a constraint that keeps every cytoskeleton vertex inside the
bilayer (Supplementary Text). E-I to E-III were also run with the shear modulus of D at the weakest strength, without
the constraint. In controls without a pressure drop, the three runs of E-III at \SI{30}{\pico\newton} were repeated up
to the push into the slit and then held for 160 time units, and for 120 time units at half the time step; two more
were equilibrated \SI{4}{\micro\meter} upstream of the slit and held there for 160 time units; and E-III at
\SI{62}{\pico\newton} and E-II at \SI{30}{\pico\newton} were held after the push for 420 time units. To measure the
stretch of the links at rest, the first run of each cell was repeated up to the end of its equilibration and followed
for 20 time units without a pressure drop (10 for D), with the link vertices written every 10 time steps (20 for D).
For each run we record the pressure step of the first failure and that of the onset. Each failed link is placed at
the position of its bilayer vertex in the last configuration saved before or at the failure (every 2 time units), and
classed by that position along the slit axis (upstream of the wall, in the slit or downstream of it) and by its
distance to the nearest wall surface (more or less than \SI{0.5}{\micro\meter}).

\clearpage

\begin{figure}[htbp]
\centering
\includegraphics[width=\textwidth]{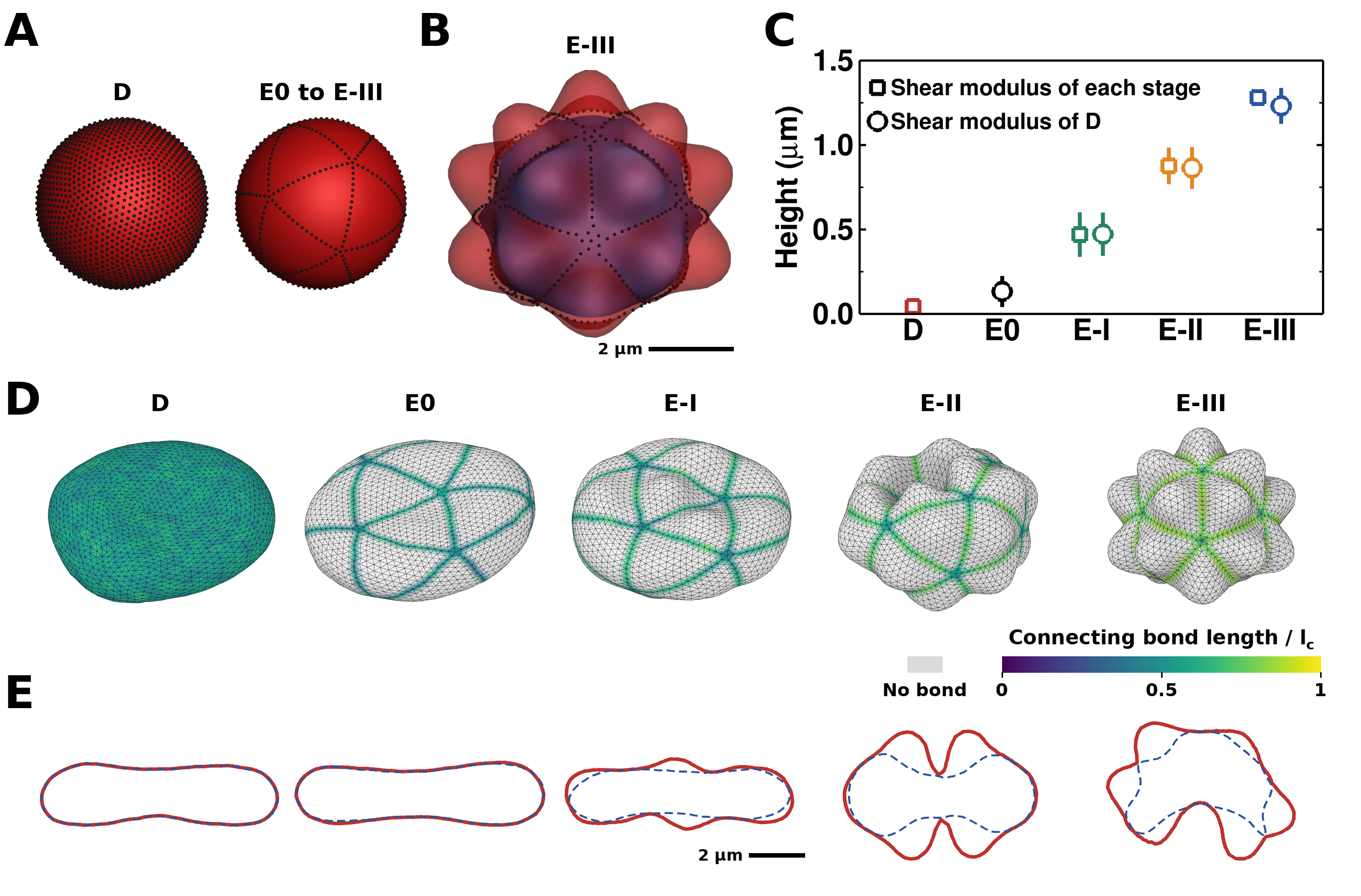}
\caption{\textbf{In the two-component model, the free bilayer patches rise higher above the cytoskeleton from E0 to E-III,
and the connecting bonds stretch further.} (A) Starting bilayer spheres with the
connected vertices as black dots: all 2562 in the discocyte D ($N_c = 16$), and the 462 on the edges of an
icosahedron in E0 to E-III ($N_c = 1$), which leave 20 triangular patches of bilayer free. (B) E-III at rest: the
bilayer (red) translucent over the cytoskeleton (blue), connected bilayer vertices as black dots; for display, the
cytoskeleton is drawn \SI{0.06}{\micro\meter} inward. (C) Height of each patch above the cytoskeleton at its highest
vertex, along the normal of the nearest cytoskeleton vertex (in D, the
same 20 sets of vertices). Squares, the cytoskeleton shear modulus of each stage; circles, that of D,
which E0 also has. Colors, here and in the following figures: D red, E0 black, E-I green, E-II orange, E-III
blue. Mean and s.d.; $n = 20$ patches. (D) The bilayer of each cell at rest, colored by the
connecting-bond length at each vertex over its contour length $l_c$, averaged over the last 20 time units (light
gray, no bond); oblique view as in Fig.~\ref{fig:exp}E. (E) Sections of the same cells
through their longest and shortest axes; bilayer red, cytoskeleton blue and dashed. In D the two layers stay within
\SI{0.06}{\micro\meter} of each other, so their lines overlap. A, D and E share the scale bar in E; scale bars,
\SI{2}{\micro\meter}.}
\label{fig:model}
\end{figure}

\begin{figure}[htbp]
\centering
\includegraphics[width=\textwidth]{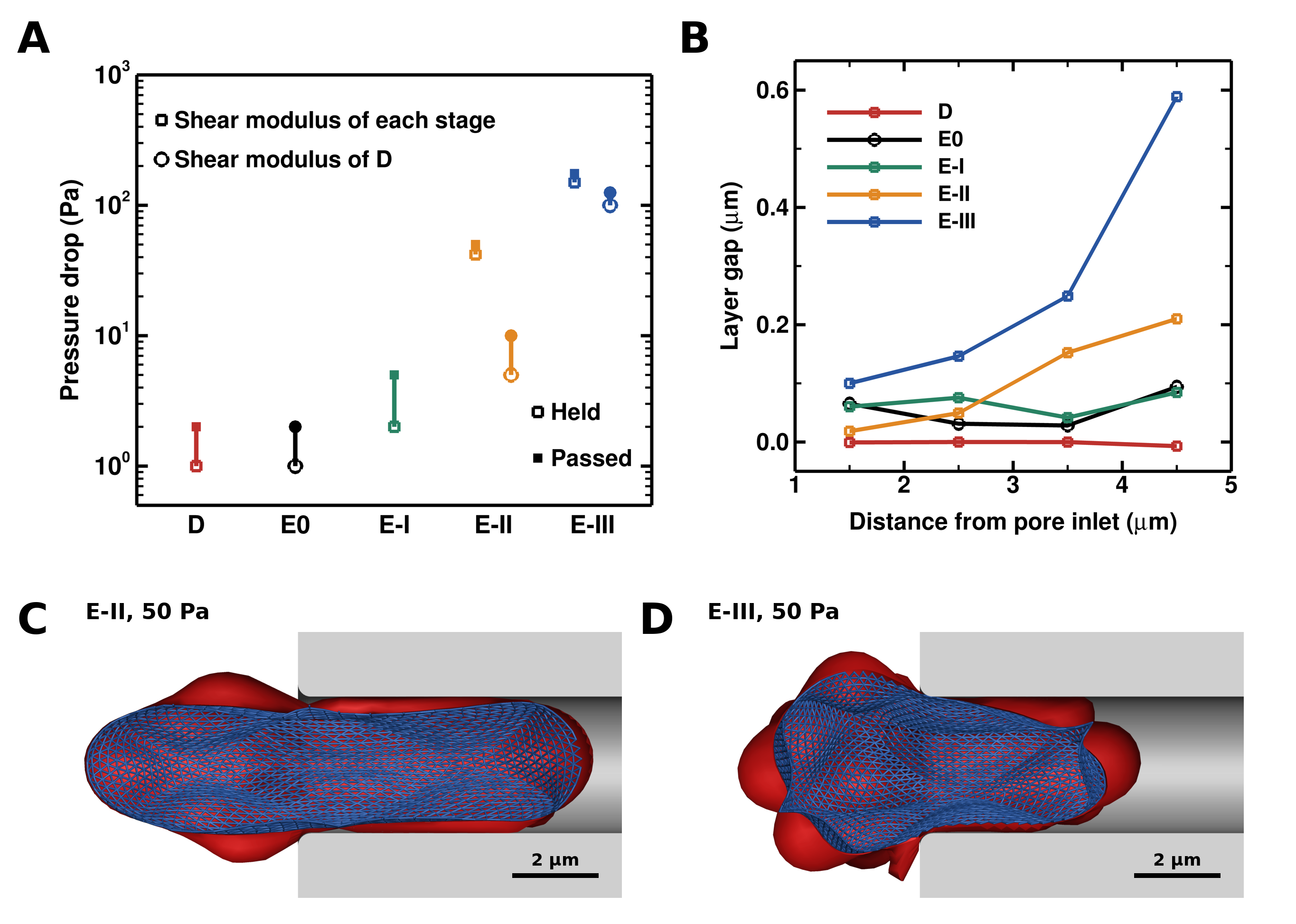}
\caption{\textbf{In a micropore, stages II and III need higher pressure drops to pass than D.} (A) Pressure drops at which single cells were held in, or
passed, a \SI{3.2}{\micro\meter} pore through a \SI{13}{\micro\meter} plate (edge radius \SI{0.3}{\micro\meter}), from
runs at fixed pressure drops: open symbol (held), the highest pressure drop at which a run was held; filled symbol
joined to it (passed), the lowest at which every run passed (at the open symbol, three of four runs of E-II and two of
four of E-III were held). Squares, each stage with its own
cytoskeleton shear modulus; circles, with the shear modulus of D (E0 has the modulus of D).
Colors as in Fig.~\ref{fig:model}. Every run is shown in fig.~\ref{fig:poreruns}. (B) Radial gap between the bilayer and the cytoskeleton in
\SI{1}{\micro\meter} slices of the pore at \SI{50}{\pascal}, with the bilayer front about \SI{5}{\micro\meter} into the
pore. (C) E-II passing and (D) E-III held at \SI{50}{\pascal}; half of each layer is cut away through the pore axis:
bilayer red, cytoskeleton drawn as a blue network of its bonds, plate gray. Scale bars, \SI{2}{\micro\meter}.}
\label{fig:micropore}
\end{figure}

\begin{figure}[htbp]
\centering
\includegraphics[width=\textwidth]{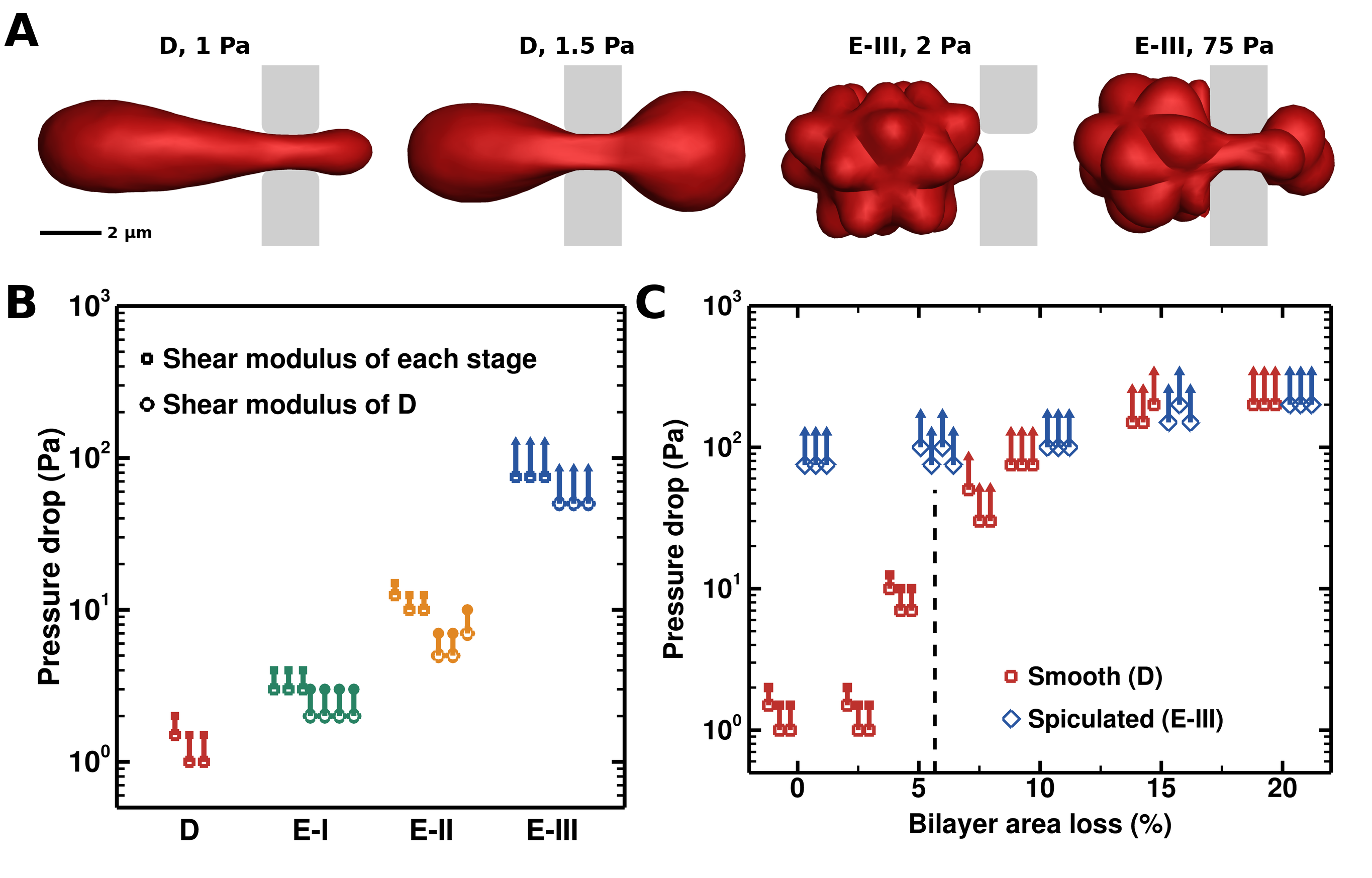}
\caption{\textbf{Spicules raise the pressure drop needed to pass the slit.} Slit \SI{4.0}{\micro\meter} wide and
\SI{1.2}{\micro\meter} high through a \SI{1.89}{\micro\meter} wall; edge radius \SI{0.3}{\micro\meter}. (A) Model
cells moving from left to right as the pressure drop is raised in steps (side views, near half of the wall
removed): D at the end of the \SI{1}{\pascal} step, where it stopped in the slit, and passing at
\SI{1.5}{\pascal}; E-III at the end of the \SI{2}{\pascal} step, held at the inlet, and at the end of the
\SI{75}{\pascal} step, the highest before its run failed at \SI{100}{\pascal}. Scale bar, \SI{2}{\micro\meter}.
(B, C) One mark per run: open symbol, the highest pressure drop that held the cell;
filled symbol joined to it, the lowest at which it passed; a line ending in an arrowhead, held at every step until
the run failed (a lower bound). Runs of one condition are side by side. (B) The stages
at the area and volume of D, with their own cytoskeleton shear modulus (squares) and with that of D (circles),
colored as in Fig.~\ref{fig:model}. (C) Against the loss of bilayer area: smooth family (D, Da025 to Da20; red
squares) and spiculated family (E-III, E3ka05 to E3ka20; blue diamonds). Dashed line, the area loss at which the smooth
cells reach $V/V_2 = 1$ (5.7\%). $n = 3$ runs per condition (E3ka05, 4; E-I with the modulus of D, 4, one started at
\SI{0.5}{\pascal}); every run in table~\ref{tab:brackets}.}
\label{fig:slit}
\end{figure}

\begin{figure}[htbp]
\centering
\includegraphics[width=\textwidth,height=0.53\textheight,keepaspectratio]{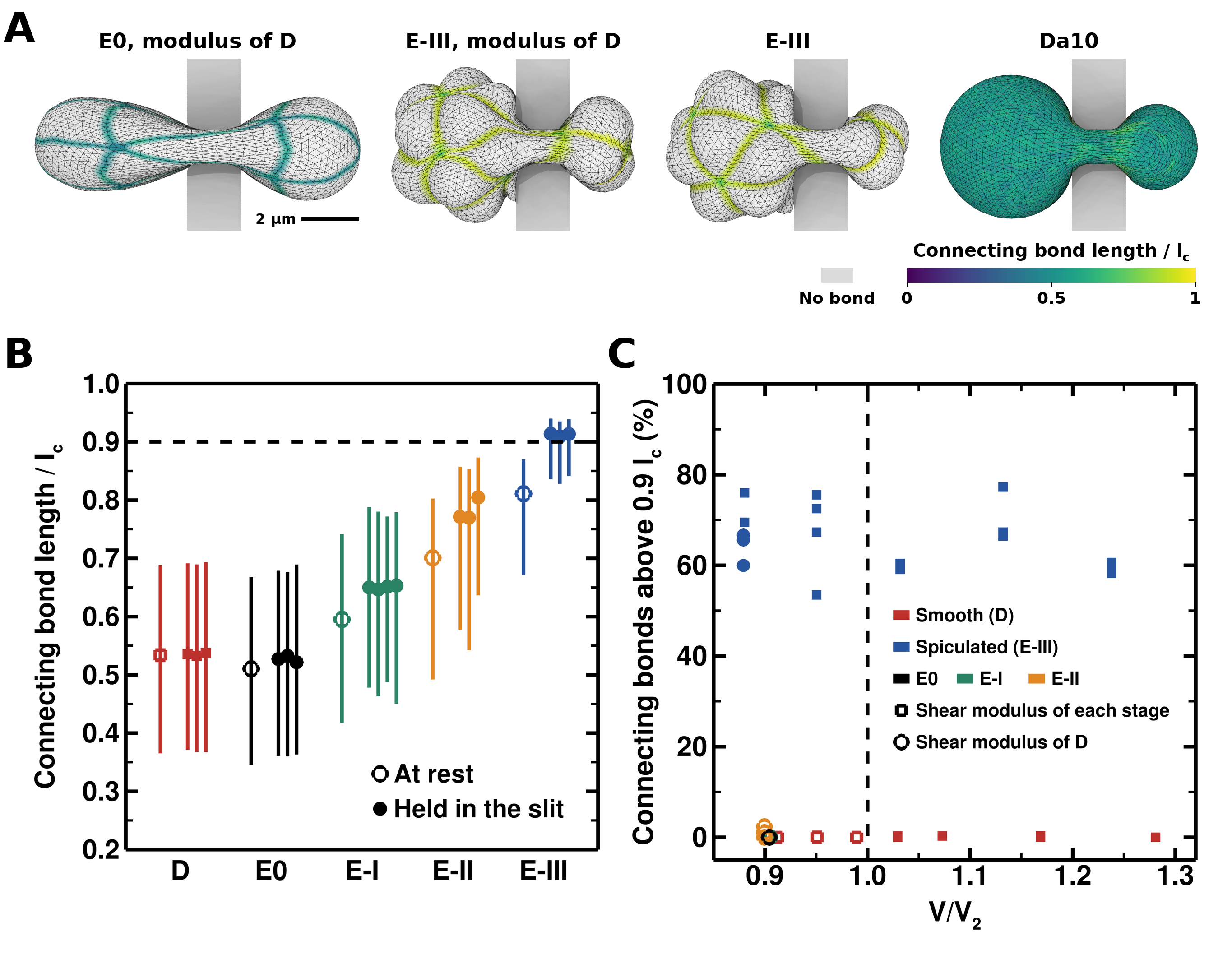}
\caption{\textbf{Spiculated cells are held with their connecting bonds near full extension, smooth cells with them
slack.} Edge radius \SI{0.3}{\micro\meter}. (A) Side views at the end of the highest held step, along the
\SI{4.0}{\micro\meter} width of the slit, near half of the wall removed; bilayer colored as in Fig.~\ref{fig:model}D.
E0 with the modulus
of D, held at \SI{1.5}{\pascal}, passed at \SI{2}{\pascal}; E-III with the modulus of D, held at \SI{50}{\pascal},
failed at \SI{75}{\pascal}; E-III and Da10, held at \SI{75}{\pascal}, failed at \SI{100}{\pascal}. Scale bar,
\SI{2}{\micro\meter}. (B) Length of the connecting bonds over $l_c$ (median; bar, 10th to 90th percentile) at rest
(open; one per cell) and at the end of the highest held step (filled; one per run), for D (squares), E0 and the
stages E-I to E-III with the modulus of D (circles). (C) Share of the connecting bonds longer than $0.9\,l_c$ at the
end of the highest held step against $V/V_2$, one mark per run: open, the run passed at the next step; filled, it
was held at every step until it failed. Squares, the modulus of each stage; circles, that of D. Colors as in
Fig.~\ref{fig:model}, with the smooth family (D, Da025 to Da20) red and the spiculated family (E-III, E3ka05 to
E3ka20) blue. Dashed lines, $0.9\,l_c$ in B and $V/V_2 = 1$ in C. $n = 3$ runs per
condition (E3ka05 and E-I with the modulus of D, 4).}
\label{fig:mechanism}
\end{figure}

\begin{figure}[htbp]
\centering
\includegraphics[width=\textwidth,height=0.5\textheight,keepaspectratio]{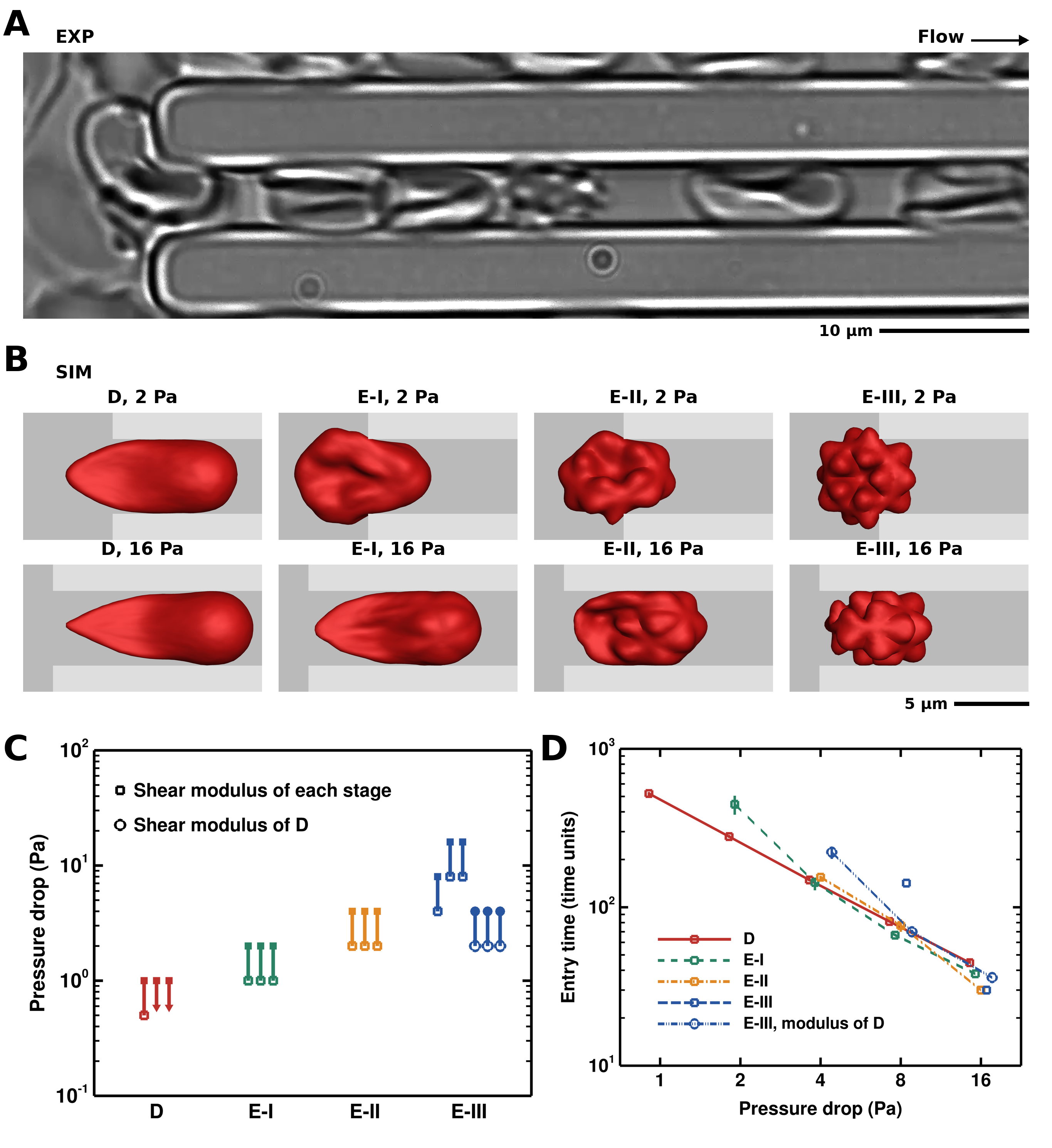}
\caption{\textbf{A constriction channel separates the echinocyte stages far less than the slit.} (A) Bright-field
frame of the device of Fig.~\ref{fig:exp}C, 90 s earlier in the same movie: an echinocyte inside a constriction
channel, between folded discocytes. Flow from left to right. Scale bar, \SI{10}{\micro\meter}. (B) Model cells at
the entrance of the model channel, 5 $\times$ \SI{5}{\micro\meter} in cross-section, seen from above (one of three
runs each). Top row, 200 time units after \SI{2}{\pascal} was applied: D and E-I, which entered (after 274 and
450 time units), and E-II and E-III, which were held. Bottom row, \SI{16}{\pascal}: the first saved frame with the
whole cell inside the channel, after 30 to 46 time units. Scale bar, \SI{5}{\micro\meter}. (C) Highest pressure
drop at which a model cell was held at the entrance (open) and lowest at which it entered (filled), joined by a
line, one per run. A line down to a triangle marks a run in which the cell was not held at any pressure drop (D, two
runs: at \SI{0.5}{\pascal} it was still entering after 1000 time units). Squares, the modulus of each stage; circles,
E-III with the modulus of D. (D) Time from the start of the pressure drop until the whole cell was inside the
channel: mean of three runs with their range; the single square at \SI{8}{\pascal} is the one run of E-III that
entered there. Symbols are shifted sideways for clarity. Colors as in Fig.~\ref{fig:model}.}
\label{fig:chip}
\end{figure}

\begin{figure}[htbp]
\centering
\includegraphics[width=\textwidth,height=0.46\textheight,keepaspectratio]{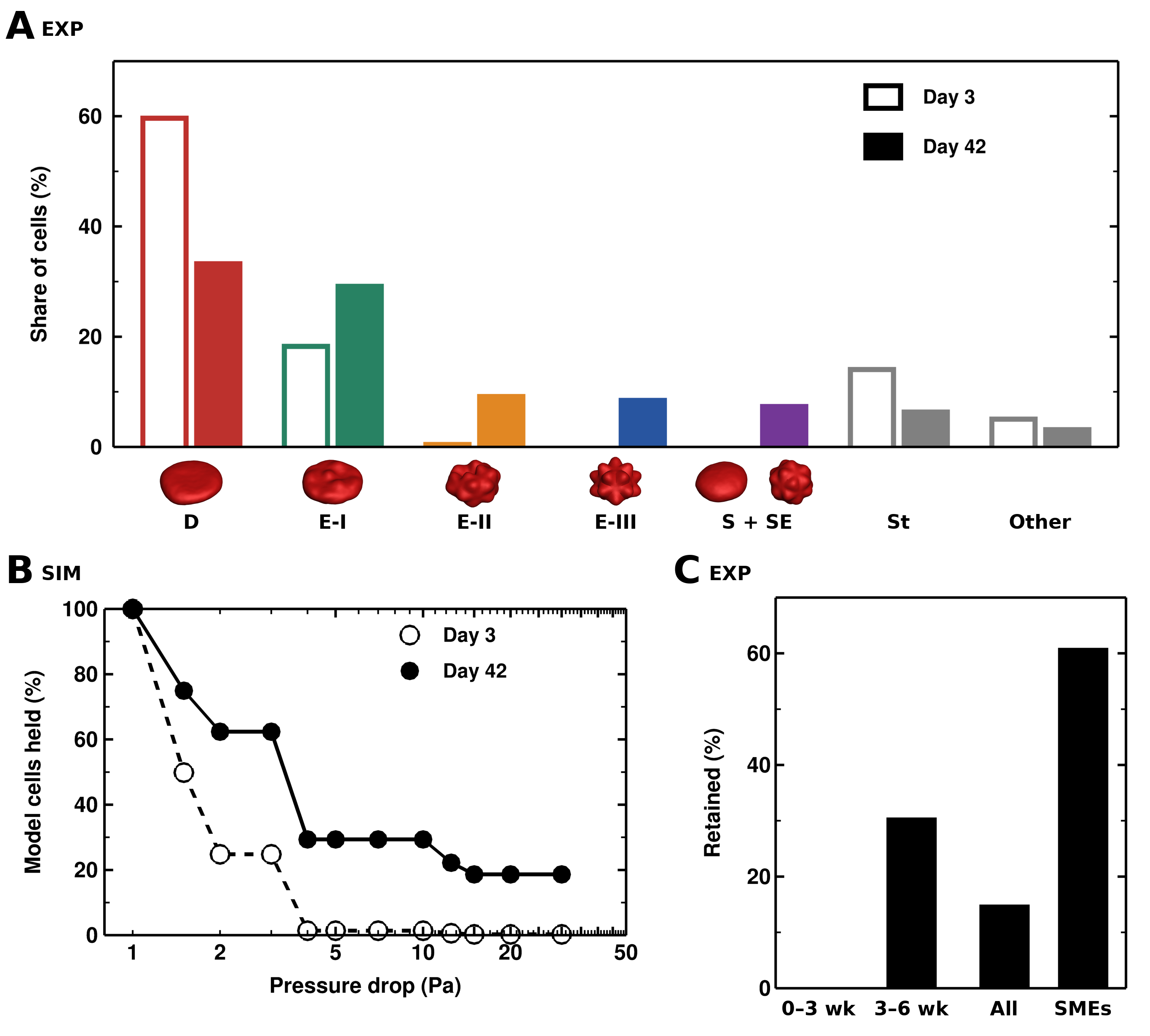}
\caption{\textbf{At higher pressure drops the slit holds only the storage-induced microerythrocytes, the class that
spleens clear most.} (A) EXP: shares of the shape classes of red cells stored for 3 (open
bars) and 42 days (filled bars), counted by differential interference contrast in units from six donors, from
Roussel et al.~\cite{Roussel2017} (D on both days and St on day 42 from their text, the rest read from their
Fig.~4C). Under each class, the model cell that stands for it, oblique view as in Fig.~\ref{fig:exp}E: D and E-I to
E-III, and Da20 and E3ka20 for the smooth (Da075 to Da20) and spiculated (E3ka05 to E3ka20) cells that stand for
spherocytes and spheroechinocytes (S + SE). Stomatocytes (St) and other cells were not modeled (gray bars). E-III, S
and SE together are the storage-induced microerythrocytes (SMEs) of Roussel et al. (B) SIM: share of the modeled cells
of A held in the slit at each pressure drop, own moduli; each class counts the fraction of its runs held
(Fig.~\ref{fig:slit}). From \SI{15}{\pascal} on, only the SME classes were held. The pressure drops end at
\SI{30}{\pascal}, above which two of the three runs of Da075 failed. (C) EXP: left pair, retention of red cells stored for 0--3
and 3--6 weeks in microsphere filters (Deplaine et al.~\cite{Deplaine2011}; none retained at 0--3 weeks); right pair, the
shares of long-stored red cells and of their SMEs cleared by perfused human spleens
within 70 minutes (Roussel et al.~\cite{Roussel2021}).}
\label{fig:population}
\end{figure}

\begin{figure}[htbp]
\centering
\includegraphics[width=\textwidth]{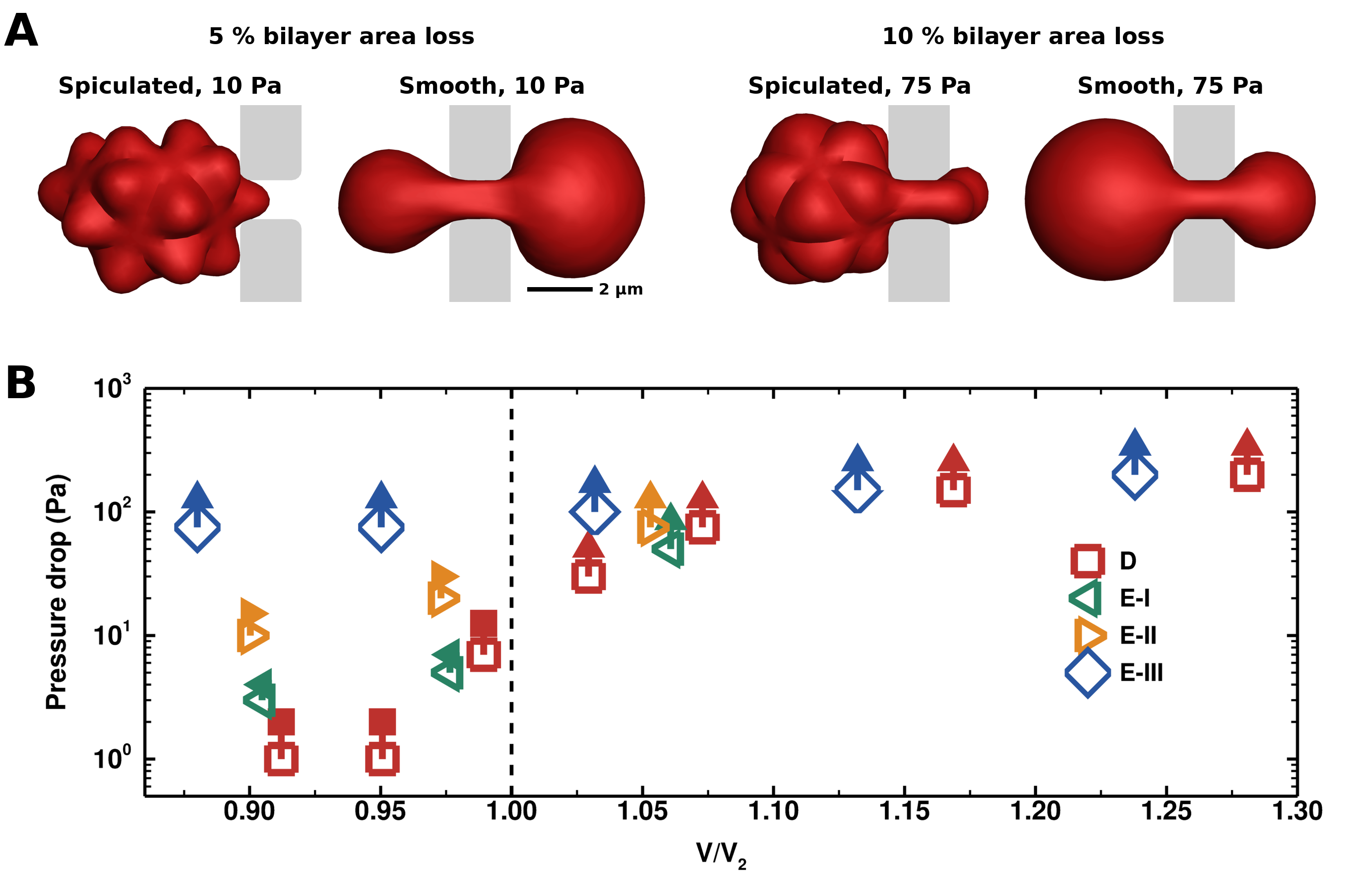}
\caption{\textbf{Below the area limit, reversing the spicules lowers the pressure drop needed to pass; above it, the
cell stays held.} (A) The spiculated cell and the smooth cell of the same bilayer
area in the slit (side views as in Fig.~\ref{fig:slit}A, one run each; scale bar, \SI{2}{\micro\meter}). With 5\%
less bilayer area, the spiculated cell (E3ka05) is held at \SI{10}{\pascal}, where the smooth cell (Da05) passes. With
10\% less, both are held at \SI{75}{\pascal} (E3ka10 and Da10), the highest pressure drop that held both. (B) Pressure drop
against $V/V_2$: D and the smooth family (red squares), E-I with 0, 5 and 10\% loss (green, left triangles), E-II with
0, 5 and 10\% loss (orange, right triangles), E-III and the spiculated family (blue diamonds); E-I and E-II with 10\%
loss ($V/V_2$ 1.058 and 1.056) are drawn 0.003 to either side. Dashed line, $V/V_2 = 1$. Marks as in
Fig.~\ref{fig:slit}, with the runs of a cell combined: for runs that passed, the open symbol is the lowest pressure drop at
which one was held and the filled symbol the highest at which one first passed, so the pass threshold of every
run lies between them; an open symbol with an arrow is the lowest pressure drop up to which every run was held before
the runs failed. $n = 3$ runs per cell (E3ka05, 4).}
\label{fig:rejuvenation}
\end{figure}

\begin{figure}[htbp]
\centering
\includegraphics[width=\textwidth]{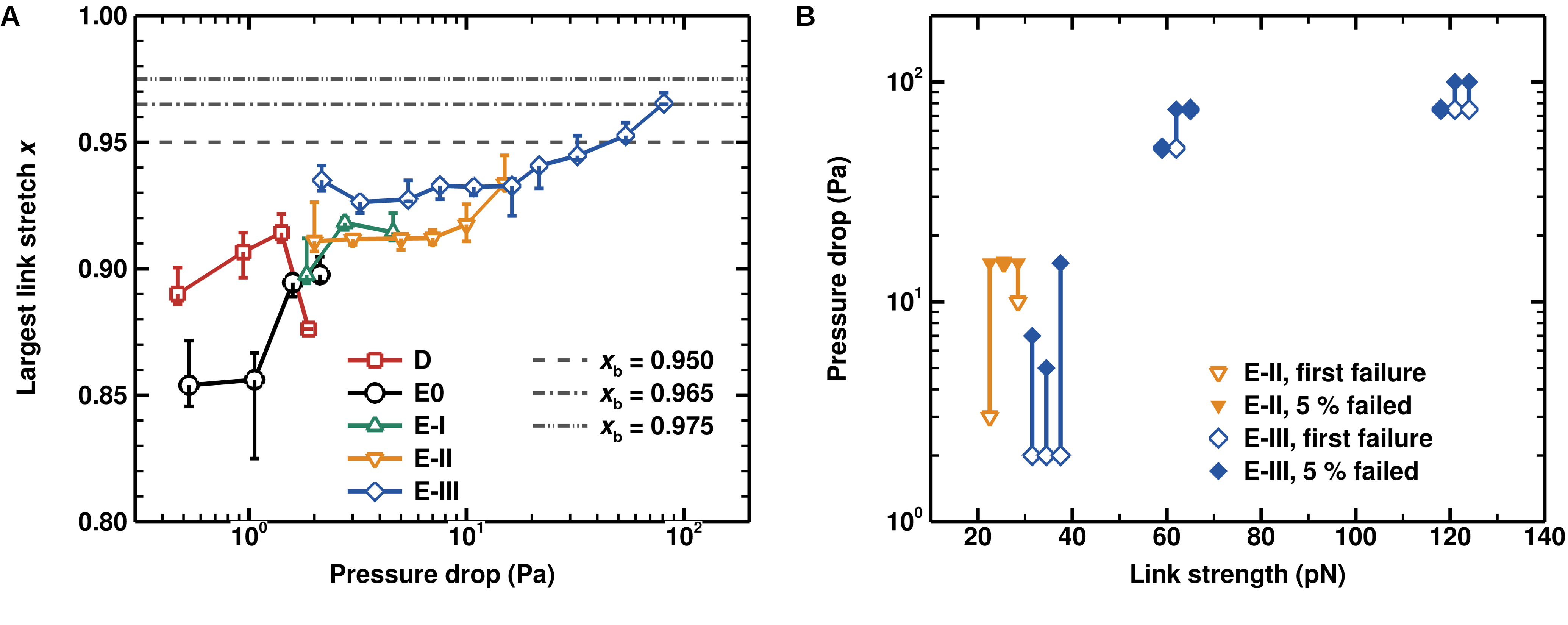}
\caption{\textbf{Stage III echinocytes lose their links at pressure drops that rise with link strength, stage II
echinocytes only when the links are weakest.} (A) Largest stretch $x$ of any link in each pressure step against the
pressure drop, with unbreakable links (the runs of Fig.~\ref{fig:slit}B), from configurations saved every 2 time
units, which miss brief peaks (Supplementary Text). Symbols, median of the three runs of each cell (one for D at \SI{2}{\pascal}, where the other two had passed);
bars, smallest and largest. Points of different cells are shifted slightly along the pressure axis so that their bars
do not overlap. Each curve ends at the pressure drop at which the cell passed or, for E-III, at the last step at which
it was held. Gray lines, the three failure stretches $x_b$. (B) Pressure drop of the first failure (open symbols) and
of the step at which 5\% of the links had failed in all (filled), for each run of E-II and E-III that reached that
count, against the link strength. The runs of one strength are drawn side by side; where the first failure came in the
step of the 5\% count, the filled symbol hides the open one. D, E0 and E-I never reached the count
(table~\ref{tab:onset}). At \SI{30}{\pico\newton}, E-III lost its links as fast without a pressure drop
(fig.~\ref{fig:restcontrols}A). Colors as in Fig.~\ref{fig:model}.}
\label{fig:onset}
\end{figure}

\begin{figure}[htbp]
\centering
\includegraphics[width=\textwidth]{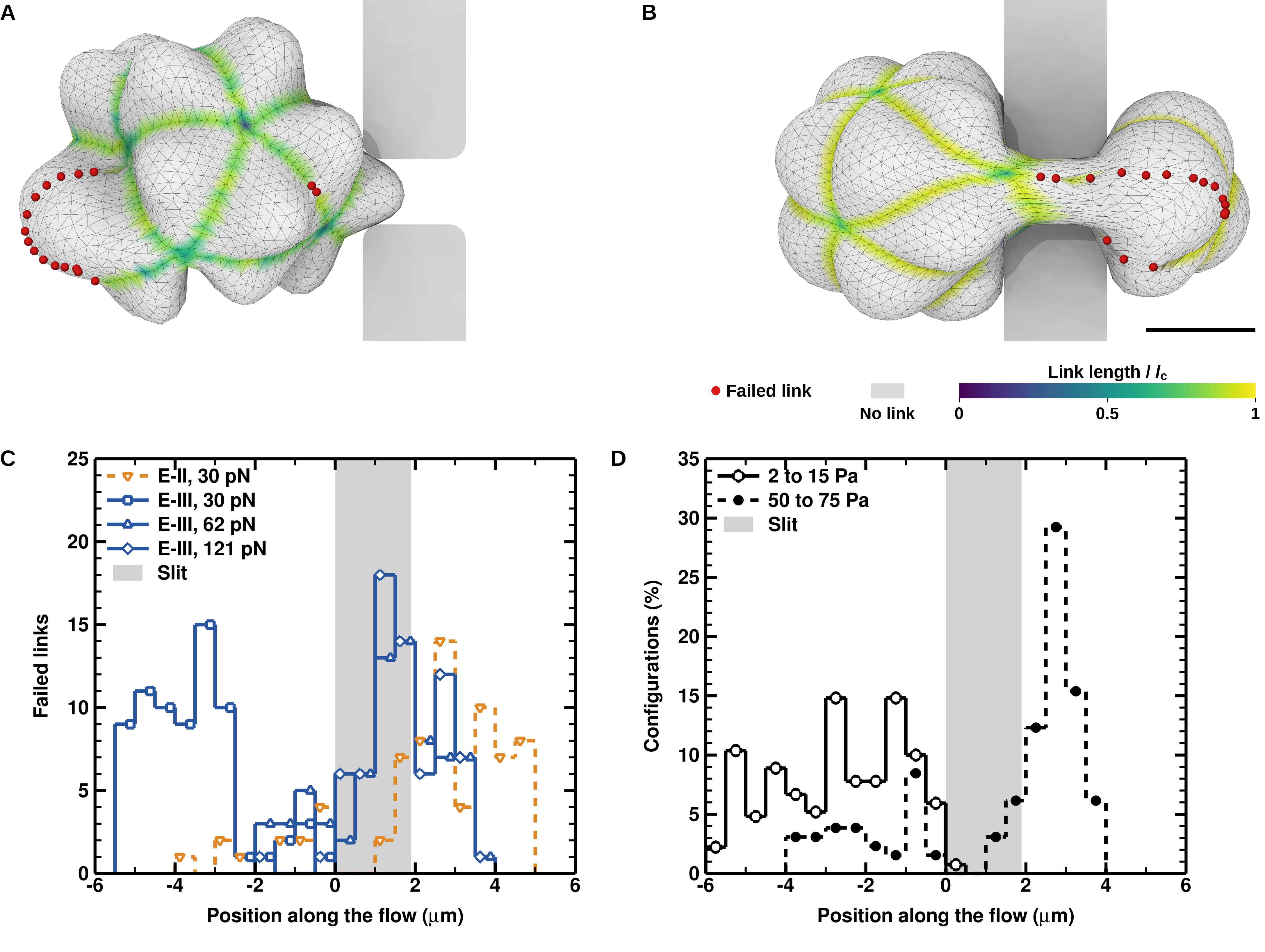}
\caption{\textbf{Weak links fail in front of the slit, stronger links in and behind it.} (A) E-III with links of
\SI{30}{\pico\newton} at \SI{5}{\pascal}, with 23 links failed, 17 of them on the side shown. (B) E-III with links of
\SI{121}{\pico\newton} at \SI{75}{\pascal}, 0.7 time units after its first failure, with 17 links failed, all on the
side shown; the membrane had warmed to 4.4 times its set temperature. In (A) and (B) the cell is seen from the
side, with the flow from left to right, at the same scale and position; bilayer mesh colored by link length over $l_c$
(gray, no intact link), wall gray. Red dots, the bilayer ends of the links failed on the side shown.
Scale bar, \SI{2}{\micro\meter}. (C) Failed links in bins of \SI{0.5}{\micro\meter} along the flow: the first 24
failures of each of the three runs of each cell and link strength that reached that count, 72 per curve. Position 0 is
the slit entrance; gray band, the slit. E-I, with four failed links in all, is left out. (D) Where the most stretched
link lay along the flow when the links could not break, in the three runs of E-III of Fig.~\ref{fig:slit}B:
percentage of the configurations saved every 2 time units at 2 to \SI{15}{\pascal} (270 configurations) and at 50 to
\SI{75}{\pascal} (130), where the weakest and the stronger links failed. Bins and band as in (C).}
\label{fig:position}
\end{figure}

\clearpage

\section*{Acknowledgments}
Simulations were carried out at the Center for Computation and Visualization of Brown University.

\paragraph*{Funding:} Research reported in this publication was supported by the National Heart, Lung, and Blood
Institute (NHLBI) and the National Institute of General Medical Sciences (NIGMS) of the National Institutes of Health
under award numbers R01HL154150 and R01GM163243. Award
R01HL154150 totals \$1,269,793 to date and award R01GM163243 totals \$394,808 to date, each with 100\% from federal
sources. The content is solely the responsibility of the authors and does not necessarily represent the official views
of the National Institutes of Health.

\paragraph*{Author contributions:} Z.C. and G.E.K. conceived the study. Z.C. designed and performed the simulations, analyzed the data, and wrote the
manuscript. J.Z. designed and performed the experiments and contributed to writing the manuscript. H.L. contributed
scientific suggestions and feedback. M.D. supervised the experimental work. G.E.K. supervised the project. All
authors discussed the results, edited the manuscript, and approved the final version.

\paragraph*{Competing interests:} There are no competing interests to declare.

\paragraph*{Data and materials availability:} All data needed to evaluate the conclusions in the paper are present in
the paper and/or the Supplementary Materials. The simulations used an in-house extension of
\textsc{lammps}~\cite{thompson2022lammps}; every model equation and parameter is given in Materials and Methods and
the Supplementary Materials. This study did not generate any new materials.

\subsection*{Supplementary materials}
Supplementary Text\\
Figs.~S1 to S13\\
Tables S1 to S4\\
Movies S1 to S4 (ancillary files)

\clearpage
\setcounter{page}{1}
\renewcommand{\thepage}{S\arabic{page}}
\setcounter{figure}{0}
\renewcommand{\thefigure}{S\arabic{figure}}
\setcounter{table}{0}
\renewcommand{\thetable}{S\arabic{table}}
\setcounter{equation}{0}
\renewcommand{\theequation}{S\arabic{equation}}

\begin{center}
{\Large Supplementary Materials for}\\[4pt]
{\Large\bfseries \papertitle}\\[10pt]
Zhaojie Chai$^{\dagger}$, Jianlu Zheng$^{\dagger}$, He Li, Ming Dao, George Em Karniadakis\\[4pt]
{\small $^\dagger$These authors contributed equally to this work.}
\end{center}

\subsubsection*{This PDF file includes:}
Supplementary Text\\
Figures S1 to S13\\
Tables S1 to S4\\
Captions for Movies S1 to S4

\subsubsection*{Other Supplementary Materials for this manuscript:}
Movies S1 to S4 (ancillary files Movie\_S1.mp4 to Movie\_S4.mp4)

\clearpage
\section*{Supplementary Text}

The parameters of the two-component model are listed in table~\ref{tab:model}.

\begin{table}[htbp]
\centering
\caption{\textbf{Parameters of the two-component model.} Values of Wen et al.~\cite{Wen2025} (their Tables~1 and 2) unless marked. E0 and the runs
with the shear modulus of D take $\mu$ of D; the area-loss families take the settings of D (Da) or E-III (E3ka) with reduced
bilayer area targets.}
\label{tab:model}
\footnotesize
\begin{tabular}{lcccc}
\hline
 & D & E-I & E-II & E-III \\
\hline
Connected vertex pairs ($N_c$) & 2562 (16) & 462 (1) & 462 (1) & 462 (1) \\
Target shrinkage ratio $r_{sh}$ & 0 & 0.113 & 0.226 & 0.377 \\
Cytoskeleton shear modulus $\mu$ (N/m) & \num{2.5e-8} & \num{2.92e-7} & \num{2.29e-6} & \num{9.1e-6} \\
Bilayer bending coefficient $k_b$ (J) & \multicolumn{4}{c}{\num{3.802e-19}} \\
Cytoskeleton bending coefficient (J) & 0 & \multicolumn{3}{c}{\num{3.56e-19}} \\
Bilayer $k_a$, $k_A$ (N/m) & \num{1.98e-5}, \num{1.98e-5} & \multicolumn{3}{c}{\num{2.27e-4}, \num{4.54e-4}$^{a}$} \\
Bilayer $k_V$ (Pa) & 19.8 & \multicolumn{3}{c}{454$^{a}$} \\
Cytoskeleton $k_a$, $k_A$ (N/m) & 0 & \multicolumn{3}{c}{\num{2.27e-4}, \num{4.54e-4}} \\
Cytoskeleton $k_V$ (Pa) & 0 & \multicolumn{3}{c}{454} \\
Cytoskeleton volume target & -- & \multicolumn{3}{c}{$0.94\,V_0(1-r_{sh})^{3/2}$\,$^{a}$} \\
Connecting bonds $k_{bs}$ (N/m) & \multicolumn{4}{c}{\num{3.96e-5}} \\
\hline
\end{tabular}
\par\smallskip
{\footnotesize $^{a}$Our choice (Materials and Methods). $N_c$ is the connecting number: the connected pairs lie on the
edges of the mesh that divides each edge of the icosahedron into $N_c$ segments. Both networks have $N_v = 2562$ vertices; the WLC contour length is 2.2
times the stress-free spring length; $A_0 = \SI{132.6}{\micro\meter\squared}$, $V_0 = \SI{92}{\micro\meter\cubed}$.
In the filtration assays the bilayer penalties are stiffened (Materials and Methods).}
\end{table}

\subsection*{Micropore sieving}

\subsubsection*{Drive force and transit times}

The drive acts on the bilayer along the pore axis with a force equal to $\Delta p$ times the mean cross-section of the
part of the cell inside the plate. A cell held at \SI{20}{\pascal} or more plugs the pore and receives 0.81--0.93 of
$\Delta p\,\pi r_p^2$ ($r_p = \SI{1.6}{\micro\meter}$); a cell held at 1 to \SI{5}{\pascal} has less of its volume
inside the plate and receives 0.39--0.84 of it. A passing cell receives on average 0.39--0.74 of it, and at one
pressure drop the
echinocyte stages receive more force than D (at \SI{50}{\pascal}: 0.69 for E-II, 0.50--0.51 for D;
fig.~\ref{fig:force}A). The impulse needed to pass, the time integral of the drive force up to the pass, is about the
same for every cell. In the 4.5 and \SI{6.9}{\micro\meter} pores at \SI{50}{\pascal}, where every cell passes, it
lies within 3\% of its median $I_0$ (seven runs of D, E0 and the echinocyte stages). In the \SI{3.2}{\micro\meter}
pore it is 0.96--1.04~$I_0$ for D, E0, E-I, and E-II and E-III with the modulus of D away from their thresholds. It is
larger at and just above the thresholds: 1.07--1.16~$I_0$ for D and E0 at \SI{2}{\pascal}, E-I at \SI{5}{\pascal}
and E-II with the modulus of D at \SI{10}{\pascal}, the lowest pressure drops at which they passed, and
1.08--1.51~$I_0$ for E-II at 42 and \SI{50}{\pascal}, E-III at 150 to \SI{200}{\pascal} and E-III with the modulus
of D at \SI{125}{\pascal}. The transit time is therefore close to $I_0/F$ for the
mean force $F$ (fig.~\ref{fig:force}B). It reports how much force the cell received, and we do not compare it with
filtration resistances. Forces are in pN for an energy unit of \SI{4.529e-20}{\joule} and a length unit of
\SI{1}{\micro\meter}; the model time has no physical calibration (Materials and Methods), so fig.~\ref{fig:force}B gives the
impulse in units of $I_0$.

\subsubsection*{Which runs give an outcome}

Each combination of cell, cytoskeleton modulus, edge radius, pressure drop and thermostat seed gives at most one
outcome, from the finished run with the smallest time step. With \SI{0.3}{\micro\meter} edges the time step is 0.0005
time units for most runs up to \SI{50}{\pascal}, 0.00025 or 0.000125 at \SI{100}{\pascal}, and 0.000125 from
\SI{125}{\pascal} and for E-II from \SI{35}{\pascal}; with 0.5 and \SI{1.0}{\micro\meter} edges (fig.~\ref{fig:edge})
it is mostly 0.0005 or 0.00025.

Bilayer vertices repel each other with a soft potential of amplitude \SI{4.5e-18}{\joule} (100 energy units) and
cutoff \SI{0.4}{\micro\meter}. Every
saved frame was checked for crossings of the bilayer surface with itself. A run with crossings was repeated with
amplitude \SI{1.36e-17}{\joule} (300 energy units; same seed and time step), and the repeat replaced it when it did not fail and had crossings in fewer
frames. One replacement changed an outcome: E-II at \SI{42}{\pascal} (one seed) passed with crossings and was held
without them. The other replacements kept the outcome. For cells whose bilayer does not fold onto itself, amplitude
300 changed nothing (D at \SI{10}{\pascal} and Da05 at \SI{20}{\pascal} passed in the same 172.5 and 96.5 time units
with either amplitude). In the renders of Fig.~\ref{fig:micropore}, C and D, every vertex lies outside the wall, but flat
faces of the mesh cut across the rounded inlet edge by up to 0.08 and \SI{0.19}{\micro\meter}; these faces are drawn along the edge.

A run that held or passed while a bond was stretched beyond its maximum length gives no outcome; its repeat with
the stronger repulsion, free of such bonds, stands in (one case, E-III at \SI{150}{\pascal}). Runs in which a bond exceeded its
maximum length before the cell held or passed are counted as failed. D at \SI{175}{\pascal} stopped with its bilayer
front past the plate exit and its rear caught at the outlet, at time steps 0.000125 and 0.0000625, and gives no
outcome. The runs of D at \SI{200}{\pascal} and of Da10 at \SI{175}{\pascal} failed, so the model does not reach the pressure drops
of the filtration experiments.

\begin{figure}[htbp]
\centering
\includegraphics[width=0.8\textwidth]{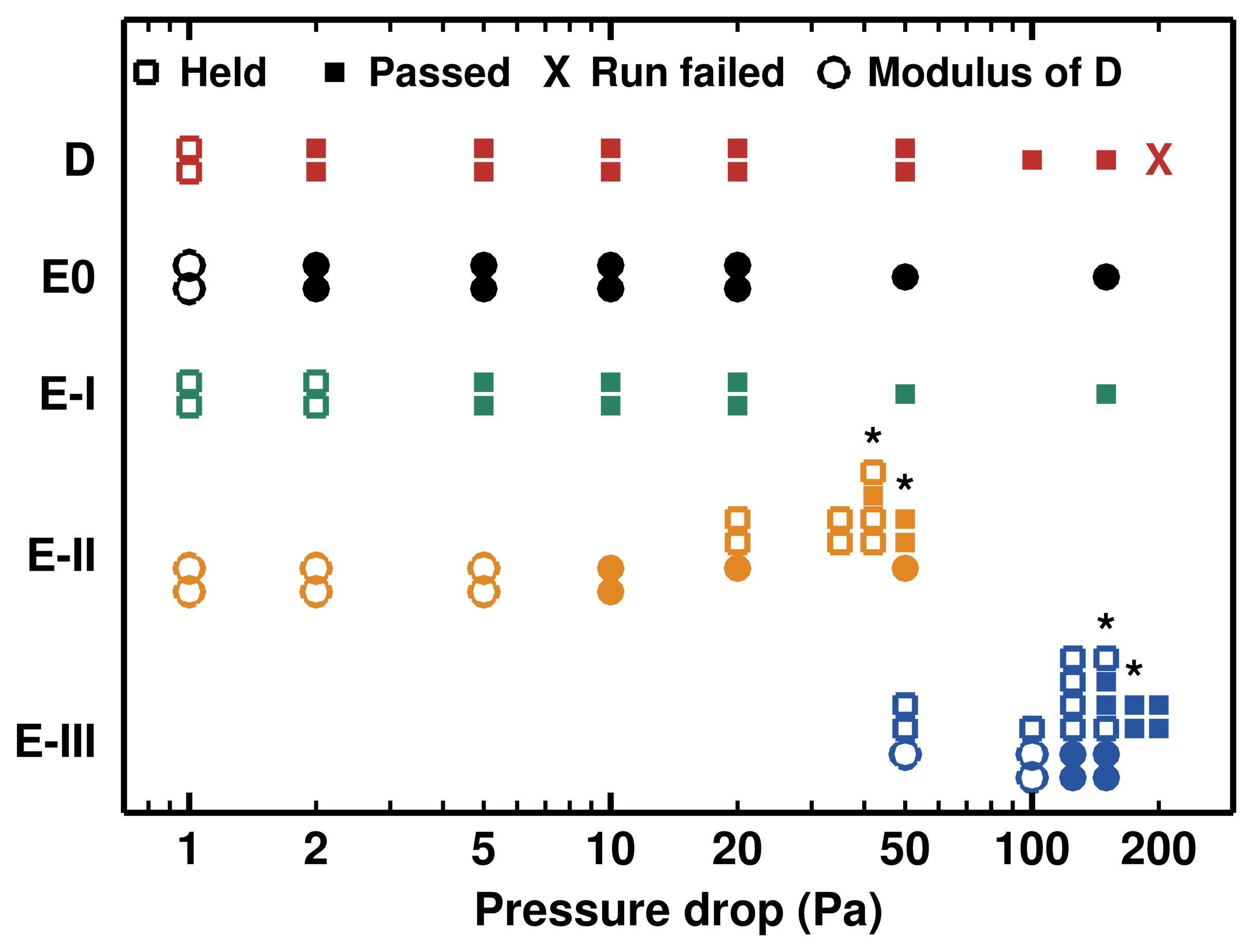}
\caption{\textbf{Every run in the \SI{3.2}{\micro\meter} pore.} Outcome of single-cell runs through a
\SI{13}{\micro\meter} plate (edge radius \SI{0.3}{\micro\meter}) at fixed pressure drops, one mark per run, from
which the brackets of Fig.~\ref{fig:micropore}A are taken: open, held; filled, passed; X, the run failed (a bond
exceeded its maximum length before the cell was held or passed). Squares, the cytoskeleton shear modulus of each stage;
circles, the modulus of D (E0 has the modulus of D). Colors as in Fig.~\ref{fig:model}. Runs at one pressure drop,
each with its own thermostat seed, are stacked, squares above and circles below the line of their stage. An asterisk
marks a stack in which a run with the stronger bilayer self-repulsion replaces a standard run whose bilayer crossed
itself or, at \SI{150}{\pascal}, had a bond stretched beyond its maximum length; the repeat is the mark next to the
asterisk. 81 runs.}
\label{fig:poreruns}
\end{figure}

\begin{figure}[htbp]
\centering
\includegraphics[width=\textwidth]{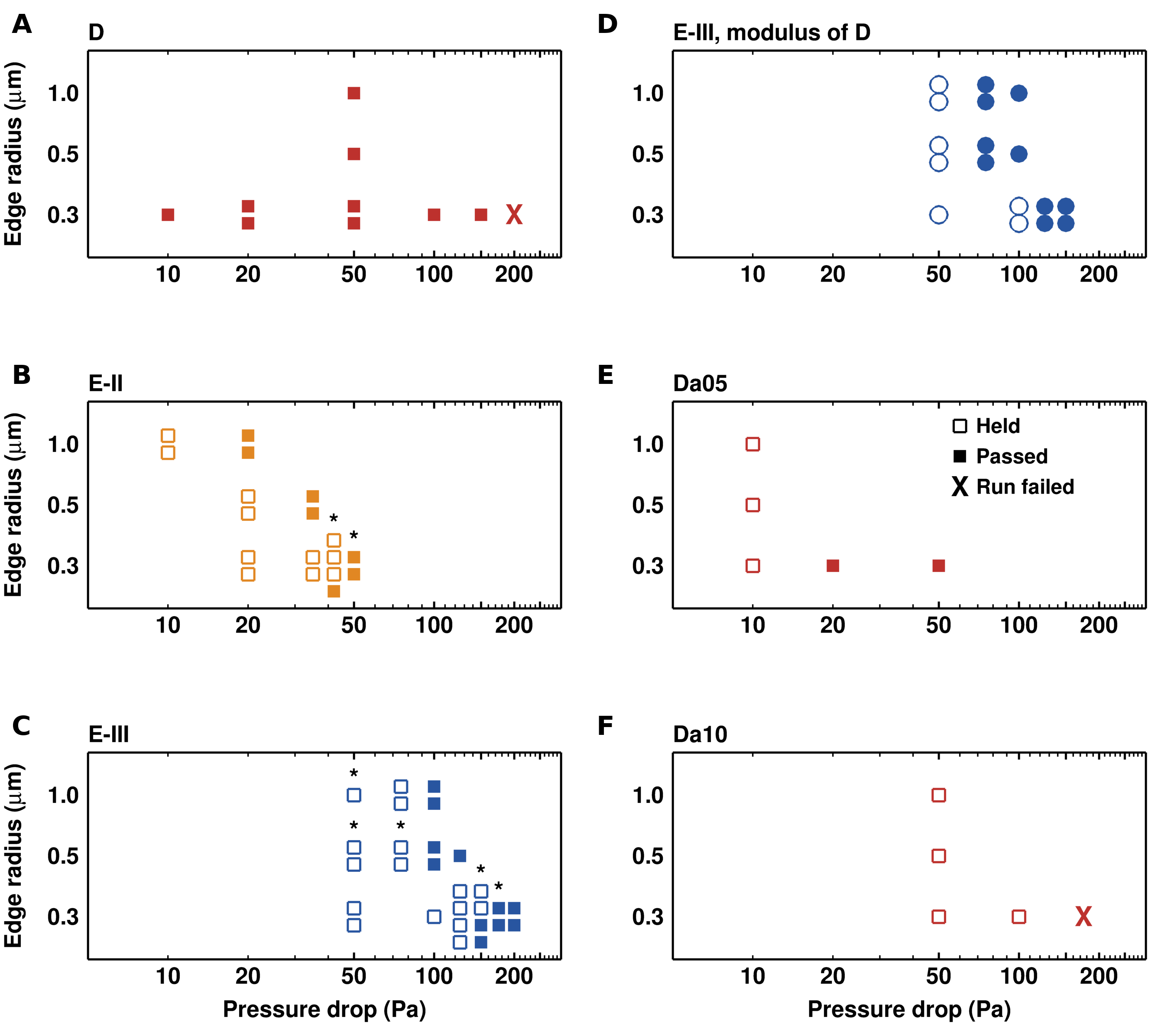}
\caption{\textbf{Micropore sieving with rounder pore edges.} Outcome against pressure drop, as in fig.~\ref{fig:poreruns}, for
pore edges rounded with radius 0.3 (as in Fig.~\ref{fig:micropore}), 0.5 and \SI{1.0}{\micro\meter}: (A) D, (B) E-II, (C) E-III, (D)
E-III with the cytoskeleton shear modulus of D, (E) Da05 and (F) Da10, discocytes with 5 and 10\% less bilayer area.
Open, held; filled, passed; X, the run failed; an asterisk marks a stack that holds a run with the stronger bilayer
self-repulsion in place of a standard run (top of the stack). With 0.5 and \SI{1.0}{\micro\meter} edges, one or two
seeds per pressure drop (two at each end of a threshold bracket).}
\label{fig:edge}
\end{figure}

\begin{figure}[htbp]
\centering
\includegraphics[width=\textwidth]{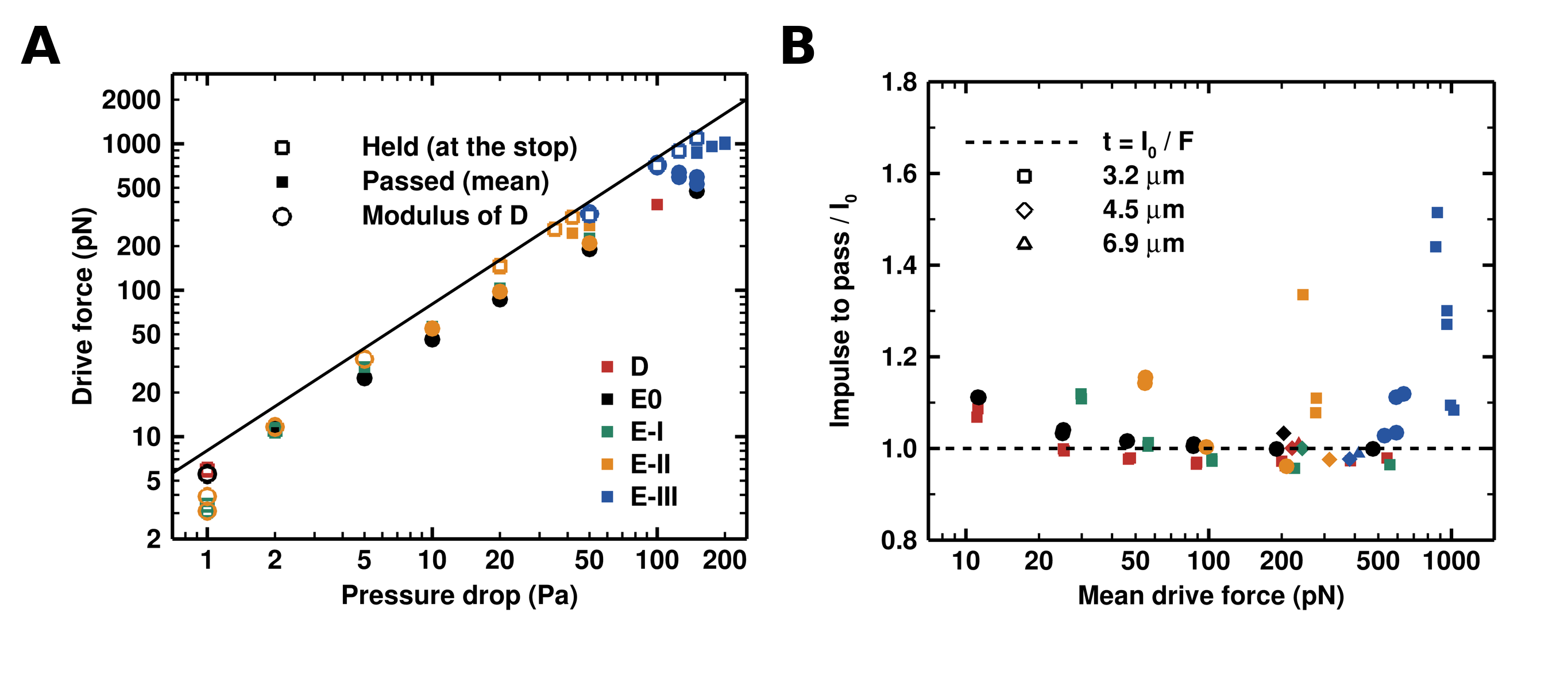}
\caption{\textbf{Drive force and transit time in the micropore runs.} (A) Drive force vs pressure drop in the
\SI{3.2}{\micro\meter} pore for the runs of fig.~\ref{fig:poreruns} (runs that failed are left out): open, held cells at the stop;
filled, mean force up to the pass; squares, the cytoskeleton shear modulus of each stage; circles, the modulus of D; line,
$\Delta p\,\pi r_p^2$, the force on a cell that fills the pore. (B) Impulse to pass (time integral of the drive force
up to the pass) over $I_0$, its median in the 4.5 and \SI{6.9}{\micro\meter} pores, vs the mean drive force up to the
pass, for the passing runs in the \SI{3.2}{\micro\meter} pore (squares and circles as in A) and at
\SI{50}{\pascal} in the 4.5 (diamonds) and \SI{6.9}{\micro\meter} (triangles) pores; dashed line, a transit time
$t = I_0/F$. Colors as in Fig.~\ref{fig:model}: D red, E0 black, E-I green, E-II orange, E-III blue. 47 passing runs in the
\SI{3.2}{\micro\meter} pore, seven in the wider pores.}
\label{fig:force}
\end{figure}

\subsection*{Interendothelial slit}
In the slit, each run raised the pressure drop in steps (Materials and Methods); below, the steps of one run are
called its ladder and each step a level. The pressure drops at which each run was held and passed are listed in
table~\ref{tab:brackets}.

\subsubsection*{Which runs give an outcome}

Every slit run that gives a mark used a time step of 0.0005 time units; repeated at half that time step,
runs failed again at the levels at which they had failed (E3ka05 at 150 and E-III at \SI{100}{\pascal}, two thermostat
seeds each). The single-step runs (Materials and Methods) started from the last saved frame of a held level, with the velocities set to zero and a new thermostat seed, and ran for 200 time units. They
gave the passes of E-I at \SI{4}{\pascal} (three seeds, after 122--130 time units) and of E-II (two seeds, after
108 and 171 time units) and Da05 (one seed, after 39 time units) at \SI{12.5}{\pascal}. Ten more repeated seven held
levels, and nine of them held for the full 200 time units; the tenth, a repeat of the \SI{75}{\pascal} hold of one
E-III run with a second thermostat seed, failed after 10 time units, and the mark keeps the hold of the ladder, which
the repeat with the first seed confirmed. The repeats of the two E-III holds started from another saved frame of the
level itself: at \SI{50}{\pascal}, 17.5 time units before its end, within the 20 time units in which the cell met the
hold criterion; at \SI{75}{\pascal}, 0.5 time units after its start.

The runs that failed stopped when the simulation lost the atoms of a bond or a triangle of the membrane. In 36 of the
37 runs that failed in Figs.~\ref{fig:slit} and~\ref{fig:rejuvenation}, a bond had first been stretched beyond its maximum length;
that bond lay within \SI{0.3}{\micro\meter} of an edge of the slit in 20 runs, elsewhere within \SI{0.3}{\micro\meter}
of the wall in 15, and \SI{0.6}{\micro\meter} downstream of the wall in one. It was a bond between the bilayer and
the cytoskeleton in 30 runs (23 of the 25 runs of spiculated cells that failed and 7 of the 12 runs of smooth cells),
a bond of the cytoskeleton in five and an edge of the bilayer mesh in one. The remaining run (E-III, one seed) stopped
without such a bond. One run (E3ka20, one seed) did not stop by itself. A bond stayed beyond its maximum length from
20 time units into \SI{300}{\pascal}, and the run was ended there and counts as failed at that level. One run of
E3ka05 was held at \SI{100}{\pascal} while a bond stayed beyond its maximum length for most of the level; as in the
pore it gives no outcome, and its repeat with the stronger repulsion (same seed and time step), which had no such bond before
it failed at \SI{150}{\pascal}, stands in. Applied from the start, the levels at which the ladders of E-III and E-III with the
modulus of D failed (100 and \SI{75}{\pascal}) made the runs fail again within 5--10 time units, at the
wall, so the failures do not come from the long holds at the lower levels.

Every saved frame of the runs in Figs.~\ref{fig:slit} and~\ref{fig:rejuvenation} and of the single-step runs that give marks was checked
for crossings of the bilayer surface with itself, and the rule of the pore was applied to every run with crossings in
a level that enters its mark. Crossings appeared in one or two frames of four runs. In E-II, seed 1, the single-step
run that passed at \SI{12.5}{\pascal} had them in 2 of its 101 frames; its repeat with the stronger repulsion held for the 200
time units without crossings and replaces it. The pass of this seed then comes from the \SI{15}{\pascal} level of its
ladder, which had crossings in 1 of its 190 frames; the ladder repeated with the stronger repulsion passed at this level without
crossings, after 63.5 time units, and replaces it, so the seed is held at 12.5 and passes at \SI{15}{\pascal}. In E-II
with the modulus of D, seed 2, the ladder had crossings in 1 of its 330 frames, during the pass at \SI{10}{\pascal};
repeated with the stronger repulsion, it passed at \SI{7}{\pascal} without crossings and replaces the run. In E-III with the
modulus of D, seed 3, the ladder had crossings in 1 of its 236 frames, in the level at which it failed
(\SI{75}{\pascal}); the repeat failed at the same level, and the run stays. The two replacements moved a mark by one
level each, in opposite directions.

Each patch of the walls reflects a vertex that crosses it within a time step but does not act on a vertex that is
already behind it, and a vertex reflected by one facet of a rounded edge can land behind the next. We therefore
measured, in every saved frame, how far each vertex lay inside the wall (its distance to the nearest surface of the
wall). With the edges of radius \SI{0.5}{\micro\meter}, no vertex lay deeper than \SI{0.001}{\micro\meter}, the
precision of the saved coordinates, in any of the 53 ladders. With the edges of radius \SI{0.3}{\micro\meter},
vertices lay deeper in five of the 70 ladders of the main text (E-III, seed 1; E-II, seeds 2 and 3; E-II with 5\% loss,
seeds 1 and 3), in both runs
of E-I with the modulus of E-III made without the option described below, and in 10 of the 16 single-step runs that give marks, always at the inlet edge of the
slit, 0.07--\SI{0.35}{\micro\meter} behind the inlet face. The deepest were bilayer vertices, up to
\SI{0.15}{\micro\meter} inside, except in E-I with the modulus of E-III, in which cytoskeleton vertices lay up to
\SI{0.33}{\micro\meter} inside in 46 and 50 frames of the level at which the cell was held (\SI{15}{\pascal}). We
repeated six ladders (E-II, E-III and E-I with the modulus of E-III, seeds 1 and 2) with an option that, after each
time step, moves any vertex inside the wall to \SI{0.001}{\micro\meter} in front of its surface and removes its
velocity into the wall; no vertex then lay deeper than \SI{0.001}{\micro\meter}. E-II passed at the same level with
seed 1 and one level higher with seed 2 (20 instead of \SI{15}{\pascal}); E-III was held up to 50 and failed at
\SI{75}{\pascal} with both seeds, one level lower, including seed 2, which had no vertex inside the wall without the
option; and E-I with the modulus of E-III was held at 10 and passed at \SI{15}{\pascal} with both seeds, one level
lower. The levels thus shifted by at most one step, in either direction, as they do between seeds, and the order of
the stages and which cells passed did not change. The outcomes of E-I with the modulus of E-III in the main text are
those of the repeats, because its hold at \SI{15}{\pascal} came with cytoskeleton vertices inside the wall and did
not recur in them; all other outcomes are those of the runs without the option.

\begin{table}[htbp]
\centering
\caption{\textbf{Pressure drops at which each run was held and passed in the slit.} Edge radius
\SI{0.3}{\micro\meter} (Figs.~\ref{fig:slit}, \ref{fig:mechanism} and~\ref{fig:rejuvenation}). For each run, the
highest pressure drop at which the cell was held and the lowest at which it passed (Pa); \dag, the run failed at this
pressure drop without the cell passing, so that the held value is a lower bound. Shear modulus of the cytoskeleton:
own, that of the stage; D, that of D; E-III, that of E-III. $^{a}$Started at \SI{0.5}{\pascal} instead of
\SI{2}{\pascal}. Single-step runs and repeats with the stronger self-repulsion as described above.}
\label{tab:brackets}
\footnotesize
\begin{tabular}{llcl}
\hline
Cell & Shear modulus & $V/V_2$ & Held/passed (Pa), one entry per run \\
\hline
\multicolumn{4}{l}{\textit{At the area and volume of D}} \\
D & D & 0.912 & 1.5/2, 1/1.5, 1/1.5 \\
E-I & own & 0.905 & 3/4, 3/4, 3/4 \\
E-II & own & 0.900 & 12.5/15, 10/12.5, 10/12.5 \\
E-III & own & 0.880 & 75/100\dag, 75/100\dag, 75/100\dag \\
E0 & D & 0.904 & 1.5/2, 1.5/2, 1.5/2 \\
E-I & D & 0.905 & 2/3, 2/3, 2/3, 2/3$^{a}$ \\
E-II & D & 0.899 & 5/7, 5/7, 7/10 \\
E-III & D & 0.879 & 50/75\dag, 50/75\dag, 50/75\dag \\
E-I & E-III & 0.906 & 10/15, 10/15 \\
\multicolumn{4}{l}{\textit{With less bilayer area}} \\
Da025 & D & 0.951 & 1.5/2, 1/1.5, 1/1.5 \\
Da05 & D & 0.989 & 10/12.5, 7/10, 7/10 \\
Da075 & D & 1.029 & 50/75\dag, 30/50\dag, 30/50\dag \\
Da10 & D & 1.073 & 75/100\dag, 75/100\dag, 75/100\dag \\
Da15 & D & 1.169 & 150/200\dag, 150/200\dag, 200/300\dag \\
Da20 & D & 1.281 & 200/300\dag, 200/300\dag, 200/300\dag \\
E3ka05 & E-III & 0.950 & 100/150\dag, 75/100\dag, 100/150\dag, 75/100\dag \\
E3ka10 & E-III & 1.032 & 100/150\dag, 100/150\dag, 100/150\dag \\
E3ka15 & E-III & 1.132 & 150/200\dag, 200/300\dag, 150/200\dag \\
E3ka20 & E-III & 1.238 & 200/300\dag, 200/300\dag, 200/300\dag \\
E1ka05 & E-I & 0.977 & 5/7, 5/7, 5/7 \\
E1ka10 & E-I & 1.058 & 50/75\dag, 50/75\dag, 50/75\dag \\
E2ka05 & E-II & 0.973 & 20/30, 20/30, 20/30 \\
E2ka10 & E-II & 1.056 & 75/100\dag, 75/100\dag, 75/100\dag \\
\hline
\end{tabular}
\end{table}

\subsubsection*{Ladder history and time window}

Applied from the start, the highest level held by every ladder of a condition held the cell again in ten of ten
conditions (one seed each: D at 1, E-I at 3, E-II at 10, E-III at 75, Da05 at 7, Da075 at 30 and E3ka05 at
\SI{75}{\pascal}; with the modulus of D, E-I at 2, E-II at 5 and E-III at \SI{50}{\pascal}), so the time spent at
the lower levels does not set the brackets. Where the pressure drops are lowest, passage is slow: E-I with the modulus of D, held at \SI{2}{\pascal} up to the limit of
500 time units, passed after 129 more in a run continued from its last frame, and E-I, held for 340 time
units at \SI{3}{\pascal}, passed after 202 more in two such runs. The cells held at higher levels did not move. In
200 more time units the rear of E-III and of E-III with the modulus of D, both held from the start at
\SI{50}{\pascal}, and of Da075 moved by at most
\SI{0.02}{\micro\meter}.

With the damping time of the thermostat at 0.2 instead of 0.5 time units, 2.5 times the friction, one ladder per
condition gave E-I (held at 3, passed at \SI{5}{\pascal}), E-II (held at 10, passed at \SI{15}{\pascal}) and E-III
(held up to 75, failed at \SI{100}{\pascal}) the levels of their three ladders at 0.5. Da05 passed at
\SI{10}{\pascal}, as in two of its three ladders. D passed at \SI{3}{\pascal}, one level above its ladders at 0.5
(1.5 or \SI{2}{\pascal}). It was still moving when the level at \SI{2}{\pascal} ended and passed after 344 time units
at \SI{3}{\pascal}.

\subsubsection*{Edge radius}

\begin{figure}[htbp]
\centering
\includegraphics[width=\textwidth]{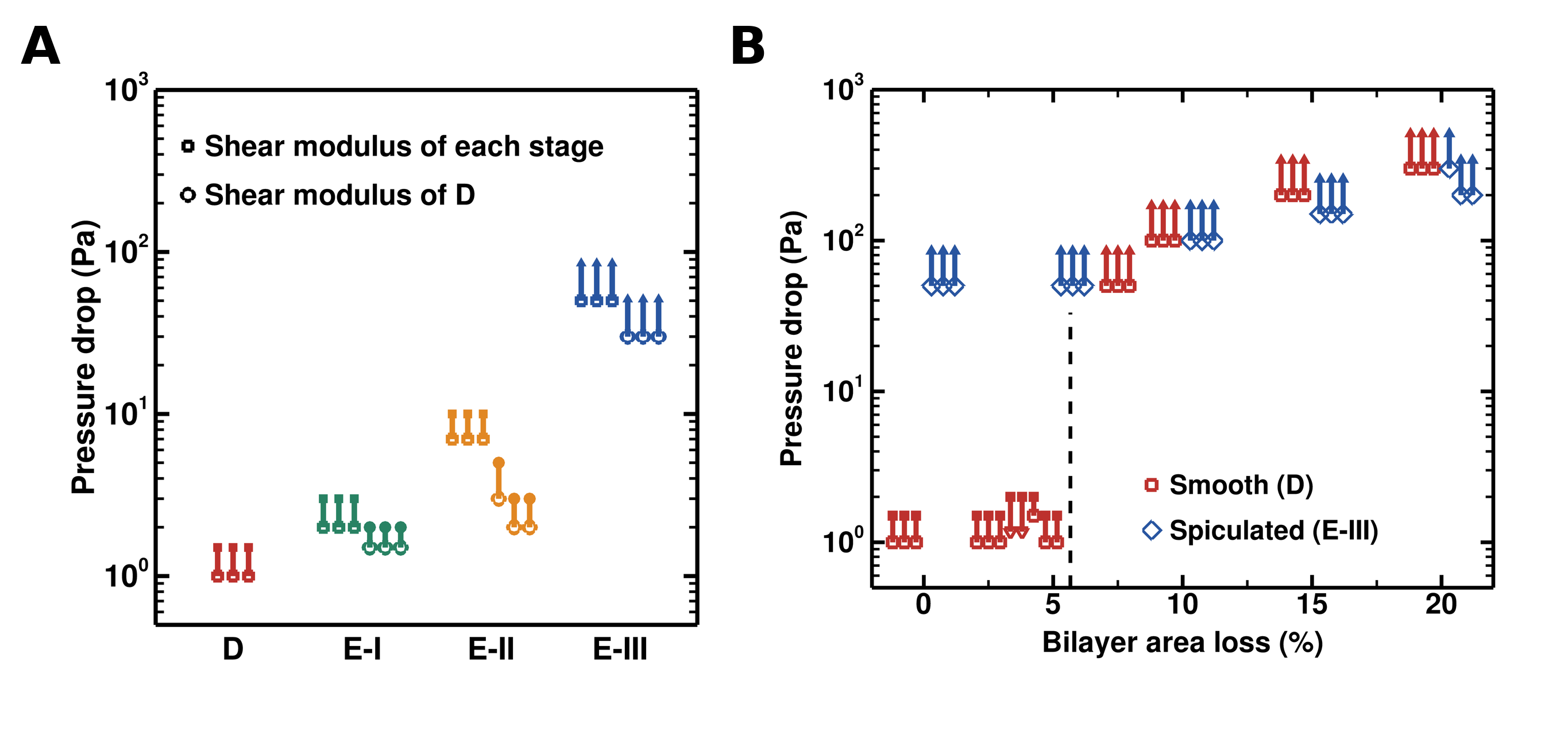}
\caption{\textbf{Passage through the slit with edges rounded with radius \SI{0.5}{\micro\meter}.} Marks as in
Fig.~\ref{fig:slit}, B and C, where the edge radius is \SI{0.3}{\micro\meter}. (A) The stages at the area and
volume of D, with their own cytoskeleton shear modulus (squares) and with that of D (circles), colored as in Fig.~\ref{fig:model}.
(B) Against the loss of bilayer area: smooth family (red squares) and spiculated family (blue diamonds); dashed line, the
area loss at which the smooth cells reach $V/V_2 = 1$. A filled symbol with a line down to an open triangle marks a
cell that passed at the first level of its ladder (an upper bound: Da05, two seeds); Da05 was also run on a ladder
from \SI{0.5}{\pascal}, and E-I with the modulus of D only on that ladder. Three seeds per condition.}
\label{fig:slitedge}
\end{figure}

\begin{figure}[htbp]
\centering
\includegraphics[width=\textwidth]{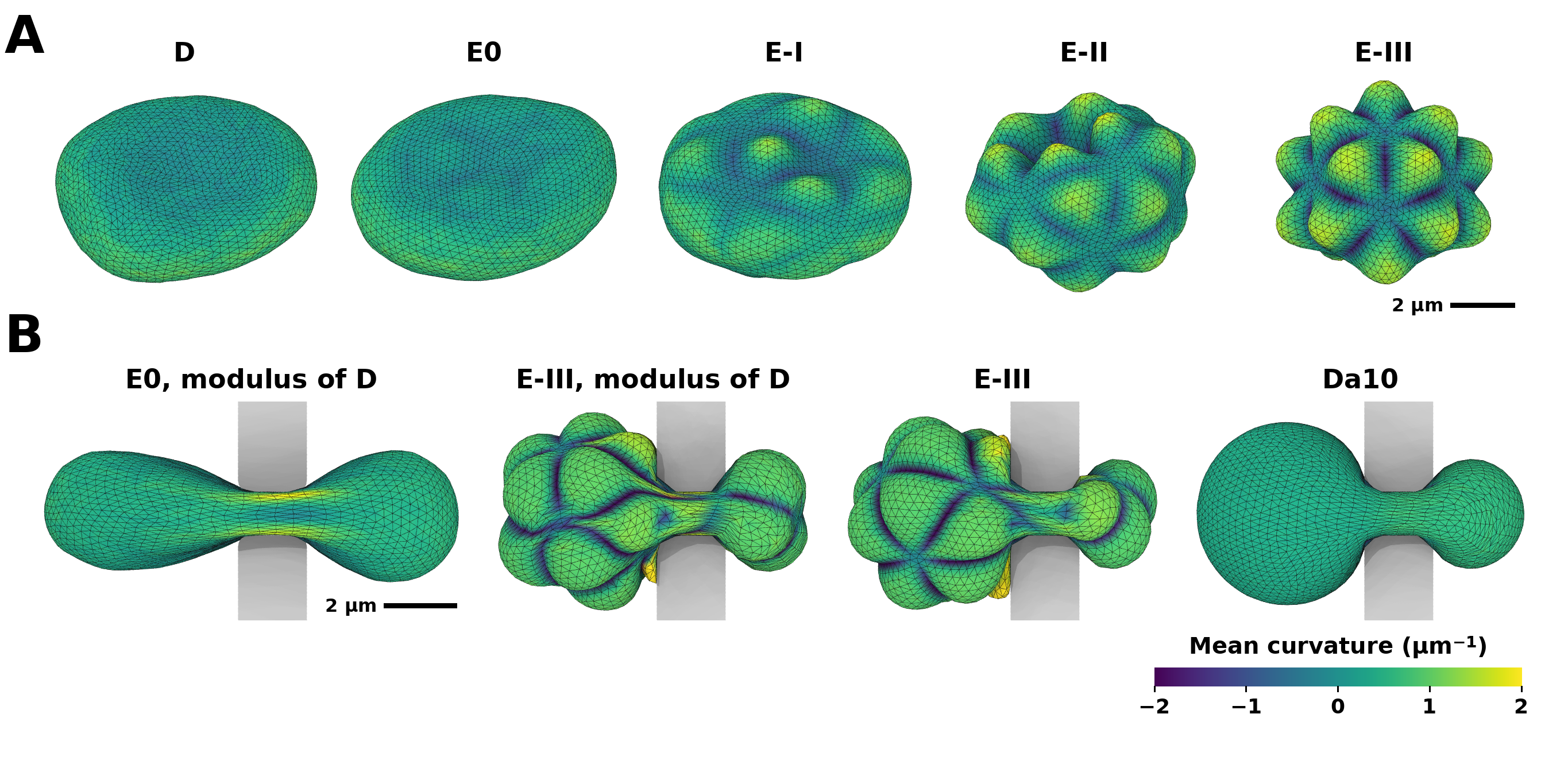}
\caption{\textbf{Mean curvature of the bilayer at rest and in the slit.} (A) The five cells at rest, as in
Fig.~\ref{fig:model}D. (B) The cells of Fig.~\ref{fig:mechanism}A at the end of the highest pressure drop at which
they were held: E0 with the modulus of D at \SI{1.5}{\pascal}, E-III with the modulus of D at \SI{50}{\pascal}, E-III
and Da10 at \SI{75}{\pascal}; side view along the \SI{4.0}{\micro\meter} width of the slit, with the near half of the
wall removed. The bilayer is colored by its mean curvature $H$ (positive where the surface is convex, $1/R$ on a sphere
of radius $R$), averaged at each vertex over the last 20 time units, as in the maps of Wen et al.~\cite{Wen2025}; thin
lines, the edges of the mesh. Scale bars, \SI{2}{\micro\meter}.}
\label{fig:curvature}
\end{figure}

With the edges of the slit rounded with a radius of \SI{0.5}{\micro\meter} instead of 0.3 (6 instead of 4 segments;
the cells of Fig.~\ref{fig:slit}, B and C, on the same ladders, three seeds each, with E-I with the modulus of D on
the ladder from \SI{0.5}{\pascal} and without single-step runs; fig.~\ref{fig:slitedge}), every cell that passed did so
at the same or a lower pressure drop: D and Da025 at \SI{1.5}{\pascal} (1.5 or \SI{2}{\pascal} with
\SI{0.3}{\micro\meter} edges), E-I at 3 (4), E-II at 10 (12.5 or 15), and with the modulus of D, E-I at 2 (3) and
E-II at 3 or \SI{5}{\pascal} (7 or 10). The largest change was for Da05, the smooth cell closest to $V/V_2 = 1$,
which passed at 1.5 or \SI{2}{\pascal} (10 or 12.5). The stages kept their order with either modulus, E-III was
again held at every level until its run failed, and the switch of the smooth family from passage to retention stayed
between Da05 ($V/V_2 = 0.99$) and Da075 (1.03); no spiculated cell passed. The levels at which the held runs failed
moved either way: lower for E-III, E-III with the modulus of D and E3ka05, the same for E3ka10, the same or lower for
E3ka15, the same or higher for Da075, Da15 and E3ka20, and higher for Da10 and Da20; as with the
\SI{0.3}{\micro\meter} edges, they are lower bounds. Two runs, both of E-III with the modulus of D, had crossings
in one frame each, seed 3 in the highest level at which it was held and seed 2 in the level at which it failed.
Repeated with the stronger repulsion, neither had crossings, and both failed at the same level as before. The repeat of seed 3
held through the level with the crossings and replaces the run; that of seed 2 failed in it, so the run stays, as for
seed 3 with \SI{0.3}{\micro\meter} edges. Neither mark changed.

\subsubsection*{Lengthening the connecting bonds of a held cell}

\begin{figure}[htbp]
\centering
\includegraphics[width=\textwidth,height=0.6\textheight,keepaspectratio]{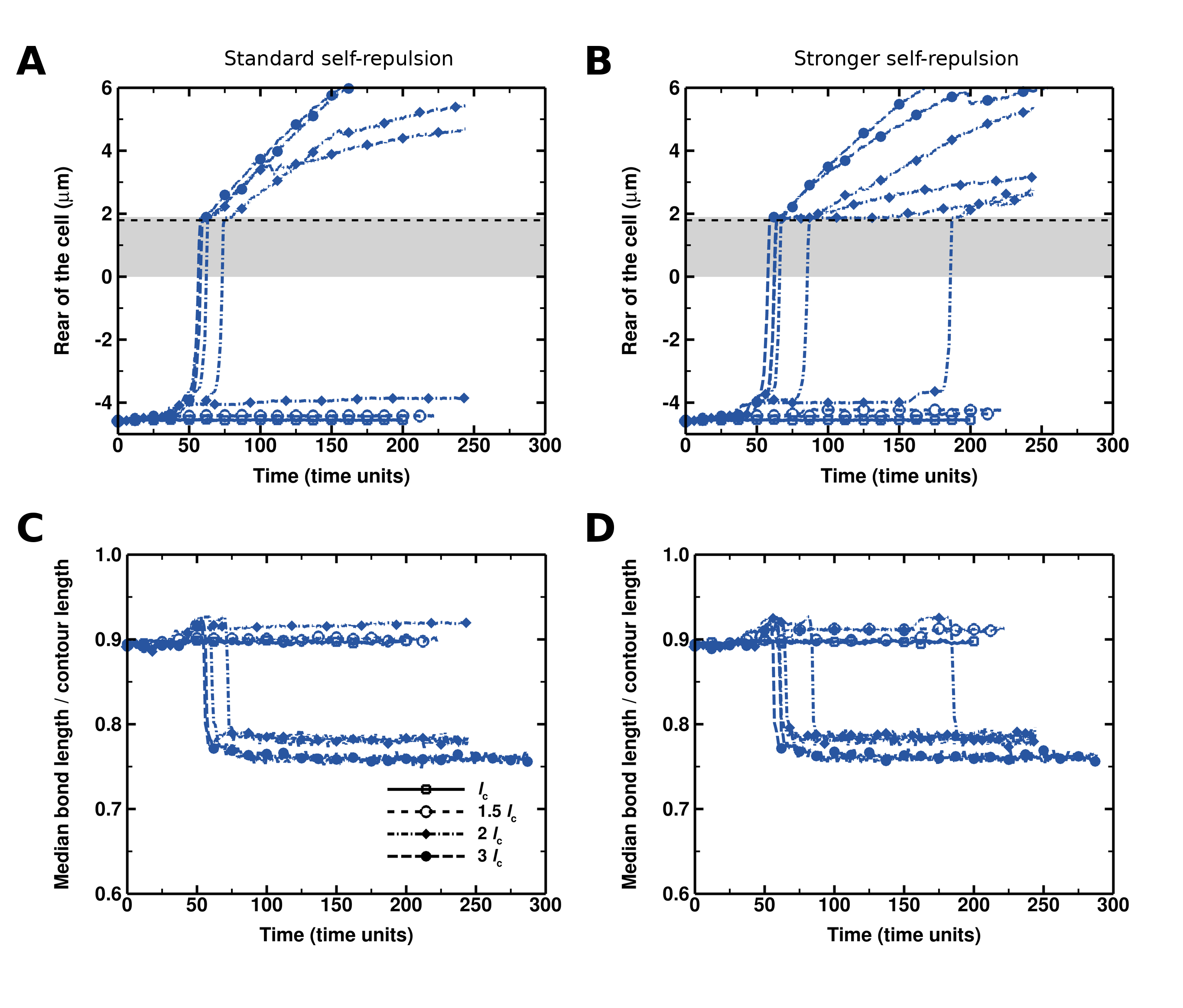}
\caption{\textbf{A held stage III echinocyte passes the slit once its connecting bonds are lengthened about twofold.}
E-III held in the \SI{50}{\pascal} step of one run of Fig.~\ref{fig:slit} (17.5 time units before that step ended),
continued at \SI{50}{\pascal} with the
contour length of its connecting bonds unchanged ($l_c$) or raised to 1.5, 2 or $3\,l_c$ in steps of $0.011\,l_c$
every 0.5 time units (over the first 22, 44 or 88 time units). (A) Position of the rear of the bilayer along the slit
axis, measured from the inlet face of the wall, with the bilayer self-repulsion of the main text. Gray band, the wall;
dashed line, the position beyond which the cell counts as passed. The pressure drop acts only inside the wall, so a
cell that has passed slows down beyond the outlet. (B) The same runs with the stronger bilayer self-repulsion. (C and
D) Median length of the connecting bonds over the contour length in force at that time, for the runs of A and B. Two
thermostat seeds per contour length, four at $2\,l_c$; symbols every 25 time units. One run with $2\,l_c$ in A failed
after 54 time units (Supplementary Text); its trace ends there.}
\label{fig:release}
\end{figure}

\begin{figure}[htbp]
\centering
\includegraphics[width=\textwidth]{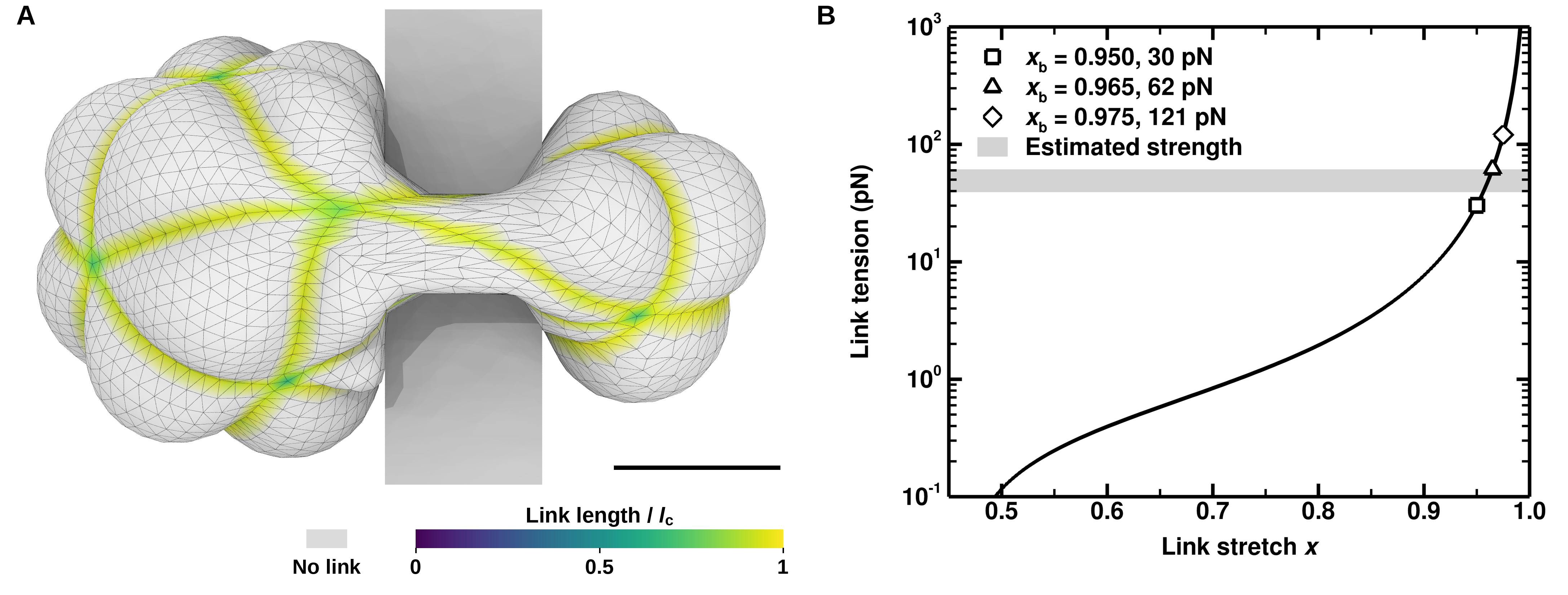}
\caption{\textbf{A held stage III echinocyte and the strength of its links.} (A) E-III held in the slit at
\SI{75}{\pascal} with unbreakable links, in one of the runs of Fig.~\ref{fig:slit}B. The cell is seen from the side, with the
flow from left to right; bilayer mesh colored by link length over $l_c$, the stretch $x$ of B (gray, no link), wall
gray. 351 of the 462 links of the cell were stretched past 90\% of their full length ($x > 0.9$), 184 of them on the
side shown. Scale bar, \SI{2}{\micro\meter}. (B)
Tension of a link against its stretch $x = r/l_c$, its length $r$ over its contour length $l_c$. The three failure
stretches, $x_b = 0.950$, 0.965 and 0.975, correspond to 30, 62 and \SI{121}{\pico\newton}; gray band, the strength
estimated from the separation energy, 39 to \SI{61}{\pico\newton}.}
\label{fig:heldlinks}
\end{figure}

In the spiculated cells that the slit held, most connecting bonds were close to their contour length $l_c$, the
largest length they can reach (Fig.~\ref{fig:mechanism}). To test whether these bonds hold the cell, we lengthened
them in a cell that was already held and changed nothing else (fig.~\ref{fig:release}). The runs are single-step runs
like those above. They start, like the repeats of that hold, from a frame saved 17.5 time units before the end of
the \SI{50}{\pascal} level of one E-III ladder (seed 1), when the cell already met the hold criterion, with the
velocities set to zero and a new thermostat seed, and keep the pressure drop at \SI{50}{\pascal}, the
highest level at which E-III was held in every run and repeat. The contour length of the 462 connecting bonds was
left at $l_c$ or raised to 1.5, 2 or $3\,l_c$, in steps of $0.011\,l_c$ every 0.5 time units (over 22, 44 or 88 time
units), and the run went on for 200 time units after the last step. The rest length and the persistence length of
these bonds, all other bonds, the wall and the time step were unchanged. The length was raised in steps because,
raised at once (to 1.5 or $2\,l_c$, at 50 or \SI{75}{\pascal}), it made seven of eight runs fail, six of them within
200 time steps. Each contour length was run with two thermostat seeds, $2\,l_c$ with four, and every run was repeated
with the stronger bilayer self-repulsion, which gives two series of ten runs (fig.~\ref{fig:release}, A and B). Runs
of one series with the same seed follow the same path for as long as their contour lengths agree, so a run toward
$3\,l_c$ repeats the run to $2\,l_c$ of its seed for 44 time units. The two runs with $l_c$ and the self-repulsion
of the main text are the repeats of this hold described above; the other 18 runs were made for this test.

With the contour length unchanged, the cell stayed in place. Its center moved by \SI{0.01}{\micro\meter} in 200 time
units (four runs). With $1.5\,l_c$ it did not pass (four runs). With the self-repulsion of the main text it advanced
by 0.23 and \SI{0.24}{\micro\meter}, nearly all of it in the first 30 time units; with the stronger one it advanced by
\SI{0.56}{\micro\meter} in both runs, in steps, the last of which fell in the final 20 time units of one run. At the
end the rear of the cell was still 6.0 to \SI{6.2}{\micro\meter} short of the position at which a cell counts as
passed. With
$2\,l_c$ and the self-repulsion of the main text, the cell passed 19 and 30 time units after the last step in two
runs. A third run failed 10 time units after the last step, when an edge of the bilayer mesh at the inlet edge of the
slit exceeded its maximum length while the cell was moving through the slit. In the fourth, the cell advanced by
\SI{1.1}{\micro\meter} in steps and had not passed after 200 time units. With the stronger self-repulsion all four
runs passed, 20, 23, 42 and 143 time units after the last step, the last one after a second hold of 110 time units
\SI{0.8}{\micro\meter} farther in. In the runs toward $3\,l_c$ the cell passed while the bonds were still being
lengthened, 57 to 63 time units after the start, at a contour length of 2.3 to $2.4\,l_c$ (four runs). While the
cell was held, the median length of the connecting bonds was 0.89 to 0.93 of the contour length then in force; after
it had passed, 0.75 to 0.80 (fig.~\ref{fig:release}, C and D). The hold thus gave way at about twice the contour
length, but not at a sharp value. At $2\,l_c$ the delay before the passage spread over a factor of seven, and two runs
gave no passage.

The runs were scored by the rules for the single-step runs above. For this test the scoring was written down when
ten of the runs were finished or under way and before the other ten started (among them seeds 3 and 4 with $2\,l_c$
and all runs toward $3\,l_c$). No bond reached its maximum length in any run but the one that failed. The bilayer crossed itself in one frame of four runs with $2\,l_c$ (seeds 1 and 2
with the self-repulsion of the main text, seeds 2 and 4 with the stronger one) and, in seed 2 toward $3\,l_c$, in two
frames of the run and one of its repeat; the other 14 runs had no crossings. By the rule above, the repeat replaces
seed 1 with $2\,l_c$ and seed 2 toward $3\,l_c$. Scored in this way, the cell did not pass with $l_c$ or $1.5\,l_c$
(two seeds each); with $2\,l_c$ it passed in seeds 1 and 2, seed 3 failed, and seed 4 did not pass; toward $3\,l_c$
it passed in both seeds. Bilayer vertices lay inside the wall at the inlet edge of the slit, up to
\SI{0.10}{\micro\meter} with $l_c$ and up to \SI{0.26}{\micro\meter} with longer bonds. With the option that moves
such vertices back to the surface of the wall, the cell did not pass with $l_c$ (one run) or $1.5\,l_c$ (two runs);
with $2\,l_c$ it passed in one of two runs, 32 time units after the last step, and in the other was held
\SI{0.9}{\micro\meter} farther in. These two runs with $2\,l_c$ were added after the first passes had been seen.

The model has no repulsion between the two networks (Materials and Methods), so only the connecting bonds keep the
cytoskeleton inside the bilayer. In every fourth saved frame we counted the cytoskeleton vertices that lay outside the
closed surface of the bilayer. While the cell was held there were at most 3 of the 2562, within
\SI{0.10}{\micro\meter} of the bilayer. While it passed, the largest number in a frame was 31 to 94, as far as 0.20 to
\SI{0.50}{\micro\meter} outside, and 13 to 77 remained in the last frame. E-III never passed with unchanged bonds, so
the reference is the passes of E-I and E-II. In two ladders and two single-step runs, the largest number in a frame
was 87 to 174, as far as 0.14 to \SI{0.40}{\micro\meter} outside. The test rests on one held cell and on runs of 200
time units.

\subsection*{Links that can fail}
The runs of this section let the connecting bonds, the links between the two layers, fail (Materials and Methods).
Four details bear on how their numbers are read.

\begin{figure}[htbp]
\centering
\includegraphics[width=\textwidth]{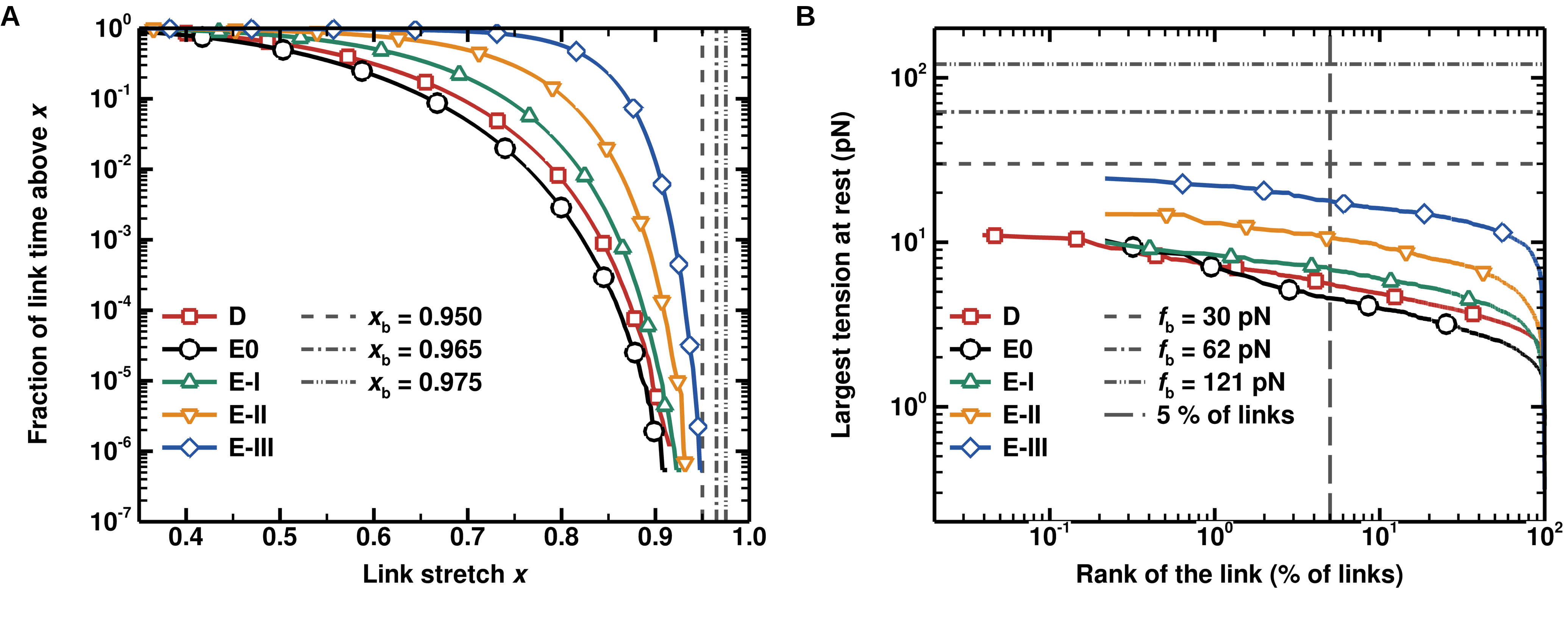}
\caption{\textbf{The links of stage III echinocytes are stretched at rest.} Links of each cell kept in front of the
slit without a pressure drop and without being pushed in, one run per cell. (A) Fraction of the time, over all links,
that a link spends above a stretch $x$, against $x$, recorded every 10 time steps over 20 time units (every 20 time
steps over 10 time units for D). Gray lines, the three failure stretches $x_b$. (B) Largest tension of each link over
the first 10 time units, from positions every 20 time steps in every cell, which misses the briefest peaks of (A); the
links are ranked from the most stretched, the rank given as a percentage of the links of the cell. Horizontal lines,
the three link strengths $f_b$; vertical line, 5\% of the links, the count we take as the onset of failure. Colors as
in Fig.~\ref{fig:model}.}
\label{fig:rest}
\end{figure}

\textit{Sampling.} The largest stretches of Fig.~\ref{fig:onset}A come from configurations saved every 2 time units
and underestimate the true largest stretch: at rest, the links of E-III reached 0.941 in such configurations but 0.951
when every 10 time steps were recorded (fig.~\ref{fig:rest}). A stretched link fluctuates with a period of about 0.014
time units at $x = 0.92$, so the records at rest give about three samples per period in the echinocytes and one and a
half in the discocyte. At its failure stretch a link oscillates with a period of about 14, 8 and 5 time steps at
$x_b = 0.950$, 0.965 and 0.975.

\textit{Warming during the push-in.} Pushing the cell into the slit before the first pressure step (the push-in) warms
the membrane to about 1.5 times the set temperature. In the runs of E-III with the strongest links, which lost no link
before \SI{75}{\pascal}, the warming had fallen below 1.1 times within one time unit of its peak, early in the first
pressure step; pushing the cell in again before the 3 or \SI{5}{\pascal} step raised it briefly to 1.18 to 1.38 times. One of the three first failures of the weakest links came within 2 time units of the peak (1.34 times the set
temperature over the 2 time units before it).

\textit{Runs that broke down.} In three of the six runs of E-III with the two stronger strengths, the run broke down
after the 24th failure. Within 451 to 2179 time steps of it, at most about one time unit, bonds of the bilayer were
stretched past their limit and the run stopped with an error. In the two runs whose kinetic temperature was logged
after the 24th failure, it had reached 13 and 25 times the set value. Many of the failures logged in this time, 62 to
247 in a run, came at stretches past the full length of a link, where its force is not defined. We do not read them as
failures of the cell, and we score these runs at the 24th failure, which came before the breakdown. In two other runs
the failures went on, and 61 and 173 links had failed when the run stopped, 1.3 time units after the 24th failure,
with the membrane at 14 and 18 times the set temperature; the sixth stopped with 26 failed links
(fig.~\ref{fig:burst}B).

\textit{Stage II echinocytes.} The links of E-II, which failed only at the weakest strength, failed mostly at
\SI{15}{\pascal}, and the 24th came 45 to 235 time units after the first. With the shear modulus of D, most of the first
24 failures of E-II again lay downstream, in the part of the cell that had gone through (12 to 16 of them).

\subsubsection*{Keeping the cytoskeleton inside the bilayer}

In the model only the links tie the cytoskeleton to the bilayer, and a link, a spring between two vertices, does not
set on which side of the bilayer its cytoskeleton end lies. Part of the cytoskeleton therefore lies just outside the
bilayer even when no link has failed, and once links fail nothing holds that part of the cytoskeleton at all. We count
a vertex as outside when the winding number of the bilayer at it (defined below) is below one half. At the end of the
runs with links that could fail but lost none, 1296 to 1355 of the 2562 cytoskeleton vertices of the discocyte lay
outside (its two layers stay within \SI{0.06}{\micro\meter} of each other; Fig.~\ref{fig:model}E), 693 to 702 of E0
and 30 to 80 of E-I and E-II, at most \SI{0.09}{\micro\meter} outside. E-III, whose runs all lost links, had 14 to 19
outside in the last configuration saved before its first failure, at most \SI{0.05}{\micro\meter} outside. We
therefore tested a constraint that returns any cytoskeleton vertex found outside the bilayer to its inside.
For each cytoskeleton vertex $\mathbf{x}$, the constraint finds the closest point $\mathbf{q}$ of the bilayer mesh, on
a face, an edge or a vertex, and the distance $s$ of $\mathbf{x}$ from it, counted positive outside the bilayer. The
sign comes from the angle-weighted normal $\mathbf{N}$ at $\mathbf{q}$: the normal of the face or, at an edge or a
vertex, the sum of the normals of the faces that meet there, weighted by their angles at that point. Where the bilayer is folded
so flat that this test is unreliable, $|(\mathbf{x}-\mathbf{q})\cdot\mathbf{N}| < 0.3\,|\mathbf{x}-\mathbf{q}|\,W$
($W = 1$ at a face, 2 at an edge and the sum of the face angles at a vertex), and for every vertex that is about to be
pushed, the sign is taken instead from the winding number of the whole bilayer at $\mathbf{x}$: the sum of the solid
angles that its faces subtend at $\mathbf{x}$, divided by $4\pi$, which is 1 inside a closed surface and 0 outside. A
vertex with $s > -d_0$, $d_0 = \SI{5}{\nano\meter}$, is pushed inward, along the line through $\mathbf{q}$, with a
force $K(s + d_0)$, $K = \SI{4.5e-3}{\newton\per\meter}$; the opposite force goes to the vertices of the
closest face, edge or vertex, shared by their barycentric weights at $\mathbf{q}$, so that the constraint exerts no
net force or torque on the cell. Vertices farther than \SI{0.15}{\micro\meter} from the bilayer are not checked. $K$
grows from zero over the first 500 time steps of a run. The constraint is not part of the potential energy. It made
each time step take 1.38 to 1.44 times as long to compute.

A constraint of this kind should change nothing while every vertex stays inside. We first checked this with
unbreakable links, repeating six ladders of Fig.~\ref{fig:slit}B: E-II and E-III, three runs each. In five of them every
pressure step ended as it did without the constraint. In the sixth, one of the E-III runs, a link reached its full
length at \SI{75}{\pascal} and the run stopped there. Without the constraint all three E-III runs stopped in the same
way one pressure step later, at \SI{100}{\pascal}, and here the two other E-III runs did too.

Such a stop at \SI{75}{\pascal} also happens without the constraint. We ran E-III with six more thermostat seeds,
each with and without the constraint and again with unbreakable links. Without the constraint one of the six runs
stopped at \SI{75}{\pascal}, when a link reached its full length; with it none did. Over nine runs each, one run
with the constraint and one without stopped at \SI{75}{\pascal}. All the other runs were held at
\SI{75}{\pascal} and stopped early in the \SI{100}{\pascal} step, each when a link reached its full length. The
seed whose run with the constraint stopped at \SI{75}{\pascal} is also the one whose run without the constraint had
the most stretched links at that level: on average 38 links above 95\% of their full length, against 7 and 8 in
the other two runs.

Finally, we repeated all 45 runs of table~\ref{tab:onset} with the constraint (table~\ref{tab:onsetcon}). The discocyte and E0 lost no link and passed
at the same pressure drops as without the constraint, 1.5 or \SI{2}{\pascal}. E-I lost one link in
each of two runs at the weakest strength, and in one of them the cell passed at \SI{3}{\pascal} instead of
\SI{5}{\pascal}. E-II reached the 5\% count at \SI{15}{\pascal} in all three runs, as without the constraint. E-III
reached it at 7, 5 and \SI{15}{\pascal} with the weakest links, at 50, 75 and \SI{50}{\pascal} with links of
\SI{62}{\pico\newton} and at 75, 100 and \SI{100}{\pascal} with the strongest: in eight of the nine runs at the same
pressure drop as without the constraint, and in the ninth one step earlier. With the weakest links, the first 24
failures of E-III all lay upstream, 23, 24 and 23 of them more than \SI{0.5}{\micro\meter} from any wall surface. With links of \SI{62}{\pico\newton}, 0, 2
and 0 of the first 24 lay upstream, 7, 5 and 4 in the slit and 17, 17 and 20 downstream, against 3 to 12 without the constraint; with the strongest links 14,
0 and 0 lay upstream, 5, 13 and 15 in the slit and 5, 11 and 9 downstream. The first 24 failures of E-II lay mostly
downstream (16, 15 and 18). All three runs of E-III with the strongest links broke down after the 5\% count. In two, 432 and 295 failures were
logged within 0.06 and 0.4 time units of the 24th failure and the run stopped with an error; the third was stopped by
the count 0.22 time units after its 24th failure, with 34 failed links, links failing past their full length and the
membrane at 128 times the set temperature. We score them by the 5\% count, which came before the breakdown.

\subsubsection*{Controls without a pressure drop and at half the time step}

\begin{figure}[htbp]
\centering
\includegraphics[width=\textwidth]{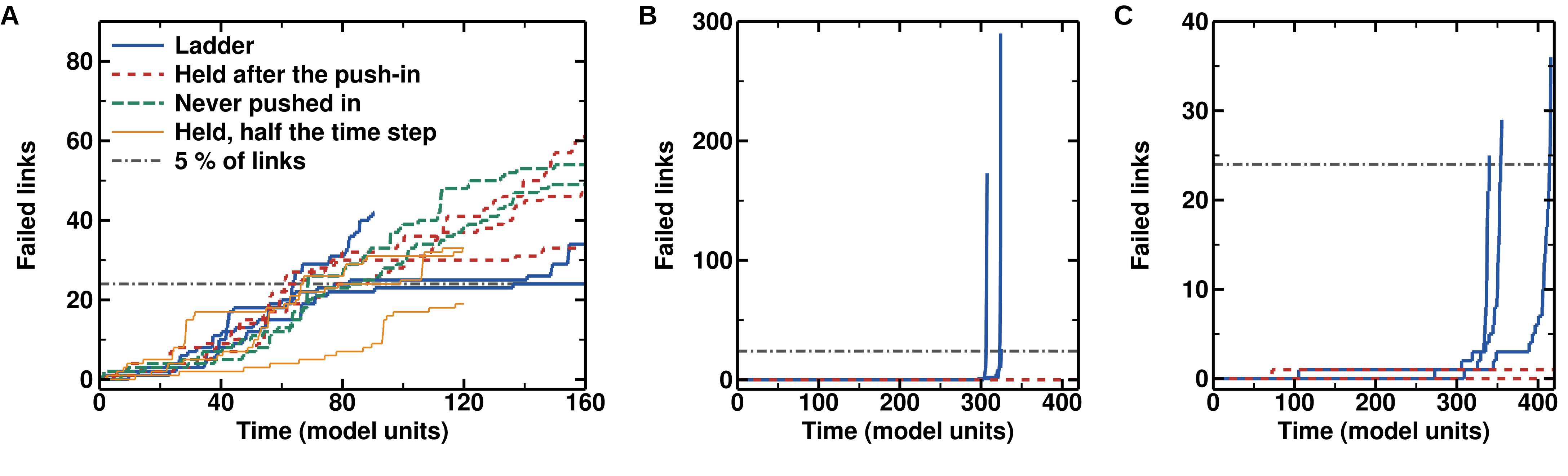}
\caption{\textbf{Without a pressure drop, E-III loses links of \SI{30}{\pico\newton} as fast as in the ladders but no
stronger ones, and E-II loses one link in three runs.} Running count of failed links against the time since the links
could fail, one line per run. (A) E-III with links of \SI{30}{\pico\newton}: the three ladders of
table~\ref{tab:onset}; the same three runs held without a pressure drop after the push into the slit, at the time step
of the ladders and at half of it; and two cells never pushed in, held \SI{4}{\micro\meter} upstream of the slit. (B)
E-III with links of \SI{62}{\pico\newton} and (C) E-II with links of \SI{30}{\pico\newton}: the ladders of
table~\ref{tab:onset} and the same three runs held after the push into the slit for 420 time units without a pressure
drop. Line styles as in the key of (A), where the dash-dot line marks 24 links, 5\% of 462. Records longer than the
axis are cut at its end, and the axes start just below zero so that runs without a failure stay visible. One ladder in
(B) ended in a numerical breakdown after 290 failures.}
\label{fig:restcontrols}
\end{figure}

To ask whether the weakest links of E-III need the slit to fail, we repeated its three runs at \SI{30}{\pico\newton}
up to the push-in, with the same seeds, and then held the cell for 160 time units without a pressure drop (fig.~\ref{fig:restcontrols}A). The first
link failed 4.8, 4.5 and 1.4 time units after the links could fail, and the 24th after 77.4, 61.8 and 60.9 time units,
against 9.6, 4.5 and 1.4 and 77.3, 63.7 and 136.1 time units in the ladders of the same seeds. By the end, 61, 33 and 47
links had failed. A cell equilibrated \SI{4}{\micro\meter} upstream of the slit and never pushed in lost its first
link after 1.0 and 7.0 time units and its 24th after 82.2 and 68.5 in runs with the first two seeds, with 49 and 54 links
failed at the end. In the three runs held after the push-in, every failure lay upstream, more than \SI{0.5}{\micro\meter} from any wall surface. The membrane of all five runs stayed near the set temperature: over the half time unit before the 24th
failure it was 0.99 to 1.12 times it. We then repeated the three held runs at half the time step, 0.00025 time units, from the moment the links could fail. They lost their first link after 11.8, 2.3 and 2.7 time units; two lost their 24th after 78.5 and 66.2 time units, and the third had lost 19 links by the end, after 120 time units (fig.~\ref{fig:restcontrols}A). By then the three had lost 19, 33 and 33 links, against 41, 30 and 37 at the time step of the ladders. Every failure lay upstream, more than \SI{0.5}{\micro\meter} from any wall surface, and over the 2 time units before the first failure and the half time unit before the 24th the membrane was at 1.00 to 1.01 times the set temperature.

E-III with links of \SI{62}{\pico\newton} and E-II with links of \SI{30}{\pico\newton} were held in the same way for 420 time units (fig.~\ref{fig:restcontrols}, B and C). In their ladders, counted from the moment the links could fail, the first link had failed after 295 to 304 time units for E-III and 105 to 309 for E-II, and the 24th after 306 to 325 and 340 to 414. Held at rest, E-III lost no link in three runs, and E-II lost one link in one of its three runs, after 72 time units, upstream and \SI{3.3}{\micro\meter} from the wall.

To test the time step, we restarted runs from saved configurations with new velocities at the set temperature, at
the time step of the ladders, 0.0005 time units (the main time step), and at half of it. Held at \SI{50}{\pascal} from a configuration
saved 2 time units before the first failure of a run with links of \SI{62}{\pico\newton}, the cell lost its first
link after 9.4 and 15.1 time units at the main time step and after 11.7 and 12.2 at half of it, and the 24th within
3.9 to 6.7 time units of the first at both. Held at \SI{75}{\pascal} from a configuration saved 3.3 time units before
the first failure of a run with the strongest links, it lost its first link after 5.6 and 6.0 time units at the main
time step and after 21.3 and 17.5 at half of it; over the 2 time units before the first failure the membrane was at
1.72 and 1.69 times the set temperature at the main time step and at 1.03 and 1.05 at half of it. At the main time
step 408 and 343 links failed before both runs broke down, 257 and 85 of them past their full length; at half the time
step 147 and 2 failed in the 30 time units, none past full length, and neither run broke down. At \SI{50}{\pascal}, one
of the two holds at the main time step broke down, with 40 of its 462 failures past full length, and neither hold at
half the time step did.

We also restarted the three runs with the strongest links from the configurations saved at the start of their \SI{75}{\pascal} step and let them climb the rest of the ladder under the same rules, at both time steps. At the main time step the restarts repeated table~\ref{tab:onset}: the first failure came at \SI{75}{\pascal} in all three, 5\% of the links had failed at 75, 100 and \SI{100}{\pascal}, and the second and third runs broke down 0.2 time units after the 24th failure. At half the time step, 5\% had failed at the same pressure drops, but the first failure came at 75, 100 and \SI{100}{\pascal}: the second and third runs lost no link at \SI{75}{\pascal}, where in table~\ref{tab:onset} each had lost one. All three broke down 0.06 to 1.2 time units after the 24th failure, with 124, 162 and 0 failures past full length. At \SI{121}{\pico\newton}, halving the time step thus delayed the first failure of a held cell and removed the lone failures at \SI{75}{\pascal}, but it changed neither the pressure drop of the 5\% count nor the burst and the breakdown that followed. Before the first failure the membrane of these restarts was at 1.17 to 1.32 times the set temperature at the main time step and at 1.07 to 1.46 at half of it. Over the
half time unit before the 24th failure, over all the holds and restarts that reached 24 failures, the membrane was at
2.0 to 6.3 times the set temperature at half the time step and at 1.2 to 4.9 at the main time step: the warming comes
with the failures.

\subsubsection*{A cytoskeleton that can tear}

In the runs above, only the links can fail. To ask whether the cytoskeleton would tear first, we let its springs
break as well from the first pressure step on, in the ladders of E-II and E-III with the constraint (table~\ref{tab:onsetcon}).
A spring breaks, as a link does, when its stretch reaches a set value, which we give as the force
the spring then carries. Its two triangles then leave the area and bending terms but keep their share of the volume
term, so that a hole does not let the network swell. Until its first spring broke, each run repeated the run of
table~\ref{tab:onsetcon} with the same seed step for step, so stronger springs cannot break earlier than weaker ones. By the
coarse-graining rules of Fedosov et al.~\cite{Fedosov2010}, which scale the spring length with the inverse square
root of the number of vertices and keep the product of persistence length and spring length fixed, a spring of our
2562-vertex network carries, at the same strain, 3.27 times the force of a spring of their 27,344-vertex network, in
which each spring is one spectrin tetramer. We know of no measured force at which a tetramer or its junction comes
apart. The nearest scale is the force that unfolds a spectrin repeat, which lengthens the chain without breaking
it. Pulled at \SI{1}{\nano\meter\per\milli\second}, repeats unfold at 21 to \SI{28}{\pico\newton} at room
temperature~\cite{Law2003a,Law2003b}, at \SI{16}{\pico\newton} at \SI{310}{\kelvin} and at \SI{11}{\pico\newton} at
\SI{315}{\kelvin}~\cite{Law2003b}. The lowest of these is \SI{36}{\pico\newton} per spring of our network.

While every link held, the largest force on any spring was 3.7 to \SI{6.9}{\pico\newton} in E-III with links of
\SI{62}{\pico\newton}, 6.5 to \SI{7.8}{\pico\newton} with links of \SI{121}{\pico\newton} and 5.5 to
\SI{6.8}{\pico\newton} in E-II with links of \SI{62}{\pico\newton}, which passed without losing a link (three runs
each, read every 0.5 time units until the first spring broke, so shorter peaks were missed). We tested springs that break at \SI{10}{\pico\newton},
about a quarter of \SI{36}{\pico\newton}, with all three seeds, and springs that break at \SI{20}{\pico\newton}, about
a half, with the first. In E-III with links of \SI{62}{\pico\newton}, the first spring broke only after a link had
failed: in the first run within the burst of link failures that ends the run, after the fifth failure and behind the
slit (at both strengths), and in the other two 0.2 and 0.8 time units after the first link failure. In all three, 5\%
of the links had failed at the same pressure drop as without tearing. With links of \SI{121}{\pico\newton}, the first
spring broke in the slit, 0.56 time units after the first link failure in the first run but 3.0 and 0.43 time units
before it in the other two. One of these two lost 5\% of its links at \SI{75}{\pascal}, one step earlier than without
tearing; the other two runs of this set lost them at the same pressure drop as without tearing. In E-II, two springs broke in one of the three runs,
in the slit, with no link failed, and every run passed at \SI{15}{\pascal}, as without tearing.

Springs of 2 to \SI{5}{\pico\newton}, 7 to 18 times below \SI{36}{\pico\newton}, were tested with the first seed.
They broke at or in the slit, in all runs but one before any link had failed. E-III then passed at 20 and \SI{30}{\pascal} with springs of 2 and
\SI{3}{\pico\newton}, instead of losing 5\% of its links at \SI{50}{\pascal}, and at \SI{50}{\pascal} with links of
\SI{121}{\pico\newton} and springs of \SI{5}{\pico\newton}, instead of being held there. The outcome did not change
with links of \SI{62}{\pico\newton} and springs of \SI{5}{\pico\newton}, which first broke 3 time units after the
first link failure, or for E-II, which passed at \SI{15}{\pascal}. Large cascades of breaking springs, like bursts of
link failures, warmed the membrane, so only the onset of tearing is quantitative. The bilayer cannot tear in the
model, and it did not lift off the torn cytoskeleton. After the first tear, the area of bilayer more than
\SI{0.5}{\micro\meter} from any cytoskeleton vertex never exceeded its largest value without tearing by more than
\SI{0.1}{\micro\meter\squared}, and where many springs broke it shrank: as E-II left the slit it was 9.0 and \SI{5.5}{\micro\meter\squared}
with springs of 2 and \SI{3}{\pico\newton}, against \SI{28.9}{\micro\meter\squared} without tearing.

\subsection*{Wall shear stress in a \SI{6}{\micro\meter} tube}
Unlike the assays, these runs include the fluid (fig.~\ref{fig:tubewss}). D, E-II and E-III travel along a straight
tube \SI{6}{\micro\meter} in diameter and periodic over \SI{40}{\micro\meter} of its length, filled with a solvent of
DPD particles of viscosity \SI{1.2}{\milli\pascal\second}, the same inside and outside the cell. The solvent is
reflected at the wall, and its axial force on the frozen particles of the wall gives the wall shear stress. Each cell
has its own shear modulus (table~\ref{tab:model}) and the stiffened bilayer penalties and bilayer self-repulsion of the
assays, at the same physical values (Materials and Methods); the thermal energy, set by the solvent, is
\SI{3.1e-22}{\joule}. A second reflecting surface keeps the membrane at least \SI{0.45}{\micro\meter} from the wall.
The bilayer exchanges dissipative forces with the solvent on both of its sides, the cytoskeleton only with the solvent
inside the cell.

Each cell was first pressed into the tube by a shrinking cylinder, with its shortest axis along the tube axis. For
\SI{2.0}{\milli\second} without flow the solvent could cross the membrane and fill the cell; the membrane was then
sealed, and a body force on the solvent, equivalent to a pressure gradient of \SI{1.40}{\pascal\per\micro\meter},
drove the flow. A solvent particle that crosses a triangle of the bilayer is sent back with the velocity
$2\mathbf{u}-\mathbf{v}$, where $\mathbf{v}$ is its own velocity and $\mathbf{u}$ that of the triangle at the crossing
point, at most three times per time step, and its change of momentum is passed to the vertices of the triangle. With
a time step of \SI{23.3}{\nano\second}, the runs of D and E-III lasted 4.38 million steps, \SI{102}{\milli\second}
from the start of the filling, and that of E-II 9.0 million steps, \SI{210}{\milli\second}, so that the two halves of
its averaging period agreed within 1\%.

The axial force of the solvent on the wall particles was summed over windows of \SI{0.47}{\milli\second}, in bins
\SI{1}{\micro\meter} long and \SI{0.99}{\micro\meter} wide (19 around the wall) that move with the cell centroid. In
each bin it was divided by the force on the same wall particles in a run without a cell and multiplied by
\SI{2.10}{\pascal}, the mean wall shear stress that the drive sets by force balance; that run gave
\SI{2.1004}{\pascal}. All values are averages over the second half of each run, 107 windows from 52 to
\SI{102}{\milli\second} for D and E-III and 222 windows from 106 to \SI{210}{\milli\second} for E-II, with standard
errors from four blocks of these windows. The peak and the mean over the cell changed by less than 1\% between the two
halves of this period.

Averaged around the wall, the stress peaked at \SI{2.73}{\pascal} beside D, \SI{3.14}{\pascal} beside E-II and
\SI{3.41}{\pascal} beside E-III, against \SI{2.10}{\pascal} without a cell (fig.~\ref{fig:tubewss}B), and single bins
reached 2.87, 3.47 and \SI{4.11}{\pascal} (fig.~\ref{fig:tubewss}C). Over the axial extent of the cell, 10.55, 9.24
and \SI{7.76}{\micro\meter}, it averaged 2.49, 2.78 and \SI{2.97}{\pascal}. Because the drive fixes the force on the
solvent, the mean over the whole wall stayed at \SI{2.10}{\pascal}, and away from the cells the stress was lower, down
to 1.83 (D), 1.70 (E-II) and \SI{1.65}{\pascal} (E-III) just ahead of them. D, E-II and E-III moved at 1.68, 1.50 and
\SI{1.37}{\milli\meter\per\second}, with mean solvent velocities of 1.21, 1.17 and \SI{1.13}{\milli\meter\per\second},
ratios of mean solvent velocity to cell velocity of 0.72, 0.78 and 0.83. These stresses are a prediction of the model,
in which the membrane stays at least \SI{0.45}{\micro\meter} from the wall.

\subsection*{Wall shear stress at a bifurcation}
The results of this section are predictions of the model. Two cells, D and E-III with the cytoskeleton shear modulus
of D, were carried by the flow onto the dividing wall of a bifurcation, and the shear stress on that wall was recorded.

\textit{Set-up.} A parent branch \SI{6}{\micro\meter} in diameter divides into two daughters of the same diameter,
whose axes lie \SI{9}{\micro\meter} apart, and the daughters join again further downstream; the unit is
\SI{53.3}{\micro\meter} long and repeats along the flow (fig.~\ref{fig:bifwss}A). The tip of the dividing wall, the
carina, lies \SI{0.5}{\micro\meter} off the axis of the parent. Walls, solvent, membrane model and time step are those
of the \SI{6}{\micro\meter} tube. The solvent was driven by a body force field taken from a lattice Boltzmann solution of
the same unit without a cell, scaled so that the straight part of the parent is driven as the tube; without a cell its
wall shear stress there was \SI{2.11}{\pascal}. Each cell was prepared as in the tube, placed on the axis of the
parent with its centroid \SI{10.5}{\micro\meter} upstream of the carina, filled for \SI{2.0}{\milli\second} without
flow and sealed. Time is counted from the start of the flow; the runs cover 22.6 (D) and \SI{20.3}{\milli\second}
(E-III with the modulus of D) of flow. In this unit, a penalty on the solvent density, evaluated on a grid of 2.7 by
3.4 by \SI{1.6}{\micro\meter}, also raised the bulk modulus of the solvent on scales above the grid by
\SI{71}{\pascal}. Its net force on the solvent stayed within 1.3\% of the drive up to 11.3 (D) and
\SI{9.9}{\milli\second}; it then rose, along the flow, and lay between 12 and 13.6\% of the drive from
\SI{16}{\milli\second} to the end of the runs.

\textit{Readouts.} As in the tube, the wall shear stress (WSS) comes from the pair forces of the solvent and the cell
on the wall particles, averaged over windows of \SI{0.47}{\milli\second}. On the curved walls of this unit, the part of
each force tangential to the wall was spread over the wall with a Gaussian kernel of width \SI{0.5}{\micro\meter}.
The carina zone is the wall within \SI{1.5}{\micro\meter} of the tip (\SI{8.2}{\micro\meter\squared}); its WSS is
the area-weighted mean of the magnitude of the stress. A cell is on the carina from the first frame (frames every
\SI{93}{\micro\second}) in which a bilayer vertex within \SI{1.5}{\micro\meter} of the tip lies within
\SI{0.05}{\micro\meter} of the reflecting surface that keeps the membrane \SI{0.45}{\micro\meter} from the wall. The
map of fig.~\ref{fig:bifwss}A is the WSS averaged over the windows with the cell on the carina, minus the WSS at the
same point of the wall without a cell.

\textit{Results.} Without a cell, the carina zone carried \SI{1.16}{\pascal}. As each cell approached, the WSS there
rose, to \SI{1.42}{\pascal} for D and \SI{1.32}{\pascal} for E-III with the modulus of D in the window that holds
first contact (7.6 and \SI{6.2}{\milli\second}; fig.~\ref{fig:bifwss}B). Neither
cell entered a daughter. Both stayed folded over the carina, with the membrane on the reflecting surface, until the
end of the runs, for at least 15.0 and \SI{14.2}{\milli\second} (fig.~\ref{fig:bifwss}C). While they sat there, the
WSS of the carina zone fell to 0.46 (D, at \SI{11.3}{\milli\second}) and \SI{0.38}{\pascal} (at
\SI{15.5}{\milli\second}) and rose again to 0.78 and \SI{0.82}{\pascal} in the last windows. Averaged over the time
on the carina, the cells lowered the WSS most on the flanks of the carina, by up to 1.37 and \SI{1.22}{\pascal}, and
raised it most on the diverging walls beside them, 3.0 and \SI{3.6}{\micro\meter} upstream of the tip, by up to 1.31
and \SI{1.79}{\pascal} (fig.~\ref{fig:bifwss}A).

\subsection*{Wall shear stress next to a flat wall}
The results of this section are predictions of the model. Each of three cells (D, E-III, and E-III with the
cytoskeleton shear modulus of D) was carried in shear flow next to a flat wall, and the force of the solvent on that
wall was recorded.

\textit{Set-up.} One sealed cell lay in solvent between two flat walls \SI{16}{\micro\meter} apart, in a box that
repeats every \SI{24}{\micro\meter} along the flow and every \SI{16}{\micro\meter} across it. The lower wall was fixed;
the upper wall was driven by a constant force, so that the solvent carried a shear stress of \SI{1.5}{\pascal}
(shear rate \SI{1235}{\per\second} without a cell). The cell started with its centroid 4.15 (D), 4.23 (E-III) and
\SI{4.22}{\micro\meter} (E-III with the modulus of D) above the lower wall and its shortest axis tilted by \ang{45}
from the wall normal toward the flow. It was filled at rest for \SI{2.0}{\milli\second}, with the bilayer open to the
solvent, and then sealed; the run started from the linear velocity profile and lasted \SI{24.3}{\milli\second}
(shear strain 30). A penalty on the solvent density, evaluated on a grid of \SI{0.67}{\micro\meter}, raised the
bulk modulus of the solvent on scales above the grid by \SI{71}{\pascal}; its net force along the flow was zero.

\textit{Readouts.} The force of the solvent on each particle of the lower wall was averaged over windows of
\SI{0.47}{\milli\second}, binned in squares of \SI{1}{\micro\meter} centered on the cell centroid of that window, and
divided by the force on the same wall particles in a run without a cell, scaled to the applied mean of
\SI{1.5}{\pascal} (that run gave \SI{1.505}{\pascal}). The wall shear stress (WSS) maps are means over the last 50\%
of each run (25 windows, \SI{12.2}{\milli\second}); errors are standard errors over five blocks of five windows.
Within one window the cell moves 3.1 to \SI{3.3}{\micro\meter} along the wall, so the maps are smoothed over that
distance along the flow. The height of a cell is that of its centroid above the lower wall. Its motion was judged
from the membrane: the bilayer vertices within \ang{30} of the two ends of the shortest axis of the first frame were
followed through the run, and the cell was taken to tank-tread when the line between their centers (the material
axis) turned by \ang{90} or more over the averaging window while the inclination of the shape stayed steady
(standard deviation \ang{5} or less).

\textit{Results.} Each cell lowered the WSS under its upstream half, in a band across the whole width of the box,
and raised it along the center line, most toward its downstream end (fig.~\ref{fig:s6}, A to C). For E-III, the WSS
fell to 1.14 $\pm$ \SI{0.03}{\pascal} at \SI{2}{\micro\meter} upstream of its centroid and rose to
1.85 $\pm$ \SI{0.01}{\pascal} at \SI{5}{\micro\meter} downstream, against \SI{1.50}{\pascal} without a cell. E-III
with the modulus of D gave 1.22 $\pm$ \SI{0.02}{\pascal} at \SI{3}{\micro\meter} upstream and
1.80 $\pm$ \SI{0.01}{\pascal} at \SI{5}{\micro\meter} downstream, and D 1.31 $\pm$ \SI{0.01}{\pascal} at
\SI{4}{\micro\meter} upstream and \SI{4}{\micro\meter} to the side and 1.69 $\pm$ \SI{0.02}{\pascal} at
\SI{6}{\micro\meter} downstream. Averaged along the flow, the WSS was 1.58 to \SI{1.59}{\pascal} on the center line
and 1.44 to \SI{1.47}{\pascal} 5 to \SI{8}{\micro\meter} to either side of it, while its mean over the whole wall
stayed at \SI{1.50}{\pascal}, as the upper wall was driven at a constant force. Because the box repeats every
\SI{24}{\micro\meter} along the flow and every \SI{16}{\micro\meter} across it, these maps are those of a regular
array of cells at these spacings.

All three cells moved away from the wall during the runs (fig.~\ref{fig:s6}D). At the end, their centroids were 6.12
(D), 6.46 (E-III) and \SI{6.45}{\micro\meter} (E-III with the modulus of D) above the lower wall; the mid-plane lies at
\SI{8}{\micro\meter}. Over the averaging window they rose by 0.80, 0.89 and \SI{0.88}{\micro\meter} (0.066 to
\SI{0.073}{\micro\meter\per\milli\second}), with mean heights of 5.75, 6.05 and \SI{6.04}{\micro\meter}. The maps
are therefore means over these heights; the maps of the first and second halves of the window give extremes that
differ by at most 5.4\%.

All three cells tank-treaded. Over the averaging window, the shape kept its inclination to the flow (long axis at
\ang{12.3} for D, \ang{21.7} for E-III and \ang{18.0} for E-III with the modulus of D; standard deviation at most
\ang{0.3}), while the membrane turned around it, in the plane of the flow and the wall normal, by \ang{175},
\ang{269} and \ang{280}. The long axis measured \SI{16.1}{\micro\meter} for D, whose cytoskeleton shear modulus is
more than two orders of magnitude below micropipette values (Materials and Methods), \SI{8.7}{\micro\meter} for
E-III and \SI{10.6}{\micro\meter} for E-III with the modulus of D.

\begin{figure}[htbp]
\centering
\includegraphics[width=\textwidth]{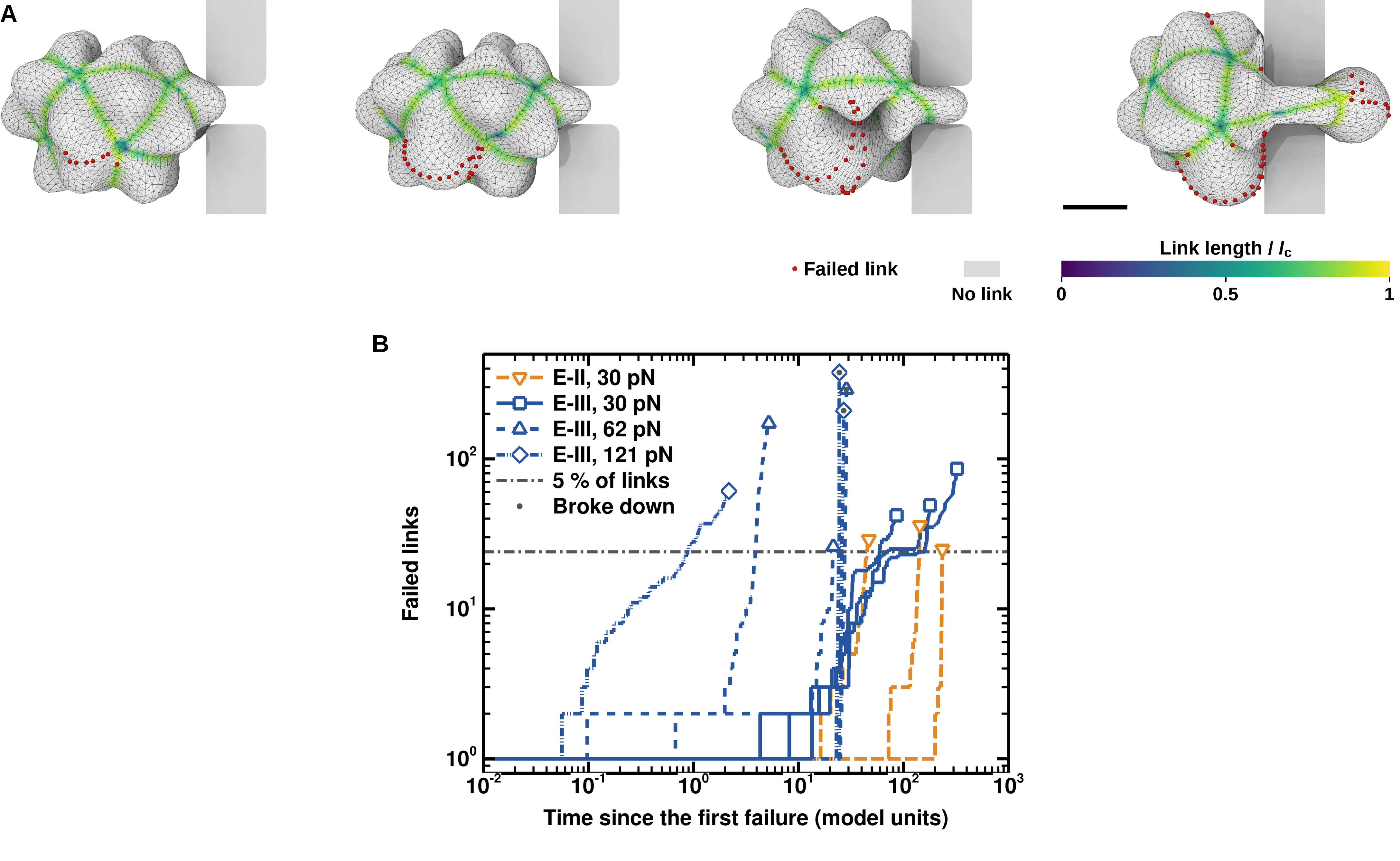}
\caption{\textbf{Weak links fail one at a time, stronger links in one burst.} (A) E-III with links of
\SI{30}{\pico\newton} at 3, 7, 10 and \SI{20}{\pascal} (left to right), with 10, 24, 34 and 81 links failed, in the
first of its three runs; the run went on past the onset of failure and stopped at \SI{20}{\pascal}, when 5\% of the
links failed within that step. The cell is seen from the side, with the flow from left to right; bilayer mesh colored by
link length over $l_c$ (gray, no intact link), wall gray. Red dots, the bilayer ends of the links failed on the side
shown. Scale bar, \SI{2}{\micro\meter}. (B) Failed links against the time since the first failure,
in every run of E-II and E-III that reached 24 failures (one curve per run). The line of each curve and the symbol at
its end give the cell and the link strength; a dot inside the symbol marks the three runs that broke down and stopped
with an error, whose last failures came as the run broke down. Gray dash-dotted line, 24 links (5\%).}
\label{fig:burst}
\end{figure}

\begin{table}[htbp]
\centering
\caption{\textbf{Pressure drop (Pa) at which 5\% of the links had failed in all, in the three runs of each cell and
link strength, with the pressure drop of the first failure in parentheses.} A dash: the count was never reached;
none: no link failed. Last column, the outcome of the ladders with unbreakable links that these runs repeat until their
first failure (Fig.~\ref{fig:slit}B): P, passed at that pressure drop (one value: the same in all three runs); H, held
up to it in all three runs, which failed at the next step. Table~\ref{tab:brackets}, which adds single-step runs,
narrows some of these outcomes: E-I passed at \SI{4}{\pascal} and E-II at 12.5 or \SI{15}{\pascal}. Runs without the
constraint that keeps the cytoskeleton inside the bilayer; table~\ref{tab:onsetcon} gives the runs with it. At
\SI{30}{\pico\newton}, E-III lost its links as fast without a pressure drop (fig.~\ref{fig:restcontrols}A).}
\label{tab:onset}
\small
\begin{tabular}{lcccc}
\hline
Cell & \SI{30}{\pico\newton} ($x_b$ 0.950) & \SI{62}{\pico\newton} (0.965) & \SI{121}{\pico\newton} (0.975) & Unbreakable \\
\hline
D     & --- (none)             & --- (none)              & --- (none)                & P at 2, 1.5, 1.5 \\
E0    & --- (none)             & --- (none)              & --- (none)                & P at 2 \\
E-I   & --- (3, none, 3)       & --- (none)              & --- (none)                & P at 5 \\
E-II  & 15, 15, 15 (3, 15, 10) & --- (none)              & --- (none)                & P at 15 \\
E-III & 7, 5, 15 (2, 2, 2)     & 50, 75, 75 (50, 50, 75) & 75, 100, 100 (75, 75, 75) & H up to 75 \\
\hline
\end{tabular}
\end{table}

\begin{table}[htbp]
\centering
\caption{\textbf{As table~\ref{tab:onset}, for the runs with the constraint that keeps the cytoskeleton inside the
bilayer.} Pressure drop (Pa) at which 5\% of the links had failed in all, in the three runs of each cell and link
strength, with the pressure drop of the first failure in parentheses. A dash: the count was never reached; none: no
link failed.}
\label{tab:onsetcon}
\begin{tabular}{lccc}
\hline
Cell & \SI{30}{\pico\newton} ($x_b$ 0.950) & \SI{62}{\pico\newton} (0.965) & \SI{121}{\pico\newton} (0.975) \\
\hline
D     & --- (none)             & --- (none)              & --- (none) \\
E0    & --- (none)             & --- (none)              & --- (none) \\
E-I   & --- (none, 3, 5)       & --- (none)              & --- (none) \\
E-II  & 15, 15, 15 (3, 15, 15) & --- (none)              & --- (none) \\
E-III & 7, 5, 15 (5, 2, 2)     & 50, 75, 50 (50, 75, 50) & 75, 100, 100 (75, 75, 75) \\
\hline
\end{tabular}
\end{table}

\begin{figure}[htbp]
\centering
\includegraphics[width=\textwidth]{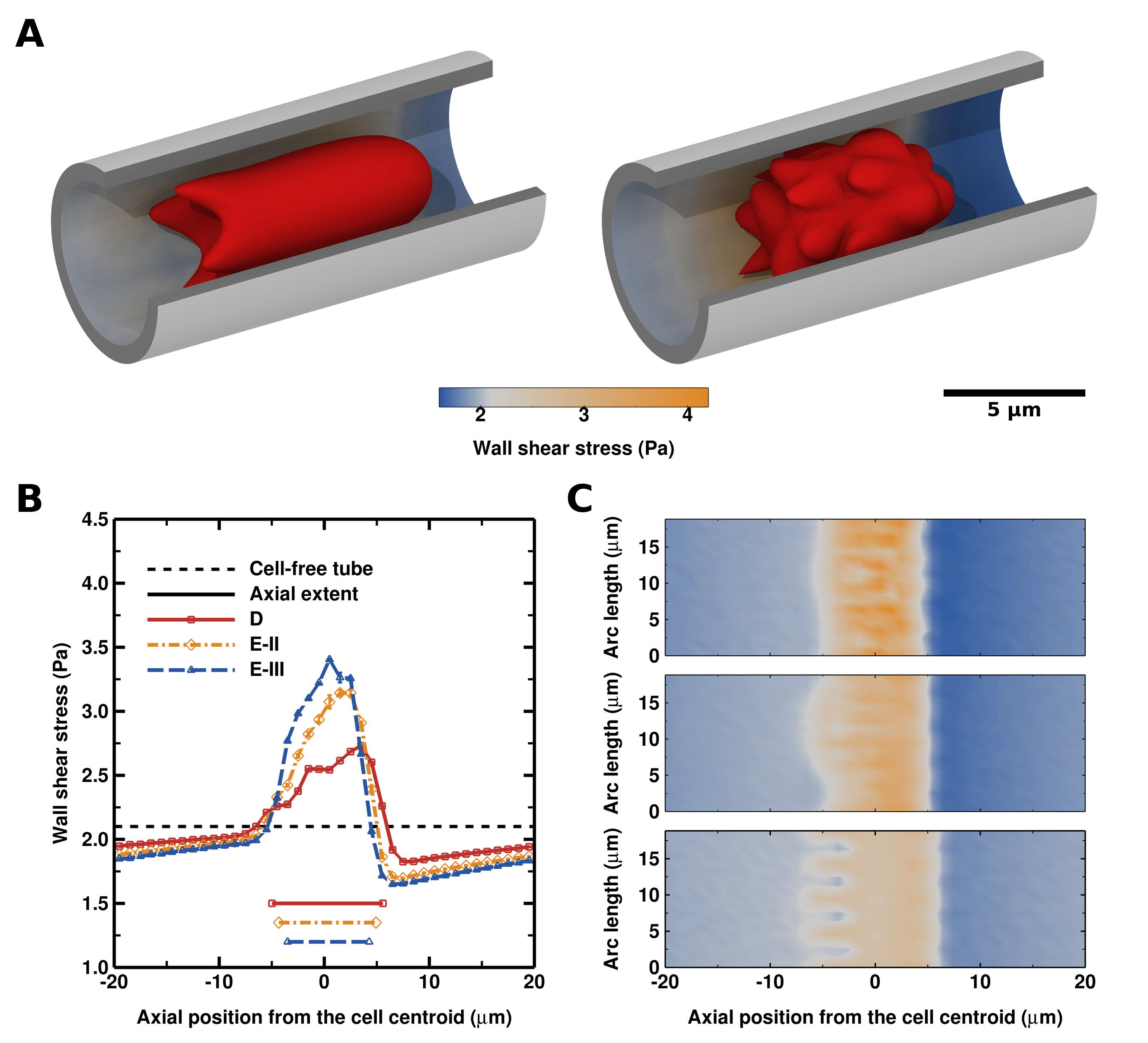}
\caption{\textbf{Wall shear stress in a \SI{6}{\micro\meter} tube peaks higher beside E-II and E-III than beside D.}
(A) D (left) and E-III (right) in frames at 60.5 and \SI{65.2}{\milli\second}, with the near quarter of the tube cut
away; bilayer red, flow from left to right; wall colored by the time-averaged wall shear stress, gray at
\SI{2.10}{\pascal}, placed around the cell of the frame. Scale bar, \SI{5}{\micro\meter}. (B) Wall shear stress averaged around the wall against the axial
position from the cell centroid, positive downstream. Dashed black line, the tube without a cell
(\SI{2.10}{\pascal}); bars below the curves, the axial extent of each cell; error bars, standard errors from four
blocks, mostly within the symbols. (C) The wall unrolled, E-III, E-II and D from top to bottom, on the color scale of
A, interpolated between the centers of bins of 1 by \SI{0.99}{\micro\meter}. Averages over 52 to
\SI{102}{\milli\second} (D, E-III) and 106 to \SI{210}{\milli\second} (E-II) of the run.}
\label{fig:tubewss}
\end{figure}

\begin{figure}[htbp]
\centering
\includegraphics[width=\textwidth]{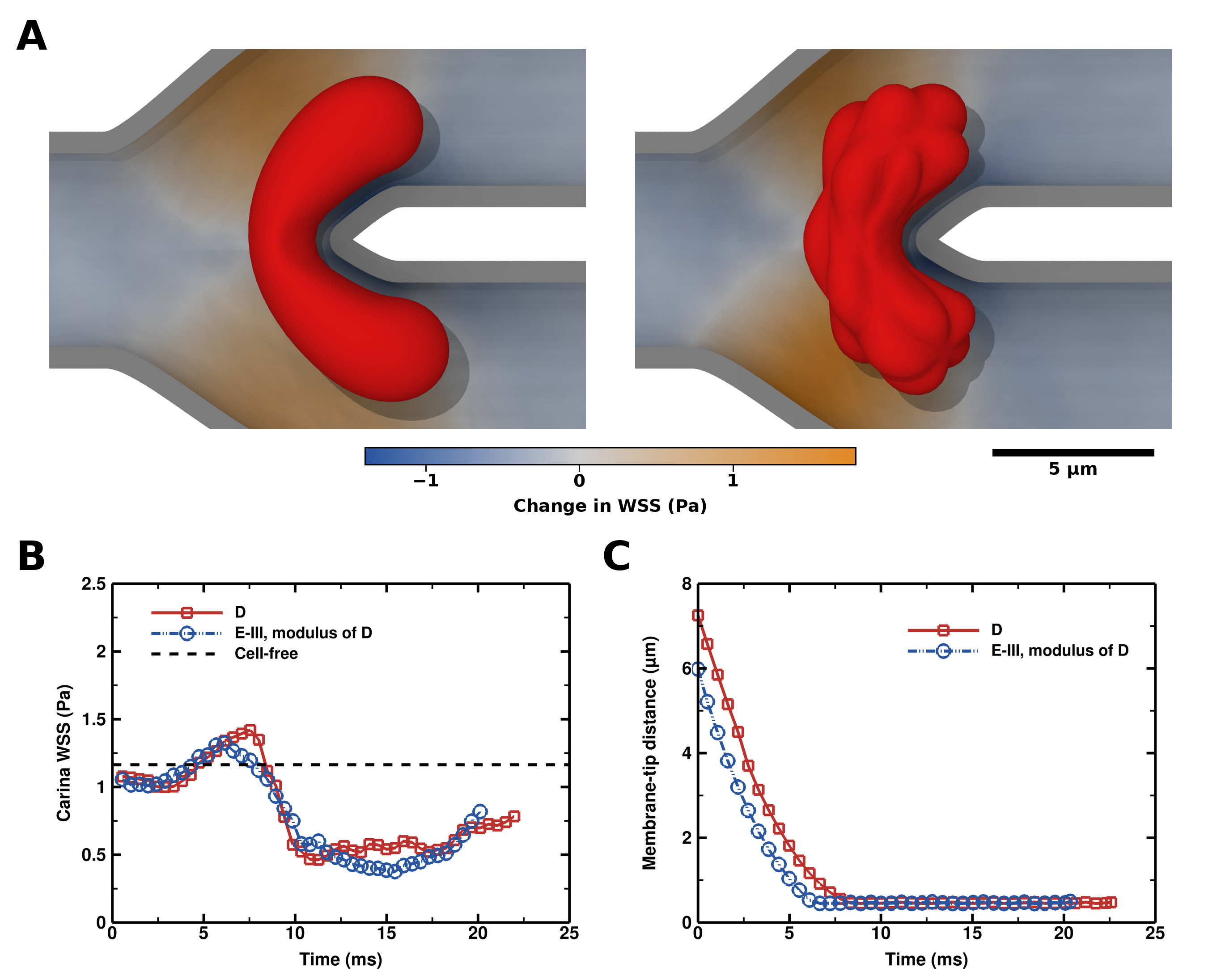}
\caption{\textbf{A cell held on the carina of a bifurcation lowers the wall shear stress there.} (A) D (left) and
E-III with the cytoskeleton shear modulus of D (right) on the carina, top views of frames at 15.1 and
\SI{13.3}{\milli\second}, wall cut at the mid-plane; bilayer red, flow from left to right. The wall is colored by the
change in the time-averaged wall shear stress (WSS) over the time on the carina, against the run without a cell
(gray, no change). Scale bar, \SI{5}{\micro\meter}. (B) WSS of the carina zone, the wall within
\SI{1.5}{\micro\meter} of the tip, in windows of \SI{0.47}{\milli\second}; dashed line, without a cell
(\SI{1.16}{\pascal}). (C) Smallest distance between the membrane and the tip; the reflecting surface keeps it at
\SI{0.45}{\micro\meter} or more.}
\label{fig:bifwss}
\end{figure}

\begin{figure}[htbp]
\centering
\includegraphics[width=\textwidth]{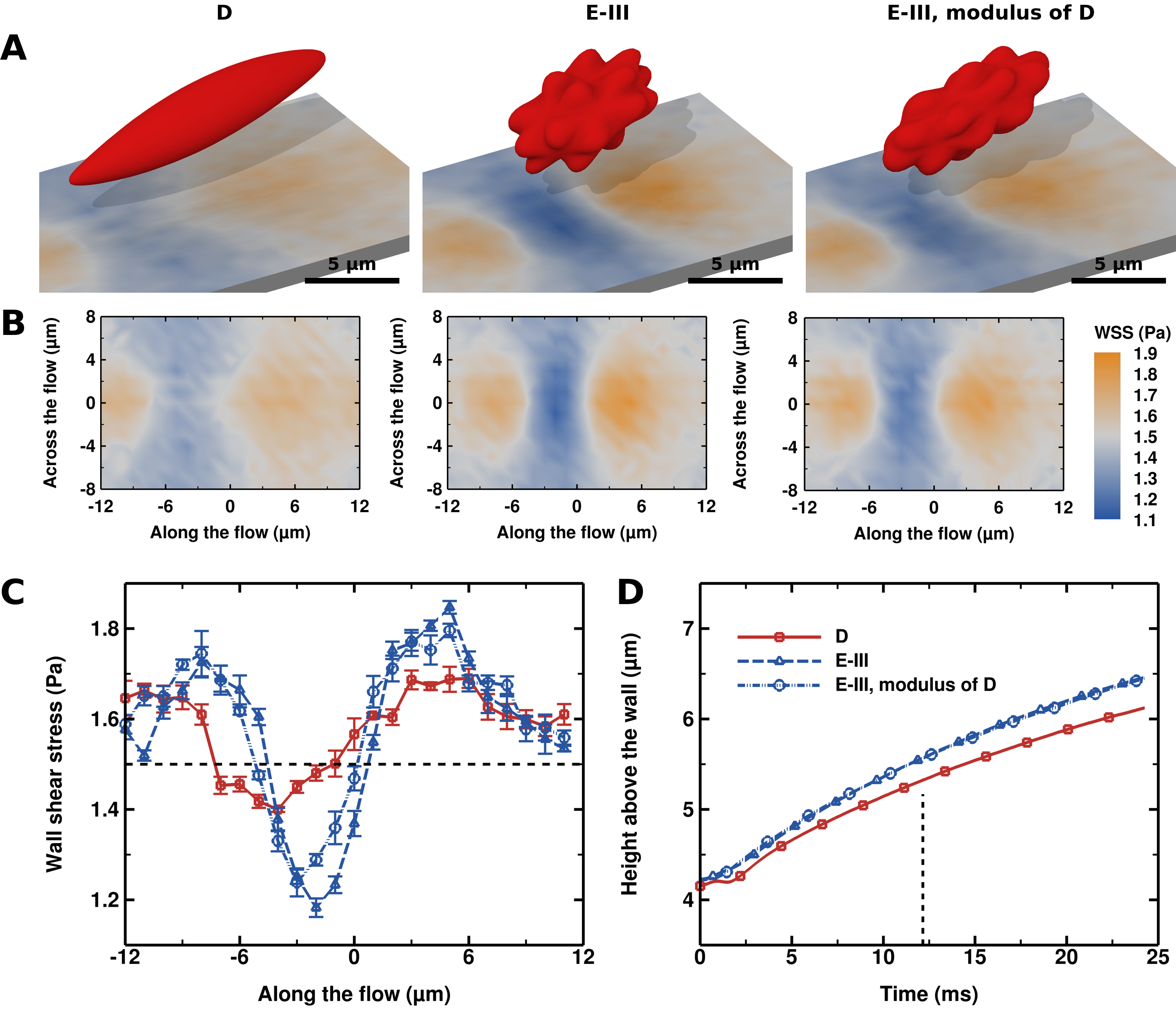}
\caption{\textbf{Cells sheared next to a flat wall change the shear stress on it.} The wall shear stress (WSS) without
a cell is \SI{1.50}{\pascal}. (A) One frame of each cell from the last 50\% of its run, above one period of the lower
wall (24 $\times$ \SI{16}{\micro\meter}), oblique view with the flow from left to right; bilayer red. The wall is
colored by the time-averaged WSS of B, gray at \SI{1.50}{\pascal}, placed at the cell centroid of that frame. The upper wall,
\SI{16}{\micro\meter} above, is not shown. Scale bars, \SI{5}{\micro\meter}. (B) Time-averaged WSS on the lower wall
in \SI{1}{\micro\meter} bins around the cell centroid, over the last 50\% of each run (\SI{12.2}{\milli\second});
one color scale for A and B. (C) WSS along the center line of B; bars, standard errors over five blocks of the
averaging window; dashed line, \SI{1.50}{\pascal}. Lines and symbols as in D. (D) Height of the cell centroid above
the lower wall; dashed line, start of the averaging window.}
\label{fig:s6}
\end{figure}

\clearpage
\paragraph*{Caption for Movie S1 (ancillary file Movie\_S1.mp4).}
\textbf{A discocyte passes the slit.} D (seed 3) on its ladder of pressure drops (Fig.~\ref{fig:slit}): held at 0.5
and \SI{1}{\pascal}, it passes at \SI{1.5}{\pascal} after 467.5 time units at that level. Side view as in
Fig.~\ref{fig:slit}A, with the near half of the wall removed; bilayer red; the cell moves from left to right. The camera
is fixed (frame \SI{21}{\micro\meter} wide). The movie starts with the cell at the slit entrance, as the ladder begins
(the 20 time units of equilibration and the push to the slit are not shown); times are in model time units
(Materials and Methods). One frame every 4 time units; 746 time units in all.

\paragraph*{Caption for Movie S2 (ancillary file Movie\_S2.mp4).}
\textbf{A stage II echinocyte passes the slit at a higher pressure drop.} As movie S1, for E-II with its own modulus
(seed 1, the repeat with the stronger bilayer self-repulsion): held at 2, 3, 5, 7 and \SI{10}{\pascal}, it passes at
\SI{15}{\pascal} after 63.5 time units at that level; the movie ends 0.5 time units later. One frame every 2 time
units; 362 time units in all.

\paragraph*{Caption for Movie S3 (ancillary file Movie\_S3.mp4).}
\textbf{A stage III echinocyte is held in the slit.} As movie S1, for E-III (seed 2): held at every level from 2 to
\SI{75}{\pascal}; the run failed at \SI{100}{\pascal} (not shown). One frame every 2 time units; 358 time units in all.

\paragraph*{Caption for Movie S4 (ancillary file Movie\_S4.mp4).}
\textbf{Links fail in a stage III echinocyte.} E-III with links of \SI{30}{\pico\newton}, in the run of
fig.~\ref{fig:burst}A. The movie starts with the cell at the slit entrance (the equilibration and the push into the
slit are not shown) and runs through the ladder of pressure drops from 2 to \SI{20}{\pascal}. The
cell is cut through the mid-plane of the slit and one half is shown, seen from the cut, with the flow from left to
right (fig.~\ref{fig:burst}A shows this half from outside, the other way up); bilayer red, cytoskeleton blue, wall gray; black dots, the bilayer ends of the links failed in this half. The
camera is fixed (frame \SI{13.5}{\micro\meter} wide). The movie ends at the last saved frame, with 81 links failed; the
run stopped 1.5 time units later, with 86. One frame every 2 time units; 330 time units in all.

\clearpage
\bibliographystyle{ScienceAdvances}
{\let\bibOld\thebibliography\renewcommand{\thebibliography}[1]{\bibOld{#1}\setlength{\itemsep}{1pt plus 1pt}}
\bibliography{ref_main}}

\end{document}